\documentclass[12pt,aps,pra,groupedaddress,amssymb,amsfonts,nofootinbib,tightenlines,floatfix]{revtex4} 
\usepackage{graphicx} 

\graphicspath{{./}{graphics/}}
\usepackage{times,color}
\definecolor{purple}{rgb}{0.625,0.125,0.9375}

\usepackage{ekqc_b}

\ignore{
rm -R /tmp/analysis; mkdir /tmp/analysis
cp `texfls analysis.log | perl -e '$a=<>; $a =~ s:/\S*(revtex4|natbib)\S*(\s|$)::g; print "$a\n"'` /tmp/analysis
find /tmp/analysis -type f -name '*.pdf' -exec pdftops -eps {} \; -exec rm {} \;
find /tmp/analysis -type f -name '*.jpg' -exec perl -e '$f=shift; $g=f$; $g=~s/.jpg/.eps/; exec("convert $f $g");' {} \; -exec rm {} \;
perl -i -n -e 's/^

cp /usr/share/texmf/tex/latex/tools/calc.sty /tmp/analysis/phcalc.sty
perl -i -n -e 's/ProvidesPackage\{calc\}/ProvidesPackage\{phcalc\}/; print;' /tmp/analysis/phcalc.sty
perl -i -n -e 's/usepackage\{calc\}/usepackage\{phcalc\}/; print;' /tmp/analysis/ekqc_b.sty

(cd /tmp/analysis; tar czvf analysis.tar.gz *)
}

\ignore{
rm -R /tmp/analysis; mkdir /tmp/analysis
cp `texfls analysis.log | perl -e '$a=<>; $a =~ s:/\S*(revtex4|natbib)\S*(\s|$)::g; print "$a\n"'` /tmp/analysis
find /tmp/analysis -type f -name '*.pdf' -exec pdftops -eps {} \; -exec rm {} \;
find /tmp/analysis -type f -name '*.jpg' -exec perl -e '$f=shift; $g=f$; $g=~s/.jpg/.eps/; exec("convert $f $g");' {} \; -exec rm {} \;
perl -i -n -e 's/^
perl -i -n -e 's/graphics\///; print;' /tmp/state/analysis.tex

cp /usr/share/texmf/tex/latex/tools/calc.sty /tmp/analysis/phcalc.sty
perl -i -n -e 's/ProvidesPackage\{calc\}/ProvidesPackage\{phcalc\}/; print;' /tmp/analysis/phcalc.sty
perl -i -n -e 's/usepackage\{calc\}/usepackage\{phcalc\}/; print;' /tmp/analysis/ekqc_b.sty

cp /usr/share/texmf/tex/latex/base/ifthen.sty /tmp/analysis/phifthen.sty
perl -i -n -e 's/ProvidesPackage\{ifthen\}/ProvidesPackage\{phifthen\}/; print;' /tmp/analysis/phifthen.sty
perl -i -n -e 's/usepackage\{ifthen\}/usepackage\{phifthen\}/; print;' /tmp/analysis/ekqc_b.sty

cp retestate_fns.math /tmp/analysis

(cd /tmp/analysis; tar czvf analysis.tar.gz *)
}

\begin{document}
\title{Fault-Tolerant Postselected Quantum Computation: Threshold Analysis}
\author{E. Knill}
\email[]{knill@boulder.nist.gov}
\affiliation{Mathematical and Computational Sciences Division,
National Institute of Standards and Technology, Boulder CO 80305} 

\date{\today}

\begin{abstract} The schemes for fault-tolerant postselected quantum
computation given in [Knill, Fault-Tolerant Postselected Quantum
Computation: Schemes, \texttt{http://arxiv.org/abs/quant-ph/0402171}]
are analyzed to determine their error-tolerance. The analysis is based
on computer-assisted heuristics. It indicates that if classical and
quantum communication delays are negligible, then scalable qubit-based
quantum computation is possible with errors above $1\;\%$ per
elementary quantum gate.
\end{abstract}


\maketitle

\ignore{
{\color{red}
Todo:
\begin{itemize}
\item How much closer to independence can I get by one
entanglement swapping step?
\end{itemize}
}
}

\section{Introduction}
\label{sect:introduction}

In~\cite{knill:qc2004a} a scheme for fault-tolerant postselected
quantum computation based on concatenating four-qubit error-detecting
codes is presented. Here, the error behavior of the scheme is analyzed
with the help of computer-assisted heuristics and shown to tolerate
errors above $1\;\%$ per elementary quantum gate.  This result
generalizes to standard quantum computation because fault-tolerant
postselected quantum computation can be used for preparing stabilizer
states with well constrained local errors. With this technique, any
sufficiently error-tolerant stabilizer code can be used with
teleported error correction~\cite{knill:qc2003b} to realize a general
fault-tolerant computation.  In particular, the analysis indicates
that the error threshold for scalable quantum computation is also
above $1\;\%$.  This substantially improves thresholds obtained
previously, both heuristically and by
proof~\cite{shor:qc1996a,kitaev:qc1996a,knill:qc1996b,aharonov:qc1996a,aharonov:qc1999a,knill:qc1997a,gottesman:qc1997a,knill:qc1998a,zalka:qc1996a,preskill:qc1998a,steane:qc1999a,gottesman:qc1999a,gottesman:qc2000b,knill:qc2000e,hughes:qc2002a,aharonov:qc2002a,duer:qc2002a,steane:qc2002a,steane:qc2002a,steane:qc2002b,steane:qc2003a}.

The analysis given below assumes familiarity with the previous two
papers in the series~\cite{knill:qc2003b,knill:qc2004a}.
See~\cite{knill:qc2003b} for motivation, background, and a brief review
of the needed stabilizer code theory and teleported
error correction. For details of the scheme analyzed here
see~\cite{knill:qc2004a}.

The following conventions are used: A \emph{postselected computation}
is one whose output is conditional on the outcomes of measurements of
ancilla qubits used during the computation. The usual model of quantum
computation is referred to as \emph{standard} quantum computation.  A
scheme for fault-tolerant computation starts with \emph{physical
qubits} and an error model for a universal set of quantum gates
(including measurement and state preparation). The schemes discussed
here involve the construction of \emph{logical qubits} to be used in a
general-purpose computation. Logical qubits are manipulated with
\emph{logical gates}.  The construction involves one or more levels of
encoding, each of which defines higher-level logical qubits in terms
of lower-level ones. In a given context, the term ``logical'' refers
to the highest-level logical qubits currently being discussed.  The
terms \emph{encoded qubit}, \emph{encoded state} and \emph{encoded
gate} are used to refer to encoded quantum information if the emphasis
is on the code used rather than on logical computation. The most
important state used for implementing fault tolerance is the encoded
Bell pair. This state consists of two encoded qubits in the standard
Bell state, ${1\over\sqrt{2}}(\ket{\biz\biz}+\ket{\bio\bio})$.  Thus,
there are two blocks of physical qubits, one for each encoded
qubit. The two blocks and their encoded qubits are referred to as the
\emph{origin} and \emph{destination} blocks and qubits, respectively.
The Bell pairs are used in teleportation by making a transversally
implemented Bell measurement on the block to be teleported and on the
origin block. Most of the states under consideration are $n$-qubit
stabilizer states characterized by an $n\times 2n$ binary generator
matrix whose rows specify a generating set of stabilizing Pauli
products.  Errors in the state are characterized by the
\emph{syndrome}, a binary $n$-vector that specifies the eigenvalues
of the stabilizing Pauli products.  The error-free state is
standardized to have a syndrome (vector) of $\mb{0}$. When talking
about encoded qubits, the generator matrix has fewer than $n$
rows. The syndrome specifies the eigenvalues of the Pauli products
determined by the rows of the generator matrix.  The following
abbreviations are used: $X=\sigma_x$, $Z=\sigma_z$,
$Y=\sigma_x\sigma_z$. The states $\ket{\biz}$, $\ket{+}$ and
$\ket{i\pi/4}$ are, respectively, the $+1$ eigenstates of $Z$, $X$ and $iY$.
$\ket{\pi/8}$ is the $+1$ eigenstate of the Hadamard
gate.

\section{Assumptions}
\label{sect:assumptions}

As is customary for theoretical threshold analyses, it is assumed that
it is possible to apply quantum gates in a parallel fashion without
geometrical constraints.  In particular, in each time step, any pair
of qubits can be acted on with a controlled not (cnot) gate (no
quantum communication latency), and all qubits may be involved in a
non-trivial operation. All gates (one-qubit state preparations,
Hadamard gates, cnots, one-qubit measurements) are assumed to take the
same amount of time. As discussed in~\cite{knill:qc2004a}, for the
postselected computation scheme used here, gate times can vary between
gate types.  In particular, the time needed for a measurement may be
longer than that for the cnot. Variations in gate time have an effect
only when prepared states are used in a standard quantum computation,
where an additional memory error may need to be accounted for.  It is
assumed that classical computation based on measurement outcomes is
instantaneous and error-free (no classical communication or
computation latency).  This condition is needed when prepared states
are to be used for standard quantum computation, as it is necessary to
determine whether all postselection criteria have been met and (for
efficient postselection) what the current syndrome is for the prepared
state.

The error model used here is based on having all errors be
probabilistic Pauli products acting independently at error locations
associated with each gate. Specifically, a quantum computation is
described in terms of its quantum network.  Each instance of a given
quantum computation is modified by probabilistically inserting Pauli
operators after each network element and before measurements. To
describe the constraints on the probabilities, an \emph{error
location} is associated with each network element.  For a
state-preparation and the Hadamard gate, the error location is in the
qubit line immediately following the state-preparation gate. For the
cnot, it extends to both qubit lines immediately after the gate. For
measurements, it is on the qubit to be measured just before the
gate. For each Pauli operator $p$ (or product of two Pauli operators
in the case of the cnot) that can act at an error location $l$, there
is an associated error parameter $e_{p,l}$. The strongest assumption
one can make is that errors are probabilistic and independent between
locations, with $e_{p,l}$ the probability that $p$ acts at location
$l$.

The heuristic used for the analysis involves a method that uses an
independent-error model to approximate the error behavior in prepared
states.  This method is used at each level of a concatenation
procedure. Given independence at the lower level, the true error
behavior is close to, but not exactly, independent.  Ideally, the
independent-error model used should not underestimate any error
probabilities after postselection.  The motivation for using such an
independent approximation comes from a classification of error models
that retain some of the crucial properties of independence.  In
particular, the independence assumption can be weakened in three
steps.  Consider sets of error locations $L$ and $K$ with $|L|+1=|K|$,
such that there is an injection $\pi$ from locations in $L$ to
equivalent locations in $K$. Locations are considered equivalent if
they are associated with identical gates.  Let $r$ be the location in
$K$ not in the image of $\pi$.  Let $p_l$ and $q_k$ be (non-identity)
Pauli operators acting at locations $l\in L$ and $k\in K$,
respectively, with the property $q_{\pi(l)}=p_l$.  The error model is
independent$(A)$ if it is always the case that the relative
probability $\textrm{Prob}(\Pi_{k\in K} q_k)/\textrm{Prob}(\Pi_{l\in L} p_l)$
is bounded by $e_{q_r,r}$, and $e_{q_r,r}$ depends only on the type of
error location and $q_r$. (The notation ``$(A)$'' appended to
``independent'' modifies the meaning of independent.)
Here, products such as $\Pi_{k\in K} q_k$ informally denote the
combination of error events where $q_k$ occurs at its location $k$.
If in this definition, $L$ is restricted to subsets of $K$, and $\pi$ is
the identity map, then the model is independent$(B)$.  In particular,
$\textrm{Prob}\Pi_{k\in K}q_k/\textrm{Prob}(\one)$ is at most
$\Pi_{k\in K} e_{q_k,k}$, where $\textrm{Prob}(\one)$ is the
probability of no error at any of the error locations. An error model
satisfying only this last property is independent$(C)$ or
\emph{quasi-independent}~\cite{knill:qc1997a}.  The heuristic analysis
reestablishes full independence at each stage and can be considered as
an attempt to capture and maintain independence$(A)$.

In the three error models of the previous paragraph, the $e_{p,l}$ are
ratios of error probabilities, so they are called \emph{error
likelihoods} for the remainder of this paper. If the errors are
independent for different locations and the error parameters are
chosen optimally, then the (marginal) probability of $p\not=\one$
acting at $l$ is given by $e_{p,l}/(1+\sum_{q\not=\one} e_{q,l})$. In
particular, the error likelihood is larger than the marginal error
probability.  The error likelihoods $e_{q,l}$ are assumed to depend
only on the type of gate with which the error location $l$ is
associated. The analysis is implemented so that any such
gate-dependent combination of $e_{q,l}$ can be used as the physical
error model.  For the examples given, a uniform-error model is used
for physical gates. The computationally determined logical-gate error
models deviate significantly from the uniform-error model.  For the
primary scheme, only cnot, preparation and measurement errors play a
role. These are parameterized by likelihoods $e_c$, $e_p$ and $e_m$.
For a cnot's error location, there are 15 possible non-identity Pauli
products, each of which has likelihood $e_c/15$. The total probability
of error is $e_c/(1+e_c)$. For preparation and measurement of $X$ and
$Z$ eigenstates, only one type of Pauli-operator error needs to be
considered ($Z$ and $X$ errors, respectively).  For universal quantum
computation, the Hadamard gate and additional state preparations are
needed.  For a Hadamard gate, each of the three non-identity Pauli
operators has likelihood $e_h/3$. The additional state-preparations'
errors can also be described as applications of one Pauli operator.
The error for preparation of $\ket{i\pi/4}$ ($\ket{\pi/8}$,
respectively) is taken to be random-$Z$ ($Y$, respectively) error with
a likelihood of $e_s$.  To reduce the parameter space, set
$e_s=e_p=e_m$ and $e_h/(1+e_h) = 3e_p/(2(1+e_p))$. This is consistent
with the assumption that errors of one-qubit gates are depolarizing
and independent of the gate.

Since the analysis assumes probabilistic errors, it does not directly
address the question of what happens when there are coherences between
different error combinations expressed as Pauli products.  Such
coherences cannot be ignored. For example, decay of $\ket{\bio}$ to
$\ket{\biz}$ cannot be expressed as a probabilistic Pauli-operator
error.  In this case, the likelihoods have to be replaced by
(relative) amplitudes.  In principle, error amplitudes can be
formalized by means of environment-labeled error sums,
see~\cite{knill:qc1997a,knill:qc2001d}.  Previous threshold analyses
that take this into account suggest that results obtained for
probabilistic errors extend to independent$(C)$ non-probabilistic
errors, but it may be necessary to square the maximum tolerable errors
in the worst case, see~\cite{terhal:qc2004a,alicki:qc2004a}.  For
example, if $0.01$ probability of error is acceptable in the
probabilistic model, then in the coherent model, this may have to be
reduced to $0.0001$. However, this conclusion seems overly
pessimistic.  First, error control based on stabilizer techniques is
designed to remove coherences between differently acting combinations
of Pauli-product errors as fast as possible, so there should be little
opportunity for coherences to add as amplitudes rather than as
probabilities. Second, at the cost of increasing the error per
operation, it is possible to deliberately randomize the computation by
applying canceling random Pauli products before and after Hadamard and
cnot gates. Random stabilizing Pauli operators may be applied after
state preparations and before measurements. This ensures that
interference between error amplitudes is well suppressed even before
error detection or correction is used. This is a version of the
twirling method used in entanglement
purification~\cite{bennett:qc1996a}. Finally, it is interesting to
note that one of the error-detecting codes used here is the same as
the amplitude damping code given on page 99 of Gottesman's
thesis~\cite{gottesman:qc1997b}.

Another type of error not considered by the probabilistic model is
leakage and qubit-loss error, whereby amplitude is lost from a qubit
to an orthogonal state space. Fortunately, the teleportation based
methods solve this problem implicitly by constantly refreshing the
qubits used. The effect of leakage on the Bell state measurements
is no different from other measurement errors. For leakage due to
qubit-loss, there is in addition the opportunity to detect the error,
which can be taken advantage of in
error-correction~\cite{knill:qc2003b}.

\section{Analysis Outline}
\label{sect:analysis outline}

Because the basic codes are CSS codes, encoding and decoding
operations require only cnots, $X$- and $Z$-postselected measurements,
and preparation of the $+1$ eigenstates of $X$ ($\ket{+}$) and $Z$
($\ket{\biz}$).  Therefore, the first task of the scheme is to use
concatenation to implement these operations with arbitrarily low error
after postselection.  Errors are controlled by purifying prepared
states at each level and by using teleported error-detection in
conjunction with each operation implemented. At all times, the total
state is a CSS state, and error behavior can be tracked by computing
the likelihoods of each possible syndrome. Subroutines that compute
likelihoods either formally or with high-precision arithmetic were
implemented using a standard computer algebra language and are
described in Sect.~\ref{sect:functions}.  Because there are
exponentially many syndromes, it is unfeasible to track them
explicitly except for small numbers of qubits at a time.  To analyze
the scheme, the error behavior at each level is abstracted by modeling
it in terms of a \emph{logical-error model} associated with encoded
gates applied at the current top level, and a \emph{terminal-error
model} of errors acting on encoded qubits at each level after the
logical errors. The heuristic involves using independent errors for
the two models.  The error likelihoods are chosen with the intention
of bounding the true error likelihoods from above after normalizing
the likelihood of having no error anywhere to $1$.  The error models
are described in Sect.~\ref{sect:current state error bounds}. The
logical- and terminal-error models are constructed level by level. The
main step of this construction is to explicitly compute the error
behavior of the purified encoded Bell pair needed to implement gates
at the next level.  Because the Bell pair involves only two blocks of
4 qubits and a purification step uses two such pairs at a time
(involving a total of 16 qubits), this can be done to arbitrary
precision with reasonable computational resources. The original scheme
suggested using one purification cycle consisting of two purification
steps to prepare a ``good'' encoded Bell pair.  Although this reduces
probabilities of undesirable syndromes (those involving an
undetectable encoded error) to second order, it is insufficient for
good independence properties. For the purpose of analysis, this is
remedied by a second purification cycle. At this point, the true error
in the encoded Bell pair has the same order behavior as independent
error applied to each qubit. The heuristic replaces the true error
model by a block-wise independent \emph{Bell-error model}. The
likelihoods for the Bell-error model are computed so as to approximate
relative likelihoods of the computed model without any obvious biases
that may reduce predicted error probabilities. The procedure is
discussed in Sect.~\ref{sect:preparation of the encoded Bell pair}.
Block-wise independence is needed to ensure that each logical gate's
error is independent and independent of the terminal-error model.
With this assumption, the teleported implementation of gates ensures
that different steps of a logical computation are isolated. The
independence of the Bell error model also makes it possible to
directly compute logical-error likelihoods for state preparation and
measurement (Sect.~\ref{sect:state preparation}). Similarly, the
logical-error likelihoods for the cnot can be computed.  However,
because the cnot is implemented directly rather than by explicitly
using lower-level encoded cnots, the computation depends on whether it
is for the physical level or for higher levels. This is explained in
Sect.~\ref{sect:controlled-not}. The main application of postselected
fault-tolerant quantum computation is to the preparation of arbitrary
stabilizer states in a fault-tolerant manner.  To do so one first
prepares the desired state in encoded form at the top level of the
concatenated error-detecting codes.  One then decodes the error-detecting
codes, accepting the result only if no errors are detected.  The error
associated with decoding is, according to the heuristic, independent
between logical qubits. How to compute it at each level of the
concatenation hierarchy is discussed in Sect.~\ref{sect:decoding a
prepared state}.  To complete the analysis, it is necessary to
determine the error of encoded Hadamard gates and of encoding the two
states needed for universal quantum computation.  This is done in
Sect.~\ref{sect:hadamard gate} and Sect.~\ref{sect:encoding an
arbitrary state}. The results for specific error models and the
implications for scalable quantum computations are given in
Sect.~\ref{sect:general threshold}. The results are discussed in
Sect.~\ref{sect:discussion}.

\section{Representing the error behavior}
\label{sect:current state error bounds}

The scheme for postselected quantum computation is to be analyzed one
level at a time. At a given level, an encoded postselected computation
is implemented. Although the entire computation occurs in parallel,
the analysis considers the computation sequentially from state
preparation onward. Thus, it makes sense to talk about the current state
$\ket{\psi}$ of the logical computation.  $\ket{\psi}$ is a state of
$n$ logical qubits, each encoded with $4^h$ physical qubits, where $h$
is the current level of the concatenation hierarchy.  (Level
$0$ is the physical level.) $\ket{\psi}$ is modified from the ideal,
error-free state $\ket{\psi_I}$ by errors. The effect of these errors
is separated into \emph{logical errors}, which can be attributed to
errors in the steps of the logical computation so far, and
\emph{terminal errors}, which act after the logical errors and directly on
encoded qubits at each level.  Most terminal errors are detectable and
are eliminated in the next step of the computation.

Logical errors are modeled by independent errors acting at error
locations associated with logical gates. Like the physical-error
model, they are characterized by likelihoods of Pauli products for
locations associated with each gate and depend only on the type of
gate, not the specific instance. logical errors on the
current state of the computation affect solely the logical qubits. They
can be detected only by implementing the next level's operations.

Terminal errors are modeled as errors acting after, and independently
of, the logical errors. Consider the blocks of four physical qubits
encoding first-level qubits. The terminal error model $T_0$ consists
of a fixed distribution of four-qubit Pauli products acting
independently on each block. The distribution can be simplified using
the stabilizer for each block. For example, if the spectator qubit is
in the state $\kets{\biz}{S}$ (see~\cite{knill:qc2004a}), then the
stabilizer is generated by $[XXXX],[ZZZZ]$ and $[IIZZ]$, and consists of $8$
Pauli products. As a result, The $2^8$ four-qubit Pauli products
consist of $32=2^5=2^{8-3}$ distinct cosets of the stabilizer, where
each coset has the same effect on the block. Thus $T_0$ is defined by
associating a likelihood to each of the $31$ non-identity cosets,
normalizing the likelihood of the identity coset to $1$.  These
likelihoods must be computed based on how the states of the blocks
are obtained. In this case, each block originates from the
destination block of an encoded Bell pair used in a teleportation
step.  The explicitly computed likelihoods of the encoded Bell state's
errors are seen to satisfy some independence criteria between the two
blocks.  In fact, the computed likelihoods are not far from what would
be obtained by acting on each qubit independently.  This ensures that
using a block-wise independent error model to bound the likelihoods is
at least reasonable.  That the computed likelihoods are related to a
qubit-independent error model can be used for choosing likelihoods for
the $31$ non-identity cosets. The likelihoods thus obtained form the
Bell-error model. The part acting on the destination blocks determines
the level-$1$ terminal-error model.  For each level $l\geq 2$,
further independent terminal errors are obtained by computing the
error likelihoods for encoded Bell states prepared at this level by use of
lower level gates. The errors of $T_{l-1}$ act on level-$(l-1)$ qubits
that form a block encoding level $l$ qubits.  The independent
combination of the errors of $T_l$ up to level $h$ constitutes the
terminal-error model for level $h$. If physical errors are
sufficiently low for the scheme to scale, each contribution $T_l$ to
the terminal error model has the property that errors with significant
likelihood change the syndrome for the code at level $l$, making them
detectable in future steps. Errors that change the encoded qubit
without affecting the syndrome are at least quadratically suppressed.
Although the actual state of the computation is always extremely noisy
due to the terminal error model, conditional on the syndrome of
the concatenated error-detecting code being $\mb{0}$, the state is
reliable. When using the state in a destructive measurement or by
decoding the error-detecting code, the relevant parts of the syndrome
are verified so that postselection eliminates the non-$\mb{0}$
syndrome events to the extent that it is necessary to do so.

\section{Computational methods}
\label{sect:functions}

There are two principal tasks that require computer assistance.  The
first is to determine the error likelihoods of syndromes for
stabilizer states given a network to prepare them and given the error
likelihoods associated with gates and other states used by the
network. The second is to approximate the error likelihoods computed
for the purified encoded Bell pairs by block-wise independent
likelihoods.  Functions to perform these tasks were implemented using
a standard computer-algebra system. The functions are included with
the electronic files submitted to \texttt{quant-ph (available at
http://arxiv.org/)} and are also available by request from the
author. For the first task, a stabilizer state with error is
represented by an array of error likelihoods with one entry for each
syndrome value. The stabilizer state is specified by the generator matrix
$Q$ for the stabilizer as defined in~\cite{knill:qc2003b}. The
syndrome is a $0-1$ vector characterizing the eigenvalues of the Pauli
product operators represented by the rows of $Q$. For a stabilizer
state on $n$ qubits, the generator matrix is $n\times 2n$ and the
syndrome is an $n$-vector.  Thus, a stabilizer state with error is
specified by the pair $Q$ and an array of $2^n$ likelihoods, one for
each of the $2^n$ possible values of the syndrome. The probabilities
can be derived from the likelihoods by dividing by the sum of the
$2^n$ likelihood values.  For flexibility, the functions were
implemented so that the likelihoods could be computed as polynomials
of a single error parameter. Typically, higher degree terms of these
polynomials contribute very little, and to take advantage of this, the
computations estimate the highest-degree term by making use of an
upper bound on the value of the error parameter.  These formal
computations were used to verify that undetected errors are indeed
suppressed to second order. However, for determining whether a given
error rate is below the threshold, high precision arithmetic was used.

To compute syndrome likelihoods for states produced by small networks,
the following four functions are needed: 

1. Adding a new qubit in the state
$\ket{\biz}$ or $\ket{+}$. This involves expanding the generator
matrix, adding a new row, and expanding the likelihood array by a
factor of two. 

2. Applying error-free cnots and Hadamard gates. This
is a transformation of the generator matrix that requires
multiplication on the right by the appropriate linear transformation.
The syndrome array is not affected. 

3. Applying errors according to an
error model. Error models include those corresponding to applying
Pauli products to an error location involving one or two qubits
according to a likelihood distribution.  This was implemented by
subroutines that compute the effect of Pauli products on syndromes
(the generator matrix is unchanged) and that are able to convolve the
error model's likelihoods with the syndrome likelihoods. 

4. Projecting
a qubit's state onto the $\ket{0}$ or $\ket{+}$ states. This is the
step involving postselection. It requires a row transformation of the
generator matrix with an associated transformation of the syndrome
array to ensure that only the last row (if any, case 1) involves an
operator not commuting with the qubit projection.  If there are no
such rows (case 2), then the row transformation is such that the first
$n-1$ rows involve the identity operator acting on the projected
qubit. In case 2, the transformation is arranged so that the last
row's operator acts only on the projected qubit.  The syndrome array
is then reduced.  In case 1, the new syndrome likelihoods are obtained
as sums of the two likelihoods associated with the two possible
eigenvalues of the operator represented by the last row.  In case 2,
the new syndrome likelihoods are given by the likelihoods associated
with eigenvalue $+1$ of the operator represented by the last row.
After the new syndrome array has been obtained, the two columns
corresponding to the projected qubit and the last row are deleted from
the generator matrix.

In addition to the four functions above, several
auxiliary functions were implemented. For example, it is convenient to
be able to apply any desired row transformation to the generator
matrix, which also permutes the syndrome array.  The mathematical
details needed to implement all these functions are given
in~\cite{knill:qc2003b}.

It is useful to observe that the likelihoods are always positive and
that all the operations used involve only addition and multiplication
of positive numbers. This ensures that numerical errors do not spread
too rapidly. The computer algebra system used has the capability to
keep track of the precision of likelihoods and likelihood ratios
obtained. That the results have sufficient precision was verified by
inspecting the recorded precision provided by the computer-algebra
system.

The method for obtaining the Bell-error model from the computed
likelihood distribution for the purified encoded Bell pair is
described in the next section.

\section{Preparation of the encoded Bell pair}
\label{sect:preparation of the encoded Bell pair}

The original scheme for postselected quantum computation is based on
concatenation of two four-qubit error-detecting codes that are used
alternately at different steps of the computation. Each step of the
computation is based on teleportation through a Bell pair encoded in
the two codes. This state is purified once before being used. For the
purpose of the computer-assisted analysis used here, the scheme is
modified in two ways.  First, only one four-qubit code is used within a
level, necessitating a change in the preparation of the encoded Bell
state. See Fig.~\ref{fig:encoded Bell}.  Second, a single purification
does not result in sufficient independence between the two codes
supporting the Bell pair, so two purification steps are used. This
results in independence to lowest order. It is expected that it is not
necessary to implement this in practice. Ideally, the purification
method used asymptotically leads to block-wise independent error
behavior.  Unfortunately, this is not the case for the purification
scheme used here, which is based on the standard entanglement
purification protocol~\cite{bennett:qc1996b}.  An alternative approach
to purification is to use the methods of Shor~\cite{shor:qc1996a} and
generalized by Steane~\cite{steane:qc1997a} to purify the encoded Bell
state by measuring syndromes with specially prepared ``cat'' states.
Although this still does not lead to asymptotic independence, it may
reduce the resources required while still approaching independence
sufficiently well for practical purposes.

\begin{herefig}
\label{fig:encoded Bell}
\begin{picture}(0,4.2)(0,-3.9)
  \nputbox{0,.1}{t}{
     \setlength{\fboxrule}{0pt}\setlength{\fboxsep}{0pt}
     \fcolorbox[rgb]{.92,.92,.92}{.92,.92,.92}{\rule{6.5in}{0pt}\rule{0pt}{3.7in}}
  }
\nputbox{-3.2,0}{tl}{\large\textbf{a.}}
\nputgr{-1.8,0}{t}{scale=.6}{enc4e4zz}
\nputbox{.4,0}{tl}{\large\textbf{b.}}
\nputgr{1.8,0}{t}{scale=.6}{enc4e4xx}
\end{picture}
\nopagebreak
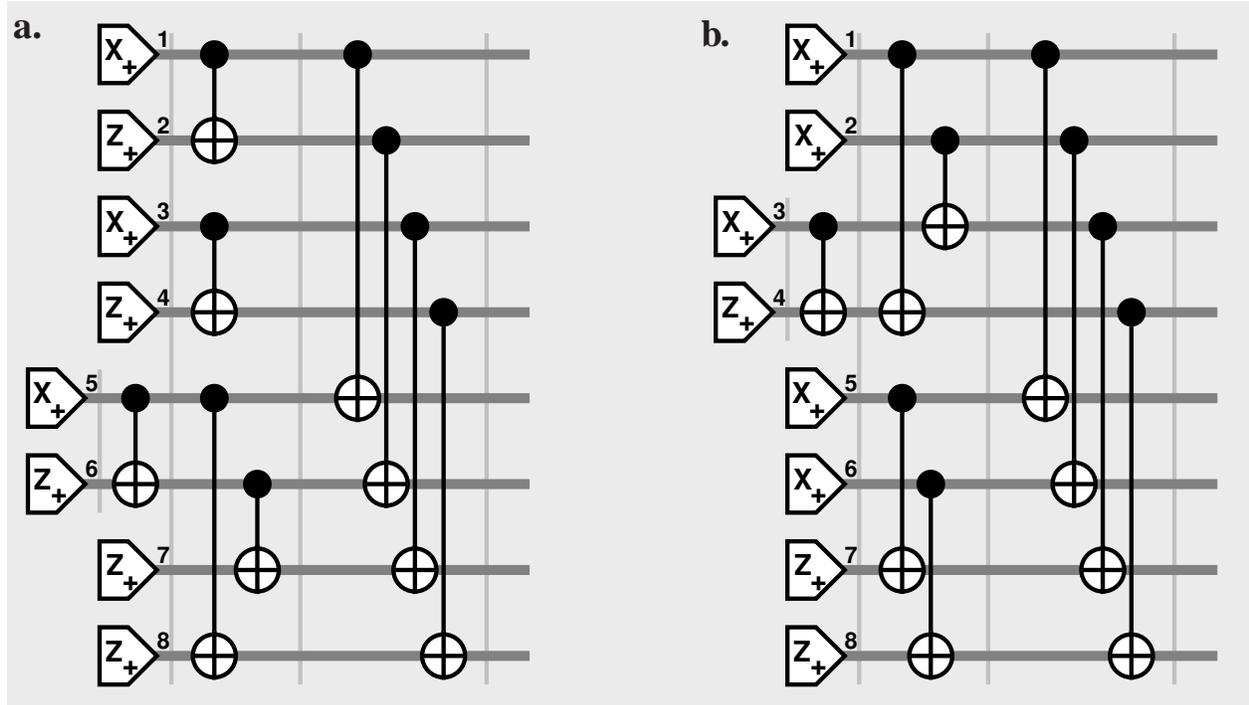
\herefigcap{States used to teleport encoded qubits.  The conventions
for network elements are as in~\cite{knill:qc2004a} At each level a
choice is made whether to have the spectator qubit in its $\kets{0}{S}$
or its $\kets{+}{S}$ state.  Network \textbf{a.} makes the encoded Bell
state with both spectator qubits in state $\kets{0}{S}$.  The output of
network \textbf{b.} has both in state $\kets{+}{S}$.
The stabilizer of \textbf{a.}'s output is
\begin{eqnarray*}
   &[XXXXIIII], [ZZZZIIII], [IIZZIIII],&\\
   &[IIIIXXXX], [IIIIZZZZ], [IIIIIIZZ],&\\
   &[XXIIXXII], [ZIZIZIZI].&
\end{eqnarray*}
The third and sixth Pauli product are the spectator qubits' $Z$ operators.
The logical qubits' operators are 
$\slb{X}{L_1}=[XXIIIIII], \slb{Z}{L_1}=[ZIZIIIII],
\slb{X}{L_2}=[IIIIXXII], \slb{Z}{L_2}=[IIIIZIZI]$. Thus the last
two operators are stabilizers of the encoded Bell pair.
Similarly, the stabilizer of \textbf{b.}'s output is
\begin{eqnarray*}
   &[XXXXIIII], [ZZZZIIII], [IXIXIIII],&\\
   &[IIIIXXXX], [IIIIZZZZ], [IIIIIXIX],&\\
   &[XXIIXXII], [ZIZIZIZI],&
\end{eqnarray*}
where the third and sixth Pauli product are the spectator qubits' $X$
operators.
}
\end{herefig}
\pagebreak

The first step of the analysis requires determining the syndrome
likelihood distribution of the encoded Bell pair if the preparation
method of Fig.~\ref{fig:encoded Bell} is used with two purification
cycles. Each purification cycle involves two steps, purifying $Z$-type
and $X$-type syndromes, respectively.  The order of the two steps is
randomized.  At each level, a choice is made of whether to use
spectators in $\kets{\biz}{S}$ or $\kets{+}{S}$ states. The choice
must be made based on the error models for gates at the current level,
as $\kets{\biz}{S}$ spectators help with suppressing $X$ errors,
whereas $\kets{+}{S}$ spectators suppress $Z$ errors.  In either case,
the syndrome distribution of the prepared encoded Bell pairs is
computed by using the functions that can simulate the preparation
networks and apply each step's error model. The randomization of the
purification order is implemented by computing likelihoods for both
possible orders and averaging them. In order for this averaging to be
correct, the normalization (sum of the likelihoods) must be identical
in the two cases. This condition is satisfied because each case
involves the same gates but in a different order.  This procedure
yields the \emph{computed error model} for the encoded Bell pairs.

If the likelihood distribution is computed formally, with cnot
preparation and measurement errors all being a multiple of a single
error parameter $e$, inspection shows that the lowest-degree
contribution to the likelihood of a syndrome is the same as the
minimum weight of the set of Pauli products that can give rise to that
syndrome. In other words, the likelihood distribution has error
behavior of the same order as that obtained if independent errors are
applied to each qubit after preparing the stabilizer state without
errors. (This is not the case if only one purification cycle is used.)
This motivates the next step, which is to heuristically bound the
error-likelihood distribution with an independent-error model. Only
independence between the origin and destination block of the encoded
Bell pair is needed. The error model is therefore designed to
independently apply certain Pauli products to each block of the
encoded Bell pair.  Consider the origin block. Because the stabilizer
consists of products of the three operators $[XXXX], [ZZZZ]$ and
$[IIZZ]$ or $[IXIX]$ (depending on whether the spectator qubit's state
is $\kets{\biz}{S}$ or $\kets{+}{S}$), and because two Pauli products
differing by an element of the stabilizer have the same effect, the
set of Pauli products can be partitioned into 32 cosets, where two
operators in the same coset have the same effect on the encoded Bell
pair. The independent-error model assigns a likelihood to each of
these 32 cosets for the origin block, and similarly for the
destination block.  Because of the encoded stabilizer operators
$[XXIIXXII]$ and $[ZIZIZIZI]$ characterizing the encoded Bell state,
each syndrome of the encoded Bell pair can be obtained in four
different ways from errors on the two blocks as classified by the 32
cosets.  The problem is to choose the origin and destination
Bell-error models to match the computed error model as closely as
possible without unintentionally improving future error behavior.
Thus, the goal is to have the Bell-error model's normalized
likelihoods exceed the computed error model's normalized
likelihoods. (The normalized likelihoods are such that the likelihood
of the $\mb{0}$ syndrome is $1$.)  Although it cannot be excluded that
increasing likelihoods in this way leads to sufficient error
cancellation to improve the error behavior, this seems unlikely.  A
second goal is to recover the same error model if indeed it is induced
by independent errors.  The computational method used to obtain the
independent Bell-error model is heuristic (see the next paragraph),
but achieves these goals to good approximation.  One way to measure
the quality of the Bell-error model is to calculate the maximum and
minimum ratios of the normalized likelihoods of corresponding
syndromes in the Bell-error model and the computed error model for the
encoded Bell state. Ideally, the minimum ratio should be $1$. It was
found to be greater than $1-10^{-6}$ in all cases and usually much
closer to $1$. The maximum ratio varies significantly and is largest
when the preparation/measurement error dominates. See
Fig.~\ref{fig:contours}.

To obtain the block-wise independent Bell-error model, each encoded
Bell pair syndrome is assigned a weight. For unbiased errors, the
weight should be the minimum weight of Pauli products that can give
rise to the syndrome.  Because errors in the models in effect at
higher levels of concatenation are highly biased, the weight
computation is modified as follows.  First a one-qubit error model is
derived from the maximum marginal error probabilities of the
likelihoods for Pauli products after the cnot. After normalizing the
identity error to $1$, the negative logarithm of the error likelihoods
of $X$, $Y$ and $Z$ are assigned as weights to each of these Pauli
operators. The weight of a Pauli product is the sum of the weights of
each one-qubit factor. Each syndrome's weight is given by the minimum
weight of Pauli products that give rise to it.  Similarly, the weights
of the 32 cosets for the origin and destination blocks are computed
and each coset $c$ is assigned to the minimum weight Pauli product
$p(c)$ that belongs to the coset.  Next, each computed error
likelihood associated with a non-zero syndrome $s$ for the encoded
Bell pair is attributed to one of the four pairs of origin and
destination block cosets that can give rise to $s$.  The choice is
made so as to minimize the sum of the weights of the two cosets. As a
result, each pair of cosets $c_1,c_2$ is assigned an error likelihood
$e(c_1, c_2)$, which is $0$ if it does not have minimum weight for the
encoded Bell-pair syndrome that is determined by the pair.  Let
$s(c_1,c_2)$ be the encoded Bell-pair syndrome determined by
$(c_1,c_2)$.  The last step is to obtain independent origin and
destination likelihoods $e_o(c)$ and $e_d(c)$ that, in a sense, bound
the likelihoods obtained for the pairs.  This is done by minimizing
the independent likelihoods of each block subject to the constraint
that if $(c_1,c_2)$ is a pair of cosets and the weight of $s(0,c_2)$
is less than the weight of $s(c_1,c_2)$, then $e_o(c_1)/e(0)\geq
e(c_1,c_2)/e(0,c_2)$, and similarly if the weight of $s(c_1,0)$ is less
than the weight of $s(c_1,c_2)$.  The idea is for the
independent-error model to have larger relative likelihoods consistent
with independence$(B)$.

The Bell-error model need not have the same error behavior for the
origin and destination blocks. However, the two blocks can be made to
behave similarly by symmetrizing the preparation procedure. This is
done by randomizing the orientation of the Bell pair, which is made
possible by using the same spectator qubit state on each block.

In the next sections, the independence of the Bell error model is used
both to simplify the computation of the next level's encoded gates and
to ensure that the desired independence properties apply throughout a
computation. Numerical results are presented in
Sect.~\ref{sect:postselected threshold}.

\section{State preparation}
\label{sect:state preparation}

Preparation of encoded states is implemented by postselected
measurements of the origin block of an encoded Bell pair.  For
preparing the logical $\kets{\biz}{L}$ ($\kets{+}{L}$) state, $Z$
($X$) measurements on each origin qubit postselected on the $+1$
eigenvalue are used.  The resulting state on the destination qubits
must be modeled by an independent preparation error acting on the
logical qubit, followed by the terminal-error model derived from the
destination block's Bell-error model. With the independence heuristic,
the effect of the measurement of the origin block's qubits depends
only on the errors acting on the origin and propagates into the
encoded qubit via the encoded entanglement. To compute the propagated
error, observe that because the origin errors are now assumed to be
independent of the destination errors, it is possible to abstract the
destination before the destination errors are applied as a single,
error-free qubit $\sysfnt{B}$ entangled with the origin's encoded
qubit.  The effect of measurement can then be computed explicitly
using the functions of Sect.~\ref{sect:functions}.  The measurement
results in either the desired state on $\sysfnt{B}$ or the orthogonal
state. The probability that the orthogonal state is obtained defines
the independent contribution to the logical error due to state
preparation.

\section{Measurement}
\label{sect:measurement}

Postselected measurement is similar to state preparation. It involves
making the same measurements, but on the destination block of an
encoded Bell pair. As a result, the analysis can be done in the same
way, which is by modeling the origin half as an ideal single qubit
entangled with the logical qubit of the destination. The effect of
measurement can be computed explicitly to determine the likelihoods
for the state that the single qubit ends up in.  Note that in an
implementation it may be the case that the origin half has already
been involved in a postselected Bell measurement. The effect of such
prior operations is taken into account by independent logical errors
active in the current state. These errors are independent of the
destination errors, making it possible to model the effects of
measurement as has been done here.  The randomized symmetrization of
the Bell pair result in logical errors for measurements that are
essentially the same as the logical errors for preparation.

\section{Cnot}
\label{sect:controlled-not}

The encoded cnot is implemented transversally by applying cnots to
corresponding qubits in the two blocks encoding the qubit to be
acted on. This is followed by a teleportation step that involves
postselected Bell measurements from the encoded qubits' blocks to the
origin blocks of two encoded Bell pairs.  To avoid delays, the cnots
of the Bell measurements can be interleaved with the transversal
cnots~\cite{knill:qc2004a}. The independence heuristic implies that
logical errors attributable to the transversal cnots and Bell
measurements are independent of earlier logical errors, terminal
errors of qubits not involved in the gate, and destination Bell errors
for the two Bell pairs used for teleportation.  As a result, the
effect of the encoded cnot can be computed by using only four blocks
whose errors are characterized by the independent destination and
origin Bell-error models. As was done for measurement and preparation
errors, each of these four blocks' encoded qubit is considered
entangled with an ideal reference qubit. In this way, the complete error
behavior of the cnot implementation can be computed.

If the qubits of the blocks are themselves encoded with the
error-detecting codes of the scheme, the cnot analysis is
modified. This is necessary because rather than implementing each cnot
according to the lower-level instructions, the cnots are implemented
as transversal physical cnots, acting on corresponding physical
qubits. This improves the error behavior of the encoded cnot because
the effect at higher levels is that the implementation of the cnot
behaves as if the Bell measurement were error-free.  To see this,
consider the logical-error model for the cnot derived at level $1$. It
is possible to virtually rearrange the cnots and errors of the cnots
and measurements so that (a) the transversal cnot is applied first
without errors, (b) all the errors occur (including those due to
origin and destination error models), and finally, (c) the Bell
measurements are implemented without errors.  The computed logical
error of the procedure is determined by those error combinations in
this virtual representation that satisfy the condition that
corresponding origin and destination syndromes are identical.
Consider the next level of encoding. The cnot implementation consists
of applying the lower-level scheme to corresponding physical origin
and destination blocks. The computed error for these cnots anticipates
the Bell-measurement errors so that it is not necessary to
re-introduce them.  That is, in computing the next-level cnot's
logical-error likelihoods, the transversal cnots come with the
previously computed logical errors, but the Bell measurements can be
taken to be error-free.

\ignore{ Here is something to puzzle over: If the postselection is
done efficiently, then instead of postselecting each Bell measurement
independently, postselection is based on parities associated with the
bits of the syndrome. Higher level parity checks involving multiple
blocks, which makes sense. However, for analysis, it is simpler to
postselect each Bell measurement on its neutral outcome. Given the
Pauli-product probabilistic error model, this works. The puzzle is
that the postselection in the implementation of the cnot (for example)
is block by block, and it is not obvious how the higher level
syndromes enter into the picture.  In particular, any non-zero higher
level syndrome implies that some individual Bell measurement must be
non-neutral.  The puzzle is to ``understand'' how these local
conditions can propagate to global ones, a puzzle that is not
unfamiliar in quantum mechanics. }

\section{Hadamard gate}
\label{sect:hadamard gate}

The Hadamard gate is not needed for scalable postselected generation
of CSS states. However, it is helpful for achieving universality.  The
computation of its contributions to the logical error is similar to
that of the cnot. A teleported implementation may be used where the
Hadamard is applied transversally to the current block of a logical
qubit, followed by a Bell measurement involving an origin block of an
encoded Bell pair. The code requires that the middle two qubits be
swapped after the transversal Hadamard, which can be taken into
account by ``crossing'' the middle two Bell measurements.  Again, at
levels other than the first, the Hadamard is implemented in a
transversal fashion rather than reimplemented using explicit
lower-level gates. As a result, the computation of higher-level
logical errors can be simplified by having error-free Bell
measurements.

\section{Decoding a prepared state}
\label{sect:decoding a prepared state}

The main application of postselected fault-tolerant quantum
computation is to the preparation of stabilizer states with errors
that are primarily local. These states can then be used in a
standard fault-tolerant quantum computation.  
They are prepared first in encoded form at a sufficiently high
level of encoding using the error-detecting codes. Once such a state
is obtained, the error-detecting codes are decoded from the bottom up.
The heuristic error model used to describe the error behavior of the
scheme implies that any encoded stabilizer state obtained before
decoding the error-detecting code is subject to logical errors due to
the gates used to prepare the encoded state, and terminal-errors that
are independently applied to each block at every level. The bottom-up
decoding process adds errors to the final state only through
undetected terminal errors and errors introduced in the decoding
process itself. Because of independence, the error introduced in this
way in the first step of the bottom-up decoding process can be
computed from the terminal error model that applies to physical
blocks. This error propagates as an additional local error in the
next-level blocks.  Since the terminal-error models at each level are
themselves independent, this local error can be added independently to
the next level's terminal error model.  The effect of the next level
of decoding can then be computed as before.  Thus, in each decoding
cycle, additional local error from below is merged with terminal
errors and decoding errors, conditioned on a successful
decoding. Scalability is verified by confirming that the error pushed
into the decoded qubit is bounded.  A bound can be estimated by
implementing this procedure for the first few levels to compute the
decoding error in each step. In the examples computed (see
Tables~\ref{tab:examples1},~\ref{tab:examples2}), the local error
introduced by decoding at each level appears to slowly decrease toward
a steady state.

\section{Encoding an arbitrary state}
\label{sect:encoding an arbitrary state}

To encode an arbitrary state, a Bell pair is encoded at the top level
of the concatenation hierarchy. The origin block of the pair is
decoded bottom-up as in the process for preparing a stabilizer state
in the previous section. The last step is to teleport a state prepared
in a physical qubit into the top-level code using the decoded origin
qubit.  This process may be referred to as \emph{injecting} a state
into a code.  The new state's error model fits the logical/terminal
error model scheme used before, where the terminal error is due to the
terminal-error model applicable to the destination block of the
encoded Bell pair. The logical error comes from the bottom-up decoding,
the (post-selected) Bell measurement used for teleportation, and
any error in the physical qubit's state.  Given the bound on the
bottom-up decoding error, it suffices to combine it with a bound on
the error in preparing the desired state in a physical qubit and the
error due to a physical Bell measurement.

\section{Postselected threshold}
\label{sect:postselected threshold}

By use of the computer-assisted heuristics described in the previous
sections, it is now possible to determine whether a given
physical-error model is scalable for postselected computation with the
error-detecting schemes under analysis.  A physical-error model is
scalable for preparing CSS states if the logical error probabilities
can be made arbitrarily small.  Universality requires logical state
preparations and purifications.  Consider preparation of logical
$\ket{\pi/8}$ states.  A version of the purification scheme for
$\ket{\pi/8}$ states given in~\cite{knill:qc2004a} is analyzed by
Bravyi and Kitaev~\cite{bravyi:qc2004a} in the context of ``magic
states distillation''. They show that magic states, which include
$\ket{\pi/8}$, are distillable given a way of preparing them with
probability of error below about $35\;\%$, assuming no error in
Clifford group operations. Encoding $\ket{\pi/8}$ by decoding one-half
of a logical Bell pair and teleporting introduces relatively little
additional error (see below). These observations apply also to
$\ket{i\pi/4}$ preparation. As a result, error in preparation of
logical $\ket{\pi/8}$ and $\ket{i\pi/4}$ states is not a bottleneck
for scalability.

To illustrate the computer-assisted heuristic analysis, the logical
error probabilities are graphed for a range of preparation/measurement
and cnot errors in Fig.~\ref{fig:contours}, which maps the maximum of
the logical-error probabilities for encoded preparation, measurement,
cnot, and Hadamard gates as a function of the physical-error
probabilities and the number of levels of encoding. As can be seen, a
small number of levels suffice to strongly suppress the logical-error
probabilities. A boundary between the region where errors are scalable
and where they are unscalable (by these schemes) emerges as the number
of levels are increased. For measurement/preparation error
probabilities well below the cnot error probability, the boundary is
between $3.5\;\%$ and $5\;\%$.

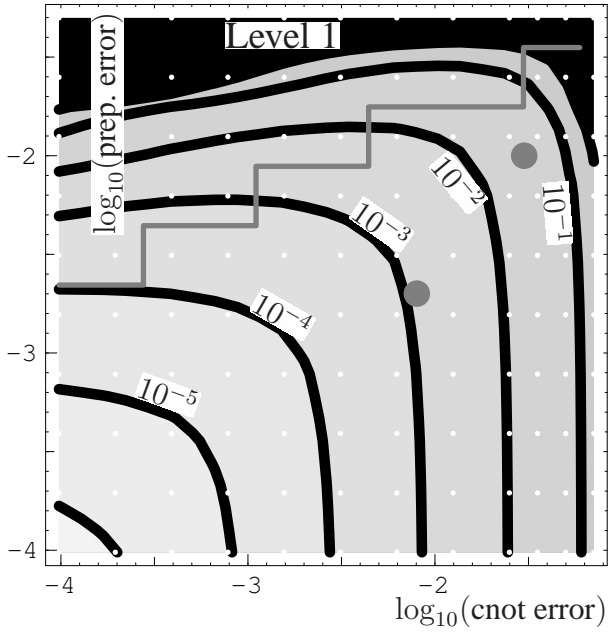
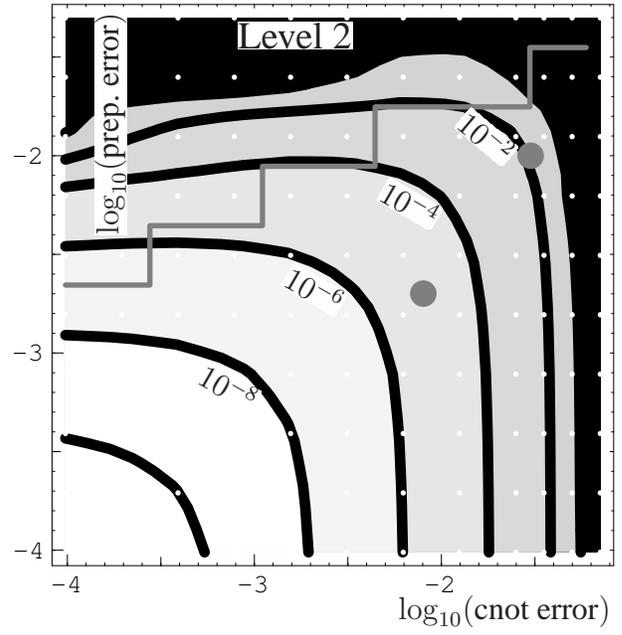
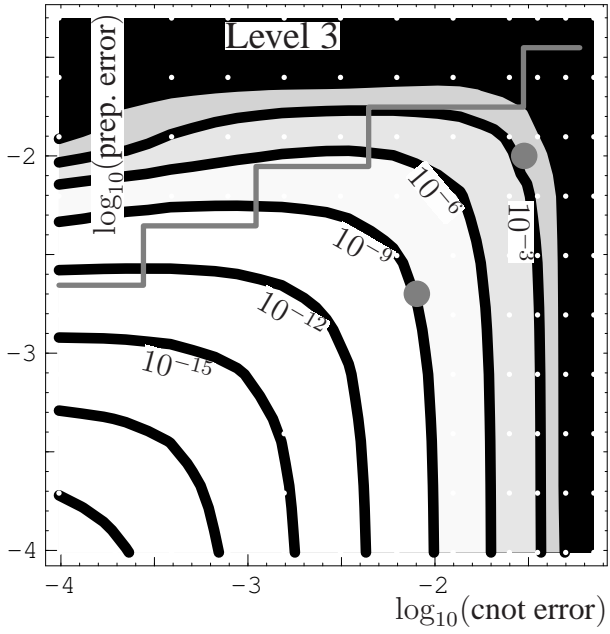
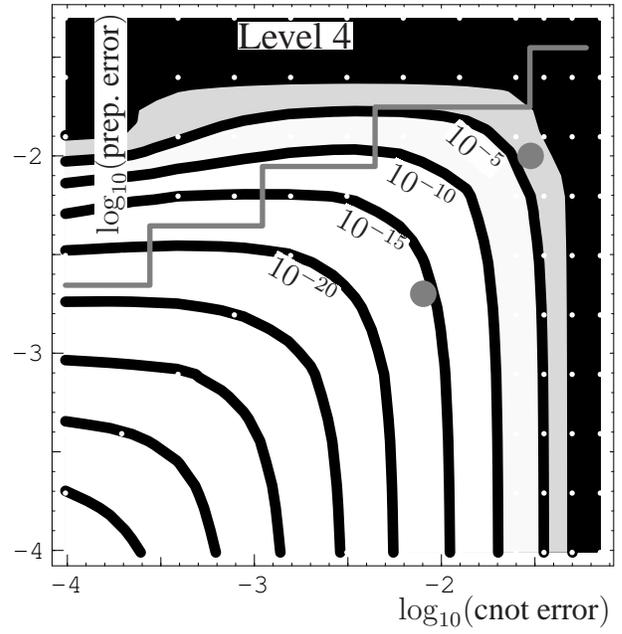
\begin{herefig} \label{fig:contours} \begin{picture}(0,6.8)(0,-6.6)
\fboxpars{0pt}{0pt} 
\nputgr{-.2,0}{tr}{}{cplotw1}
\nputbox{-2.2,-.1}{tl}{\fcolorbox[rgb]{1,1,1}{1,1,1}{\large Level 1}}
\nputbox{-.2,-3.1}{tr}{$\log_{10}(\textrm{cnot error})$}
\nputbox{-2.9,-0.05}{tl}{\rotatebox{90}{\fcolorbox[rgb]{1,1,1}{1,1,1}{$\log_{10}(\textrm{prep.~error})$}}}
\nputbox{-.46,-1.1}{c}{\rotatebox{-85}{\fcolorbox[rgb]{1,1,1}{1,1,1}{$10^{-1}$}}}
\nputbox{-.98,-.94}{c}{\rotatebox{-50}{\fcolorbox[rgb]{1,1,1}{1,1,1}{$10^{-2}$}}}
\nputbox{-1.38,-1.16}{c}{\rotatebox{-35}{\fcolorbox[rgb]{1,1,1}{1,1,1}{$10^{-3}$}}}
\nputbox{-1.9,-1.63}{c}{\rotatebox{-30}{\fcolorbox[rgb]{1,1,1}{1,1,1}{$10^{-4}$}}}
\nputbox{-2.5,-2.05}{c}{\rotatebox{-20}{\fcolorbox[rgb]{1,1,1}{1,1,1}{$10^{-5}$}}}

\nputgr{.2,0}{tl}{}{cplotw2}
\nputbox{1.4,-.1}{tl}{\fcolorbox[rgb]{1,1,1}{1,1,1}{\large Level 2}}
\nputbox{3.35,-3.1}{tr}{$\log_{10}(\textrm{cnot error})$}
\nputbox{.65,-0.05}{tl}{\rotatebox{90}{\fcolorbox[rgb]{1,1,1}{1,1,1}{$\log_{10}(\textrm{prep.~error})$}}}
\nputbox{2.7,-.7}{c}{\rotatebox{-40}{\fcolorbox[rgb]{1,1,1}{1,1,1}{$10^{-2}$}}}
\nputbox{2.3,-1.05}{c}{\rotatebox{-30}{\fcolorbox[rgb]{1,1,1}{1,1,1}{$10^{-4}$}}}
\nputbox{1.8,-1.5}{c}{\rotatebox{-30}{\fcolorbox[rgb]{1,1,1}{1,1,1}{$10^{-6}$}}}
\nputbox{1.35,-2.01}{c}{\rotatebox{-30}{\fcolorbox[rgb]{1,1,1}{1,1,1}{$10^{-8}$}}}

\nputgr{-.2,-3.5}{tr}{}{cplotw3}
\nputbox{-2.2,-3.6}{tl}{\fcolorbox[rgb]{1,1,1}{1,1,1}{\large Level 3}}

\nputbox{-.2,-6.6}{tr}{$\log_{10}(\textrm{cnot error})$}
\nputbox{-2.9,-3.55}{tl}{\rotatebox{90}{\fcolorbox[rgb]{1,1,1}{1,1,1}{$\log_{10}(\textrm{prep.~error})$}}}
\nputbox{-.65,-4.7}{c}{\rotatebox{-90}{\fcolorbox[rgb]{1,1,1}{1,1,1}{$10^{-3}$}}}
\nputbox{-1.1,-4.49}{c}{\rotatebox{-50}{\fcolorbox[rgb]{1,1,1}{1,1,1}{$10^{-6}$}}}
\nputbox{-1.48,-4.78}{c}{\rotatebox{-30}{\fcolorbox[rgb]{1,1,1}{1,1,1}{$10^{-9}$}}}
\nputbox{-1.85,-5.13}{c}{\rotatebox{-30}{\fcolorbox[rgb]{1,1,1}{1,1,1}{$10^{-12}$}}}
\nputbox{-2.45,-5.4}{c}{\rotatebox{-15}{\fcolorbox[rgb]{1,1,1}{1,1,1}{$10^{-15}$}}}

\nputgr{.2,-3.5}{tl}{}{cplotw4}
\nputbox{1.4,-3.6}{tl}{\fcolorbox[rgb]{1,1,1}{1,1,1}{\large Level 4}}
\nputbox{3.35,-6.6}{tr}{$\log_{10}(\textrm{cnot error})$}
\nputbox{.65,-3.55}{tl}{\rotatebox{90}{\fcolorbox[rgb]{1,1,1}{1,1,1}{$\log_{10}(\textrm{prep.~error})$}}}
\nputbox{2.64,-4.26}{c}{\rotatebox{-40}{\fcolorbox[rgb]{1,1,1}{1,1,1}{$10^{-5}$}}}
\nputbox{2.36,-4.46}{c}{\rotatebox{-30}{\fcolorbox[rgb]{1,1,1}{1,1,1}{$10^{-10}$}}}
\nputbox{2.1,-4.7}{c}{\rotatebox{-30}{\fcolorbox[rgb]{1,1,1}{1,1,1}{$10^{-15}$}}}
\nputbox{1.74,-4.95}{c}{\rotatebox{-30}{\fcolorbox[rgb]{1,1,1}{1,1,1}{$10^{-20}$}}}

\end{picture} 
\herefigcap{Plots of error probabilities achieved after one, two,
three and four levels of concatenation.  The spectator qubits for the
first level are in state $\kets{+}{S}$.  For the next three levels
they are in state $\kets{\biz}{S}$.  The $x$- and $y$-axes show the
$\log_{10}$ of the total error probabilities of the physical cnot and
the physical preparation/measurement, respectively. The Hadamard
gate's error probability is $3/2$ times the preparation/measurement
error probability, consistent with the assumption that preparation and
measurement error is comparable to other one-qubit gate errors except
that one of the Pauli operators has no effect.  The probability of
each non-identity Pauli product contributing to the error is assumed
to be identical.  The contours show the maximum error probability of
the encoded cnot, Hadamard and preparation/measurement gates at
logarithmic intervals. The highest contour corresponds to a maximum
error probability of $.25$. The contour intervals are adjusted from
level to level. Darker shadings correspond to higher error
probabilities. Black indicates the region where scalability clearly
fails due to the error probabilities having exceeded $.25$.  The
contours were obtained by computing the error probabilities at $100$
points in a $10\times 10$ logarithmic grid.  The separation between
adjacent grid points corresponds to a change in error probabilities by
a factor of $2$. An additional $20$ points were computed for
logarithmic cnot error probabilities of $\log_10(0.071)=-1.15$. and
$\log_10(0.0354)=-1.45$. The positions of the points at which the
probabilities were computed are shown as white dots.  Interpolation
was used to derive the contours. Note that the grid point separation
corresponds to logarithmic intervals of about $0.3$ ($0.15$ near the
right scalability boundary). The contour positions, particularly the
one at the scalability boundary, should therefore be regarded as
having positional uncertainties of that order. The contours indicate
decreasing error with increasing cnot error probabilities at high,
constant measurement/preparation error rates. This is an artifact of
the method used for obtaining the independent Bell-error error
model. In particular, the quality of approximation of the
independent-error model is not very good when preparation/measurement
error is high compared to cnot error. This is probably because the
method is based on cnot error and does not take
preparation/measurement error into account. When the approximation is
bad, the computed logical errors grow.  The region where the maximum
normalized likelihood ratio between the independent Bell-error model
and the computed error model (see Sect.~\ref{sect:preparation of the
encoded Bell pair}) exceeds 100 is above the gray staircase line in
each plot. To determine the behavior of specific examples more
precisely requires further computation. Two examples are given in
Tables~\ref{tab:examples1} and~\ref{tab:examples2}. The locations of the
error probabilities used in the examples are indicated by the two gray
points.} \end{herefig}

\pagebreak

Two example error combinations are examined in detail in
Tables~\ref{tab:examples1} and~\ref{tab:examples2}.  The first is near the
scalability boundary for the scheme analyzed here. The second is lower
by about
a factor of $5$. The measurement/preparation error probability
is taken to be a factor of 3-4 below the cnot error probability. This
choice is motivated by an estimate of how well measurement error can
be suppressed by using ancillas and
cnots~\cite{divincenzo:qc2000a}. As in Fig.~\ref{fig:contours}, the
Hadamard-gate error probability is $3/2$ of the
measurement/preparation error. The first example has all gate-error
probabilities above $1\;\%$. Specifically, the measurement/preparation
error is $1\;\%$ and the cnot error is $3\;\%$. Four levels are required
to reach logical gate-error probabilities near or below $10^{-4}$.
One more level reduces them below $10^{-7}$. The choice of spectator
qubit state at each level significantly affects the logical gate error
distributions. Note that $Y$ errors are always significantly
suppressed, which is a feature of CSS codes. Also, having the
spectator qubit be in state $\kets{\biz}{S}$ strongly suppresses $X$
errors compared to $Z$ errors, except for the Hadamard gate. This is
because the implementation of the Hadamard gate voids the ability
of the teleportation step to detect errors affecting only the part of
the syndrome due to the spectator qubits' states. The biases suggest
that it may be worth using different codes adapted to this bias at the
higher levels.  The decoding error is clearly stable as more levels
are added. In fact, it is monotone decreasing.  By extrapolation, the
total decoding error is below $2.3\;\%$ no matter how many levels are
used. This implies that if a $\ket{\pi/8}$ state is injected into the top
level as discussed above, the newly encoded $\ket{\pi/8}$ state should
have error probability of at most $9\;\%$, bounding the sum of the
measurement/preparation error ($1\;\%$), the decoding error ($2.3\;\%$),
and the Bell measurement error ($5\;\%$, due to the cnot and two
measurements).  This is well below the purification
threshold~\cite{bravyi:qc2004a}.

The second example involves measurement/preparation error of $0.2\;\%$
and cnot error of $0.8\;\%$, and exhibits similar features except that
only two levels are needed to suppress the maximum error probability
below $10^{-5}$. The decoding error probability decreases to a value
below $0.55\;\%$.

\begin{heretab}
\label{tab:examples1}
{\small\begin{tabular}{|l||l|l|l|ll||}
\hline
Error probabilities:  & \multicolumn{5}{l||}{$e_p=e_m=0.01, e_c=0.03$} 
\\\hline\hline
Level 1 &\multicolumn{5}{l||}{Spectator:\ $\kets{+}{S}$}
\\\hline
Independence quality & 
   \multicolumn{5}{l||}{3.20} 
\\\hline
Prep./Meas. error&  
    $X: 4.14*10^{-3}$ & $Z: 9.82*10^{-4}$ &  &
    Max: & $4.14*10^{-3}$
\\\hline
Cnot marginal error& 
    $X: 1.83*10^{-2}$ & $Z: 6.11*10^{-3}$ & $Y:3.70*10^{-4} $ &
    Total: & $2.98*10^{-2}$
\\\hline
Hadamard error & 
    $X: 1.10*10^{-2}$ & $Z: 1.10*10^{-2}$ & $Y: 2.31*10^{-3}$  &
    Total: & $2.44*10^{-2}$
\\\hline
Decode error & 
    $X: 5.61*10^{-3}$ & $Z: 1.30*10^{-2}$ & $Y: 4.75*10^{-3}$ &
    Total: & $2.34*10^{-2}$
\\\hline\hline
Level 2 &\multicolumn{5}{l||}{Spectator:\ $\kets{\biz}{S}$}
\\\hline
Independence quality & 
   \multicolumn{5}{l||}{8.59} 
\\\hline
Prep./Meas. error&  
    $X: 4.87*10^{-4}$ & $Z: 4.34*10^{-4}$ &  &
    Max: & $4.87*10^{-4}$
\\\hline
Cnot marginal error& 
    $X: 5.03*10^{-3}$ & $Z: 3.39*10^{-3}$ & $Y: 8.33*10^{-6} $ &
    Total: & $9.53*10^{-3}$
\\\hline
Hadamard error & 
    $X: 6.02*10^{-3}$ & $Z: 6.02*10^{-3}$ & $Y: 1.24*10^{-4}$  &
    Total: & $1.22*10^{-2}$ 
\\\hline
Decode error & 
    $X: 1.30*10^{-2}$ & $Z: 4.86*10^{-3}$ & $Y: 4.56*10^{-3}$ &
    Total: & $2.24*10^{-2}$
\\\hline\hline
Level 3 &\multicolumn{5}{l||}{Spectator:\ $\kets{\biz}{S}$}
\\\hline
Independence quality & 
   \multicolumn{5}{l||}{2.64} 
\\\hline
Prep./Meas. error&  
    $X: 5.10*10^{-6}$ & $Z: 2.42*10^{-4}$ &  &
    Max: & $2.42*10^{-4}$
\\\hline
Cnot marginal error& 
    $X: 9.37*10^{-5}$ & $Z: 1.22*10^{-3}$ & $Y: 4.19*10^{-8} $ &
    Total: & $1.57*10^{-3}$ 
\\\hline
Hadamard error & 
    $X: 7.27*10^{-4}$ & $Z: 7.27*10^{-4}$ & $Y: 7.39*10^{-7}$  &
    Total: & $1.45*10^{-3}$ 
\\\hline
Decode error & 
    $X: 1.23*10^{-2}$ & $Z: 4.55*10^{-3}$ & $Y: 4.34*10^{-3}$ &
    Total: & $2.12*10^{-2}$ 
\\\hline\hline
Level 4 &\multicolumn{5}{l||}{Spectator:\ $\kets{\biz}{S}$}
\\\hline
Independence quality & 
   \multicolumn{5}{l||}{1.90} 
\\\hline
Prep./Meas. error&  
    $X: 3.11*10^{-10}$ & $Z: 2.70*10^{-5}$ &  &
    Max: & $2.70*10^{-5}$
\\\hline
Cnot marginal error& 
    $X: 2.08*10^{-8}$ & $Z: 1.32*10^{-4}$ & $Y: 1.07*10^{-12} $ &
    Total: & $1.59*10^{-4}$ 
\\\hline
Hadamard error & 
    $X: 3.60*10^{-5}$ & $Z: 3.60*10^{-5}$ & $Y: 1.31*10^{-9}$  &
    Total: & $7.19*10^{-5}$ 
\\\hline
Decode error & 
    $X: 1.22*10^{-2}$ & $Z: 4.24*10^{-3}$ & $Y: 4.22*10^{-3}$ &
    Total: & $2.07*10^{-2}$ 
\\\hline\hline
Level 5 &\multicolumn{5}{l||}{Spectator:\ $\kets{+}{S}$}
\\\hline
Independence quality & 
   \multicolumn{5}{l||}{1.94} 
\\\hline
Prep./Meas. error&  
    $X: 3.65*10^{-15}$ & $Z: 7.37*10^{-9}$ &  &
    Max: & $7.37*10^{-9}$
\\\hline
Cnot marginal error& 
    $X: 2.35*10^{-14}$ & $Z: 6.12*10^{-8}$ & $Y: 3.74*10^{-22} $ &
    Total: & $6.87*10^{-8}$ 
\\\hline
Hadamard error & 
    $X: 1.76*10^{-8}$ & $Z: 1.76*10^{-8}$ & $Y: 3.10*10^{-16}$  &
    Total: & $3.51*10^{-8}$ 
\\\hline
Decode error & 
    $X: 4.16*10^{-3}$ & $Z: 1.22*10^{-2}$ & $Y: 4.16*10^{-3}$ &
    Total: & $2.05*10^{-2}$ 
\\\hline\hline
\end{tabular}}
\vspace*{.2in} \heretabcap{Details for physical error probabilities of
$1\;\%$ for state preparation and measurement and $3\;\%$ for cnot.  The
independence quality is the maximum ratio of the normalized
likelihoods of the independent Bell and the computed error models
(see Sect.~\ref{sect:preparation of the encoded Bell pair}).
The minimum ratio found was $0.99999957$.  Errors are given in
probabilities.  Preparation and measurement errors are identical, due
to randomization of the encoded Bell pair preparation to ensure
bilateral symmetry of the errors.  For preparation and measurement,
the probabilities are assigned to only one operator that affects the
state or measurement result. For example, a $\ket{\biz}$ preparation
is affected only by $X$ errors.  The cnot errors shown are the maximum
marginal probabilities for the given Pauli operators.  The total error
probabilities are computed by summing the probabilities of each Pauli
error. The number of physical qubits per logical qubits at
level $5$ is $1024$.
}
\end{heretab}

\begin{heretab}
\label{tab:examples2}
{\small\begin{tabular}{|l||l|l|l|ll||}
\hline
Error probabilities:  & \multicolumn{5}{l||}{$e_p=e_m=0.002, e_c=0.008$} 
\\\hline\hline
Level 1 &\multicolumn{5}{l||}{Spectator:\ $\kets{+}{S}$}
\\\hline
Independence quality & 
   \multicolumn{5}{l||}{1.11} 
\\\hline
Prep./Meas. error&  
    $X: 1.71*10^{-4}$ & $Z: 3.61*10^{-5}$ &  &
    Max: & $1.71*10^{-4}$
\\\hline
Cnot marginal error& 
    $X: 8.09*10^{-4}$ & $Z: 2.35*10^{-4}$ & $Y:2.76*10^{-6} $ &
    Total: & $1.27*10^{-3}$
\\\hline
Hadamard error & 
    $X: 4.46*10^{-4}$ & $Z: 4.47*10^{-4}$ & $Y: 6.62*10^{-5}$  &
    Total: & $9.59*10^{-4}$
\\\hline
Decode error & 
    $X: 1.15*10^{-3}$ & $Z: 3.26*10^{-3}$ & $Y: 1.11*10^{-3}$ &
    Total: & $5.52*10^{-3}$
\\\hline\hline
Level 2 &\multicolumn{5}{l||}{Spectator:\ $\kets{\biz}{S}$}
\\\hline
Independence quality & 
   \multicolumn{5}{l||}{1.14} 
\\\hline
Prep./Meas. error&  
    $X: 2.58*10^{-7}$ & $Z: 5.32*10^{-7}$ &  &
    Max: & $5.32*10^{-7}$
\\\hline
Cnot marginal error& 
    $X: 2.56*10^{-6}$ & $Z: 3.63*10^{-6}$ & $Y: 1.85*10^{-11} $ &
    Total: & $7.01*10^{-6}$
\\\hline
Hadamard error & 
    $X: 3.82*10^{-6}$ & $Z: 3.82*10^{-6}$ & $Y: 1.07*10^{-8}$  &
    Total: & $7.66*10^{-6}$ 
\\\hline
Decode error & 
    $X: 3.21*10^{-3}$ & $Z: 1.08*10^{-2}$ & $Y: 1.08*10^{-3}$ &
    Total: & $5.38*10^{-3}$
\\\hline\hline
Level 3 &\multicolumn{5}{l||}{Spectator:\ $\kets{\biz}{S}$}
\\\hline
Independence quality & 
   \multicolumn{5}{l||}{1.52} 
\\\hline
Prep./Meas. error&  
    $X: 5.95*10^{-13}$ & $Z: 1.16*10^{-10}$ &  &
    Max: & $1.16*10^{-10}$
\\\hline
Cnot marginal error& 
    $X: 1.86*10^{-11}$ & $Z: 9.03*10^{-10}$ & $Y: 9.29*10^{-21} $ &
    Total: & $1.04*10^{-9}$ 
\\\hline
Hadamard error & 
    $X: 3.52*10^{-10}$ & $Z: 3.52*10^{-10}$ & $Y: 2.48*10^{-16}$  &
    Total: & $7.04*10^{-10}$ 
\\\hline
Decode error & 
    $X: 3.21*10^{-3}$ & $Z: 1.08*10^{-3}$ & $Y: 1.08*10^{-3}$ &
    Total: & $5.37*10^{-3}$ 
\\\hline\hline
\end{tabular}}
\vspace*{.2in}
\heretabcap{Details for physical error probabilities of $0.2\;\%$ for state
preparation and measurement and $0.8\;\%$ for cnot. See Table~\ref{tab:examples1}
for a description of the entries. The number of physical qubits
per logical qubit at level $3$ is $64$.}
\end{heretab}

\section{General threshold}
\label{sect:general threshold}

A strategy for implementing a standard quantum computation is to use
postselected quantum computation for preparing the encoded states
needed for operations and error-correction based on a code $C$ with
good error-correction properties. Provided the logical errors
($C$-logical errors) for gates acting on states encoded with $C$ are
sufficiently small, standard schemes can be applied to $C$-encoded
qubits to achieve scalability. These schemes are most efficient if the
starting error rates (in this case the $C$-logical error rates) are
already small. However, the cost of small $C$-logical errors can be a
substantial increase in overhead for postselected state
preparation. The tradeoffs involved are yet to be investigated.  The
states needed for computation with $C$-encoded qubits are encoded
stabilizer states, encoded Bell pairs with cnot or Hadamard gates
pre-applied, and encoded states such as $\ket{\pi/8}$.  The latter may
be noisy and purified later if desired. State preparation proceeds by
first obtaining the desired state encoded at a suitable level of the
concatenated error-detecting codes, then decoding the error-detecting
code hierarchy. Three types of error affect the prepared state. The
first is error associated with the encoded network used to prepare the
state in encoded form. These errors are estimated by the logical error
model at the top level of the error-detecting code hierarchy.  The
second is local error from decoding.  It is, to a good approximation,
independent from qubit to qubit and can be computed as shown in
Tables~\ref{tab:examples1},~\ref{tab:examples2}.  The third is due to
any memory errors arising from delays while waiting for the
postselected measurements to complete.  The contribution of the first
error type to $C$-logical errors can be estimated as the product of
the number of logical gates used in the preparation of the needed
states and the maximum error probability for top-level encoded
gates. For scaling to work straightforwardly with $C$-encoded qubits
and gates, the contribution of the first error type needs to be small
compared to the effect on the logical qubits of local error due to
decoding. This can be achieved by using sufficiently many levels of
the error-detecting code.

To be specific, assume that the maximum $C$-logical error must be
below $10^{-4}$ for standard scalability schemes to apply and consider
the first example, with memory/preparation error of $1\;\%$ and cnot
error of $3\;\%$. The decoding error is less than $2.3\;\%$ per qubit at
any level after the second. Choose $C$ such that given an error
probability per qubit of $11\;\%$ or less, the probability that the
error is not correctable is less than $0.5*10^{-4}$. Such codes exist.
Let $L$ be the length of an encoding network for pairs of encoded Bell
pairs with a cnot pre-applied. Choose a level of encoding with the
error-detecting code such that the top level logical error rate
is much less than $0.5*10^{-4}/L$ per gate. This implies that the
intrinsically uncorrectable error per $C$-logical gate implemented by
error-correcting teleportation is well below $0.5*10^{-4}$.  It
suffices to show that error that cannot be corrected due to excessive
local errors arising from decoding the error-detecting code and other
operations needed for the $C$-logical gate implementation is also
below $0.5*10^{-4}$. Referring to~\cite{knill:qc2003b}, the scheme
works if the total error probability from a Bell measurement, a
possible memory delay and twice the $2.3\;\%$ decoding error is less
than $11\;\%$.  The Bell measurement error can be estimated as $5\;\%$
(one cnot, two measurements). Therefore, if the contribution from
memory error is less than about $0.4\;\%$, the scheme is expected to be
successful.  However, the resources required for state preparation as
outlined are daunting: at least five levels of concatenation of the
error-detecting codes are required and the code $C$ is likely to be
complex.  If the error rates of the second example apply, the number
of levels of error-detecting codes and the required error-tolerance of
$C$ can both be significantly reduced. In general, a large reduction
in overhead results from having lower physical error rates.

\section{Discussion}
\label{sect:discussion}

The evidence presented above suggests that it is possible to quantum
compute with error rates above $1\;\%$ per gate.  The next steps are to
improve the efficiency of the scheme and to better determine the
resources required as a function of error rates. Some aspects of the
scheme as used in the analysis above are chosen for the purpose of
simplifying the analysis. For example, the two purification steps in
each encoded-Bell-state preparation can probably be replaced by one,
or by a cat-state syndrome-extraction procedure as mentioned in
Sect.~\ref{sect:preparation of the encoded Bell pair}.  Further
improvements can be obtained by adapting the code used to each level's
error behavior.  An important contribution to resource overheads is
the probability of successful state preparations.  This probability
decreases rapidly with more levels of error-detecting code or with the
complexity of the first level codes used for standard quantum
computation. Without using special techniques, this probability
decreases exponentially with the number of qubits involved.  As a
result, one should switch from a purely postselected scheme to one
that permits recovery from error as soon as possible.  One idea is to
combine the advantages of both for state preparation at higher levels.

An interesting question is to consider the relationship between
postselected and standard quantum computation in the presence of
errors. Is it true that if postselected quantum computation is
scalable for an error model, then so is standard quantum computation?
Alternatively, it is possible that there are error models for which 
one can quantum compute arbitrarily accurately, but only with
exponentially small probability of success. In the calculations shown
here, scalability of CSS gates seems to imply that universal quantum
computation is scalable. Is this always the case?

\begin{acknowledgments}
This work was supported by the U.S. National Security Agency. It is a
contribution of the National Institute of Standards and Technology, an
agency of the U.S. government, and is not subject to U.S. copyright.
\end{acknowledgments}

\bibliography{journalDefs,qc,analysis}

\ignore{
\pagebreak
\appendix
\section{Implementation of the computer assisted heuristic analysis}
\label{sect:mathematica functions}

{
\footnotesize
(*\begin{verbatim}
 * Mathematica code.
 *
 * Computation of syndrome probability distributions.
 * Two methods are given.
 *  1. Formal method, where the error model is parametrized
 *    by one parameter so that likelihoods are expressible
 *    as "error polynomials".
 *  2. Numeric method, using high precision as needed
 *    to ensure that values obtained are accurate enough.
 * 
*)
<<LinearAlgebra`MatrixManipulation`;
Off[LinearSolve::nosol];

progressdisplay = True;
pdinterval = 2^14;

epmaxe = 1/500; (* Maximum value for the error paramter to be 
  * considered. This is used for truncating, that is
  * rigorously bounding from above the likelihoods of 
  * syndromes. *)
eptruncate = -1; (* Maximum degree term to compute exactly. 
  * Terms of degree 6 are maintained, but their coefficient
  * is a rigorous upper bound assuming that the error parameter
  * is at most epmaxe. A lower bound is obtained by omitting
  * the degree 6 term. *)
eprec = 45;

(* Multiply two one-variable polynomials expressed as a list of
 * integers with reduction to degree eptruncate+1 using epmaxe. 
 * Ceiling[] operates on rational numbers to
 * give a hard upper bound. 
 * If the polynomials have degree 0, just multiply as numbers.*)
truncmul[x_,y_] := Block[{tv},
    If[Length[x] == 1,
      tv={x[[1]]*y[[1]]},
      tv=ListConvolve[x,y,{1,-1},0];
      trunc[tv,Min[Length[x],Length[y]]]
    ]
  ];
trunc[x_,l_] := Append[Take[tv,l-1],
                   Ceiling[Apply[Plus,
                         MapIndexed[(#1*epmaxe^(#2[[1]]-1))&,
                           Take[tv,l-1-Length[tv]]]]]];
truncfill[x_,l_] := (* Fill or bound polynomial-as-list 
                     * x to l terms. *)
    If[Length[x]> l,
       trunc[x],
       Join[x,Table[0,{l-Length[x]}]]
      ];
truncfill[x_] := truncfill[x,eptruncate+2];

(* Quadratic forms needed for stabilizer manipulations. 
 * Stabilizer operators are represented by binary row vectors
 * of length 2*n, for n qubits. *)
Sl[n_] := BlockMatrix[Outer[Times, IdentityMatrix[n], {{0,0},{1,0}}]];
Su[n_] := BlockMatrix[Outer[Times, IdentityMatrix[n], {{0,1},{0,0}}]];
S[n_] := Sl[n]+Su[n];

utones[l_] := Table[If[i<j,1,0],{i,1,l},{j,1,l}];
ut[m_] := m*utones[Length[m]]; (* The strictly upper triangular 
                                  part of a matrix. *)

hd = {{0,1},{1,0}}; (* Hadamard transform's effect on Pauli products
                       in the Z_2 representation. *)
  hdpos[a_] = (* Auxiliary function used to apply 
               * Hadamard to qubit a. *)
      Flatten[Outer[List,{2*a-1,2*a},{2*a-1,2*a}],1];
cnt = {{1,0,1,0},{0,1,0,0},{0,0,1,0},{0,1,0,1}};  
              (* Cnot in Z_2 representation. *)
  cntpos[a_,b_] = (* Auxiliary function for applying cnot 
                   * to qubits a,b *)
      Flatten[Outer[List,{2*a-1,2*a,2*b-1,2*b},
                         {2*a-1,2*a,2*b-1,2*b}],1];
   (* Target is qubit 2. The convention is to multiply 
    * row vectors on the right.*)
had[a_,l_]  := (* Hadamard acting on qubit a among l qubits. *)
    ReplacePart[IdentityMatrix[2*l],hd,
                hdpos[a],hdpos[1]];
cnot[a_,b_,l_] := (* Cnot acting on qubits a (control) 
                   * and b (target) among l qubits. *)
    ReplacePart[IdentityMatrix[2*l],cnt,
                cntpos[a,b],cntpos[1,2]];

(* List of binary numbers in list form. 
 * The least significant digit first: *)
allbins[l_] := Table[Reverse[IntegerDigits[i,2,l]],{i,0,2^l-1}];
(* An auxiliary function: *)
binpwrs[l_] := (binpwrs[l]=Table[2^r,{r,0,l-1}]); 
(* Binary in list form to number: *)
bintonum[x_] := x.binpwrs[Length[x]]; 
(* Number to binary in list form: *)
numtobin[x_,l_] := Reverse[IntegerDigits[x,2,l]]; 

(* The following nicely format error polynomials for printing. 
   It also specifies the conventions used for operators. *)
xyzs[x_] := StringJoin[Partition[x,2] /. {{1,0}->"X", {0,1}->"Z", 
                                          {1,1}->"Y", {0,0}->"I"}];
xyzs[List[List[x__],y__]] := Map[xyzs[#]&, List[List[x],y]];
xyzs[List[List[x__]]] := Map[xyzs[#]&, List[List[x]]];
xyzs[ep[x_,g_]] := 
  sep[
    Apply[Plus,
          MapIndexed[
           (Apply[Plus,
             MapIndexed[(#1*e^(#2[[1]]-1))&,#1]]
             *sy[StringJoin[numtobin[#2[[1]]-1,
                            Length[g]]/.{0->"0",1->"1"}]])&,x]],
    xyzs[g]
  ];
prxyzs[ep[x_,g_]] :=
  (* Print error polynomial. *)
  Block[{p1,d},
    (* p1=x[[1]]; If[Length[p1] == 1, d=p1[[1]], d=1]; *)
    d=1;
    Print["sep["];
    MapIndexed[(Print["  sy[\"", 
                StringJoin[numtobin[#2[[1]]-1,
                           Length[g]]/.{0->"0",1->"1"}], 
                      "\"]*(", 
                      InputForm[Apply[Plus,
                        MapIndexed[Function[{x1,x2}, (x1*e^(x2[[1]]-1))],#1]]/d],"),"])&,x];
    Print["  ", InputForm[xyzs[g]]];
    Print["];"];
  ];

(* Converting X,Y,Z,I strings to binary format. *)
pprs[List[x__]] := Map[pprs,List[x]];
pprs[x_] := If[StringQ[x],Flatten[Table[StringTake[x,{i}],
                          {i,1,StringLength[x]}] /. 
                   {"X"->{1,0}, "Y"->{1,1}, "Z"->{0,1}, "I"->{0,0}}],
               x];
syrs[x_] := Table[StringTake[x,{i}],
                  {i,1,StringLength[x]}] /. {"0"->0, "1"->1};

(* Converting the printed output of prxyzs back to internal error polynomial
 * form, useful when reading back data saved in a text file. *)
sep[x__,g_] :=
  Block[{v,tg,l,c,tc,k,sye,es,sn},
    tg = pprs[g]; l=Length[tg];
    tc = List[x];
    c = Table[0,{2^l},{eptruncate+2}];
    For[k=1,k<=Length[tc],k=k+1,
      sye=tc[[k]]; 
       (* Has the form:
        * sy["01000111"]*(2*e^2+284*e^3+8986*e^4+115344*e^5+832515*e^6) *)
      es = CoefficientList[sye/.{sy[_]->1},e];
      sn = bintonum[syrs[sye/.{sy[r_]*u_ -> r}]];
      c[[sn+1]] = es;
    ];
    ep[c,tg]
  ];

(*
 * Error polynomials.
 * ep[sypoly,g]: 
 *    - sypoly is the table of table of coefficients of the syndromes.
 *      Each coefficient is stored as a table of coefficients up
 *      to degree truncate+1 (length=eptruncate+2).
 *    - g is the generator matrix for the code whose syndromes appear.
 *      In particular, Length[sypoly] = 2^(Length[g]).
 *      g must be non-singular and define a maximal stabilizer (i.e.
 *      its codes are one-dimensional).
 * syp[sym[syterm,sy],...]: Sparse syndrome polynomial.
 *    - syterm is the coefficient of sy (error polynomial as
 *      list of coefficients or numeric likelihood.
 *    - sy is a 0-1 vector representing a syndrome.
 *   
*)

(* Evaluate likelihoods for a given error parameter. *)
evep[ep[x_,g_],e_] :=
  ep[Map[evep[#1,e]&,x],g];
evep[syp[y_]] := Map[evep[#1,e]&,syp[y]];
evep[sym[a_,b_]] := sym[evep[a,e],b];
evep[x_,e] := {Apply[Plus,MapIndexed[Function[x1,x2,(x1*e^(x2[[1]]-1))],x]]};

(* Sum rule for error polynomials: *)
epplusep[ep[x_,g_],ep[y_,g_]] := ep[x+y,g];

(* Multiply an error polynomial by a sparse syndrome polynomial. *)
eptimes[ep[x_,g_],syp[sym[a_,b_]]] := eptimes[ep[x,g],sym[a,b]];
eptimes[ep[x_,g_],syp[sym[a_,b_],y__]] :=
  epplusep[eptimes[ep[x,g],sym[a,b]],eptimes[ep[x,g],syp[y]]]
eptimes[ep[x_,g_],sym[a_,b_]] :=
  Block[{l,nx},l=Length[g];nx=Table[0,{2^l},{Length[x[[1]]]}];
    If[progressdisplay && Length[x]>pdinterval,
       Print["eptimes: sym[", StringJoin[b/.{0->"0",1->"1"}], "]..."],
       Null];
    MapIndexed[(nx[[bintonum[Mod[numtobin[#2[[1]]-1,l]+b,2]]+1]] 
                = truncmul[a,#1];#1)&,x];
    ep[nx,g]
  ];


(* Apply Pauli Product. *)
eptimes[ep[x_,g_], pp[y_]] := 
  eptimes[ep[x,g], sym[1,pptosy[y,g]]];
pptosy[y_, g_] := Mod[(g.S[Length[y]/2]).y,2];

(* Error polynomial generator matrix row transformation. *)
eptimes[ep[x_,g_], btr[m_]] :=
  Block[{l,nx,q},l=Length[g];nx=Table[0,{2^l},{Length[x[[1]]]}];
    q=btrq[m,g]; 
    MapIndexed[(nx[[bintonum[Mod[(m.numtobin[#2[[1]]-1,l])+q,2]]+1]] 
                = #1)&,x];
    ep[nx,Mod[m.g,2]]
  ];
btrq[m_,g_] := 
  Table[0, {i,1,Length[g]}];
  (* Now simplified, 
   * without effect on CSS codes. Originally it was: 
   * Mod[Table[m[[i]].ut[g.Sl[Length[g[[1]]]/2].Transpose[g]].m[[i]],
   *        {i,1,Length[g]}],2];
   *)

(* Error polynomial gate application. *)
eptimes[ep[x_,g_], gm[m_]] := ep[x,Mod[g.m,2]];
eptimes[ep[x_,g_], had[a_]] := 
  eptimes[ep[x,g],gm[had[a,Length[g[[1]]]/2]]];
eptimes[ep[x_,g_], cnot[a_,b_]] := 
  eptimes[ep[x,g],gm[cnot[a,b,Length[g[[1]]]/2]]];

(*
 * Sparse polynomials representing errors to be applied.
*)

errgen[p_,g_] := (* p is a list of pairs consisting
  * of Pauli products and error polynomials as lists.
  * g is the stabilizer generator matrix to use. 
  * Output is a sparse error polynomial with the identity error
  * included at a multiple of 1. *)
  Block[{l,v,c},
    l=Length[g[[1]]]/2; 
    v=Table[0,{2*l}];c=Table[0,{eptruncate+2}];
    Apply[syp,Table[sym[If[i==0,ReplacePart[c,1,1],truncfill[p[[i]][[2]]]],
          If[i==0,pptosy[v,g],
          pptosy[pprs[p[[i]][[1]]],g]]],
          {i,0,Length[p]}]]
  ];
errone[a_,p_,g_] := (* An error model to be applied to qubit a. 
  * p and g are as in errgen[]. *)
  Block[{xp,v},xp={};v=Table[0,{Length[g[[1]]]}];
    Map[(xp=Append[xp,{ReplacePart[v,pprs[#1[[1]]],
                     {{2*a-1},{2*a}},{{1},{2}}],#1[[2]]}];0)&,p];
    errgen[xp,g]
  ];
errtwo[a_,b_,p_,g_] := (* An error model to be applied to qubits a,b. *)
  Block[{xp,v},xp={};v=Table[0,{Length[g[[1]]]}];
    Map[(xp=Append[xp,
      {ReplacePart[v,pprs[#1[[1]]],
        {{2*a-1},{2*a},{2*b-1},{2*b}},{{1},{2},{3},{4}}],#1[[2]]}];0)&,p];
    errgen[xp,g]
  ];

hsymone[p_] := (* Hadamard symmetrize a one qubit error specification. *) 
  Block[{xpt,i,k},
    xpt=Table[truncfill[{0}],{i,0,3}];
    For[i=1,i<=Length[p],i=i+1,
      k=bintonum[pprs[p[[i]][[1]]]]+1;
      xpt[[k]]=xpt[[k]]+(1/2)*truncfill[p[[i]][[2]]];
      k=bintonum[pprs[p[[i]][[1]]].hd]+1;
      xpt[[k]]=xpt[[k]]+(1/2)*truncfill[p[[i]][[2]]];
    ];
    Delete[MapIndexed[({numtobin[#2[[1]]-1,2],#1})&,xpt],1]
  ];

hcsymtwo[p_] := (* Symmetrize a two qubit error specification
                 * under parallel H and any cnot. *)
  Block[{xpt,xptt,i,k},
    xpt=Table[truncfill[{0}],{i,0,15}];
    (* Parallel H: *)
    For[i=1,i<=Length[p],i=i+1,
      k=bintonum[pprs[p[[i]][[1]]]]+1;
      xpt[[k]]=xpt[[k]]+(1/2)*truncfill[p[[i]][[2]]];
      k=bintonum[pprs[p[[i]][[1]]].had[1,2].had[2,2]]+1;
      xpt[[k]]=xpt[[k]]+(1/2)*truncfill[p[[i]][[2]]];
    ];
    xptt=Table[truncfill[{0}],{i,0,15}];
    For[i=2,i<=16,i=i+1,
      k=i;
      xptt[[k]]=xptt[[k]]+(1/6)*xpt[[i]];
      k=bintonum[Mod[numtobin[i-1,4].cnot[1,2,2],2]]+1;
      xptt[[k]]=xptt[[k]]+(1/6)*xpt[[i]];
      k=bintonum[Mod[numtobin[i-1,4].cnot[2,1,2],2]]+1;
      xptt[[k]]=xptt[[k]]+(1/6)*xpt[[i]];
      k=bintonum[Mod[numtobin[i-1,4].cnot[1,2,2].cnot[2,1,2],2]]+1;
      xptt[[k]]=xptt[[k]]+(1/6)*xpt[[i]];
      k=bintonum[Mod[numtobin[i-1,4].cnot[2,1,2].cnot[1,2,2],2]]+1;
      xptt[[k]]=xptt[[k]]+(1/6)*xpt[[i]];
      k=bintonum[Mod[numtobin[i-1,4].cnot[1,2,2].cnot[2,1,2].cnot[1,2,2],2]]+1;
      xptt[[k]]=xptt[[k]]+(1/6)*xpt[[i]];
    ];
    
    Delete[MapIndexed[({numtobin[#2[[1]]-1,4],#1})&,xptt],1]
  ];

(* Specify the error polynomials for operators in
 * each equivalence class: *)
symerrsone[x_,y_] := hsymone[{{"X",2*x},{"Y",y}}];
(* Specify the error polynomial for operators in each equivalence class: *)
symerrstwo[xi_,yi_,xz_] := 
  hcsymtwo[{{"XI",6*xi},{"YI",6*yi},{"XZ",3*xz}}];



(* Rules for applying error models to an error polynomial. *)
eptimes[ep[x_,g_],errone[a_,e_]] := 
    eptimes[ep[x,g],errone[a,e,g]];
eptimes[ep[x_,g_],errtwo[a_,b_,e_]] := 
    eptimes[ep[x,g],errtwo[a,b,e,g]];
eptimes[ep[x_,g_],errgen[e_]] := 
    eptimes[ep[x,g],errgen[e,g]];

(* Applying gates without error. *)
tcnot[ep[x_,g_], a_, b_] :=
   eptimes[ep[x,g],cnot[a,b]];
thad[ep[x_,g_], a_] :=
   eptimes[ep[x,g],had[a]];
tprep[op_] := (* Makes a one qubit code in the specfied state. *)
    ep[{Join[{1},Table[0,{eptruncate+1}]],
                  Table[0,{eptruncate+2}]},{pprs[op]}];
tprep[ep[x_,g_], op_] := (* Adds one qubit at the end,
   * prepared in +1 eigenstate of op. Follows with error merr. *)
   epjoin[ep[x,g], tprep[op,Length[x[[1]]]]]; 
tprep[op_,l_] := ep[{Join[{1},Table[0,{l-1}]],Table[0,{l}]},{pprs[op]}];
tprepn[op_] := (* Makes a one qubit code in the specfied state. *)
    ep[{Join[{SetPrecision[1,eprec]},
             Table[SetPrecision[0,eprec],{eptruncate+1}]],
                  Table[SetPrecision[0,eprec],{eptruncate+2}]},{pprs[op]}];
tprepn[ep[x_,g_], op_] := (* Adds one qubit at the end,
   * prepared in +1 eigenstate of op. Follows with error merr. *)
   epjoin[ep[x,g], tprepn[op,Length[x[[1]]]]]; 
tprepn[op_,l_] := ep[{Join[{SetPrecision[1,eprec]},
                           Table[SetPrecision[0,eprec],{l-1}]],
                      Table[SetPrecision[0,eprec],{l}]},{pprs[op]}];
tmeas0[ep[x_,g_],a_,op_] :=
  meas0[ep[x,g], a, op];

(* Applying gates with error. *)
ecnot[ep[x_,g_], a_, b_, cerr_] :=
   eptimes[eptimes[ep[x,g],cnot[a,b]],errtwo[a,b,cerr]];
ehad[ep[x_,g_], a_, merr_] :=
   eptimes[eptimes[ep[x,g],had[a]],errone[a,merr]];
eprep[op_, merr_] := (* Makes a one qubit code in the specfied state. *)
    eptimes[ep[{Join[{1},Table[0,{eptruncate+1}]],
                  Table[0,{eptruncate+2}]},{pprs[op]}], errone[1,merr]];
eprep[ep[x_,g_], op_, merr_] := (* Adds one qubit at the end,
  * prepared in +1 eigenstate of op. Follows with error merr. *)
  eptimes[epjoin[ep[x,g], eprep[op,Length[x[[1]]]]], 
           errone[Length[g]+1,merr]]; 
eprep[op_,l_, merr_] := 
  eptimes[ep[{Join[{1},Table[0,{l-1}]],Table[0,{l}]},{pprs[op]}],
          errone[1,merr]];
eprepn[op_, merr_] := (* With high precision floats. *)
    eptimes[ep[{Join[{SetPrecision[1,eprec]},
                  Table[SetPrecision[0,eprec],{eptruncate+1}]],
                  Table[SetPrecision[0,eprec],{eptruncate+2}]},
               {pprs[op]}], errone[1,merr]];
eprepn[ep[x_,g_], op_, merr_] := (* Adds one qubit at the end,
  * prepared in +1 eigenstate of op. Follows with error merr. *)
  eptimes[epjoin[ep[x,g], eprepn[op,Length[x[[1]]]]], 
           errone[Length[g]+1,merr]]; 
eprepn[op_,l_, merr_] := 
  eptimes[ep[{Join[{SetPrecision[1,eprec]},
              Table[SetPrecision[0,eprec],{l-1}]],
              Table[SetPrecision[0,eprec],{l}]},{pprs[op]}],
          errone[1,merr]];
emeas0[ep[x_,g_],a_,op_,merr_] :=
  meas0[eptimes[ep[x,g], errone[a,merr]], a, op];
    
    

(*
 * Direct join of two codes and their error polynomials.
*)

epjoin[ep[x_,g_],ep[y_,h_]] :=
  Block[{l,k,r,s},
    l=Length[g];k=Length[h];
    r=Length[g[[1]]]/2;s=Length[h[[1]]]/2;
    v=Table[0,{2^k},{2^l},{Length[x[[1]]]}];
    ep[Flatten[MapIndexed[(truncmul[x[[#2[[2]]]],y[[#2[[1]]]]])&,
                          v,{2}],1],
         Join[Transpose[Join[Transpose[g],Table[0,{2*s},{l}]]],
              Transpose[Join[Table[0,{2*r},{k}],Transpose[h]]]]]
  ];

(*
 * Measurement of a qubit.
*)

meas0[ep[x_,g_],l_,op_] :=
  Block[{r,j,k,gl,trsf,lc,q,v,ng,b}, 
    If[Head[op] === String, r=pprs[op], r=op];
    gl=Length[g];
    (* Determine a basis change of g such that only one row anticommutes
       with op. *) 
    (* Find the first row that has an anticommuting op. *)
    k=0; trsf=IdentityMatrix[gl];
    For[j=1, (k==0)&&(j<= gl), j=j+1,
      If[(g[[j]][[{2*l-1,2*l}]].S[1]).r != 0,k=j,Null]
    ];
    If[k>0,
      (* Since g defines commuting ops, the measured operator 
       * is not in the span of g. *)
      b=1;
      For[j=k+1, j<= gl, j=j+1,
        If[g[[j]][[{2*l-1,2*l}]].S[1].r != 0,
          trsf[[j,k]]=1,Null]
      ],
      (* The measured operator is in the span of g, 
       * bring it to the top. *)
      b=0;
      lc = LinearSolve[Transpose[g], 
                       ReplacePart[Table[0,{Length[g[[1]]]}], r, 
                          {{2*l-1},{2*l}}, {{1},{2}}], 
                       Modulus->2];
      k==0;
      For[j=1, (k==0)&&(j<= gl), j=j+1,
        If[lc[[j]]==1, k=j, Null];
      ];
      trsf[[k]]=lc
    ];
    q=btrq[trsf,g]; v=Table[0,{2^(gl-1)},{Length[x[[1]]]}]; 
    ng=Mod[trsf.g,2];
    If[progressdisplay && Length[x]>pdinterval,
       Print["meas0: ", op, "..."],
       Null];
    MapIndexed[(
        Block[{c,ni},
          ci=Mod[(trsf.numtobin[#2[[1]]-1,gl])+q,2];
          ni=bintonum[Delete[ci,k]]+1;
          If[b==0 && ci[[k]]==1, 0, v[[ni]] += #1];
        ]
      )&,x]; 
    ep[v,Transpose[Delete[Transpose[Delete[ng,k]],{{2*l-1},{2*l}}]]]
  ];


(*
 * Purify two of the same.
 * Works for codes with a basis of pure X, or pure Z stabilizers
 * (CSS codes). Else the 7-qubit Hamming code is needed to
 * purify by applying this code's decoding algorithm transversally.
*)

purify[ep[x_,g_],"Z", merr_, cerr_] :=
  Block[{y,l,j},
    l=Length[g[[1]]]/2;
    y=epjoin[ep[x,g],ep[x,g]];
    For[j=1,j<=l,j=j+1,
      If[progressdisplay,Print["purify: Qubit ", j, "..."],Null];
      y=emeas0[ecnot[y,1+l,1,cerr],1,"Z",merr];
    ];
    y
  ];
purify[ep[x_,g_],"X", merr_, cerr_] :=
  Block[{y,l,j},
    l=Length[g[[1]]]/2;
    y=epjoin[ep[x,g],ep[x,g]];
    For[j=1,j<=l,j=j+1,
      If[progressdisplay,Print["purify: Qubit ", j, "..."],Null];
      y=emeas0[ecnot[y,1,l+1,cerr],1,"X",merr];
    ];
    y
  ];

(* The following must be after all other eptimes definitions to
 * avoid unintended use. *)
eptimes[ep[x_,g_],List[a__]] := ep[Map[(truncmul[List[a],#])&,x],g];
eptimes[ep[x_,g_],a_] := ep[a*x,g];

(*
 * Algorithms needed to represent independent error models
 * and compute near-independent parameters given syndrome distributions.
 *)

(*
 * Independent qubits error model data structure.
 * ie[{{ex,ez,ey},...},{{xp,zp},...}, {w,...}]
 *   The first list gives the error likelihoods, the second
 *   the Pauli products associated with the likelihoods.
 *   Each pair is the Pauli product induced by one 
 *   independent contribution,
 *   with xp the associated X Pauli product and zp the 
 *   Z Pauli product.
 *   The last parameter assigns integer weights to each error operator.
 *   It is used to determine at what order the operator should occur.
 *)

(*
 * Algorithm for determining near-independent
 * error parameters given a syndrome error distribution and an independent
 * qubits error model.
 *)
nialg[ie[erp_,ps_, wps_], ep[x_,g_]] :=
  (* Returns {ie[newerp,ps,wps],mins}, where
   * mins is the minimum weight Pauli product for each syndrome. *)
  Block[{lg,lp,ln,mw,fp,nsp,rsp,tech,etos,mins,ws,iee,oerp,j,k,t1,t2,t3,t4,t5},
    If[Length[x[[1]]] > 1, 
      Print["nialg: Can't operate on non-numeric error polynomials."];
      Return[Null], 
      Null
    ];

    lg = Length[g]; lp=Length[ps];
    fp = Map[(pptosy[#1,g])&,Map[pprs,Flatten[ps,1]]];

    (* Construct etos and ws:
     * etos[i] is the list of minimum weight error combinations
     * for the i'th syndrome, ws[[i]] is the minimum weight. *)    
    mw = Apply[Plus,wps];
    ws = Table[mw+1,{2^lg}];
    For[k=0,k<2^(2*lp),k=k+1,
      If[progressdisplay && (lp>7) && (Mod[k,2^14] == 0),
        Print["nialg: Compiling etos at ", k, "..."]
        ,
        Null
      ];

      t1=numtobin[k,2*lp];
      t2=Mod[t1.fp,2];
      t3=bintonum[t2]+1;
      t4=minw[t1,wps];
      If[t4<ws[[t3]],
        etos[t3]={};ws[[t3]]=t4;
        ,
        Null;  
      ];

      If[t4==ws[[t3]],
        etos[t3] = Append[etos[t3],t1];
        ,
        Null;
      ];
    ];

    mins = Table["",{2^lg}];
    For[k=0,k<2^lg,k=k+1,
      t1 = etos[k+1][[1]];
      t2 = Table[0,{Length[pprs[ps[[1,1]]]]}];
      For[j=1,j<=lp,j=j+1,
        If[t1[[2*j-1]] == 1,
          t2 = t2+pprs[ps[[j,1]]];
          ,
          Null;
        ];
        If[t1[[2*j]] == 1,
          t2 = t2+pprs[ps[[j,2]]];
          ,
          Null
        ];
      ];
      mins[[k+1]] = xyzs[Mod[t2,2]];
    ];

    (* Distribute error probability to minimal error sets. 
     * iee is to contain the distributed error probabilities. *)
    iee = Table[0,{2^(2*lp)}];
    For[k=0,k<2^lg,k=k+1,
      If[ws[[k+1]] > mw,
        Print["nialg: Syndrome without associated errors found."];
        Return[Null]
        ,
        Null
      ];
      t1 = Length[etos[k+1]];
      t2 = x[[k+1]][[1]];
      For[j=1,j<=t1,j=j+1,
        iee[[bintonum[etos[k+1][[j]]]+1]] = t2/t1;
      ];
    ];

    (* Determine minimum independent error parameters for which
     * likelihood ratios are bounded. *)
    oerp = Table[{0,0,0},{lp}];
    For[k=0,k<2^(2*lp),k=k+1,
      If[progressdisplay && (lp>7) && (Mod[k,2^14] == 0),
        Print["nialg: Compiling oerp at ", k, "..."]
        ,
        Null
      ];
      t1=numtobin[k,2*lp];
      For[j=1,j<=lp,j=j+1,
        t2 = t1[[{2j-1,2j}]];
        If[!(t2 === {0,0}),
          t3 = bintonum[t2];
          t4 = bintonum[ReplacePart[t1,{0,0},{{2j-1},{2j}},{{1},{2}}]];
          t5 = If[iee[[t4+1]] > 0, iee[[k+1]]/iee[[t4+1]],
                  If[iee[[k+1]] > 0, Infinity,0]];
          If[oerp[[j,t3]]< t5, oerp[[j,t3]]=t5, Null]
          ,
          Null
        ];
      ];
    ];

    {ie[oerp,ps,wps],mins}
  ];


minw[x_] :=
  Block[{w,j},
    w=0;
    For[j=1,j<=Floor[Length[x]/2],j=j+1,
      If[x[[2j-1]]+x[[2j]]>0,w=w+1,Null];
    ];
    w
  ];

minw[x_,wx_] :=
  Block[{w,j},
    w=0;
    For[j=1,j<=Floor[Length[x]/2],j=j+1,
      If[x[[2j-1]]+x[[2j]]>0,w=w+wx[[j]],Null];
    ];
    w
  ];

(*
 * Algorithm for determining near-independent
 * error parameters given a syndrome error distribution and 
 * a list of independent error sets, each error set
 * is a list of Pauli products, each to be given its own
 * likelihood within the set.
 *)
nialgl[ie[erp_,ps_, wps_], ep[x_,g_], marg_] :=
  (* Note: erp[[j]] here has the same length as ps[[j]]. 
   * erp can be given as {}. *)
  Block[{lg,lp,np,nps,mw,lmg,
         ln,nsp,rsp,tech,etos,ws,iee,oerp,j,k,t1,t2,t3,t4,t5},
    If[Length[x[[1]]] > 1, 
      Print["nialgl: Can't operate on non-numeric error polynomials."];
      Return[Null], 
      Null
    ];

    lg = Length[g]; lp=Length[ps];
    nps = Map[(Length[#1]+1)&,ps];
    np = Apply[Times, nps]; 
      (* Number of error combinations to consider. *)

    (* Construct etos and ws:
     * etos[i] is the list of minimum weight error combinations
     * for the i'th syndrome, ws[[i]] is the minimum weight. *)    
    ws = Table[Infinity,{2^lg}];
    lmg = Prepend[Map[(-Log[2,#1[[2,1]]])&, marg],0];
    For[k=0,k<np,k=k+1,
      If[progressdisplay && (lp>7) && (Mod[k,2^14] == 0),
        Print["nialgl: Compiling etos at ", k, "..."]
        ,
        Null
      ];
      (**)
      t1=numtostr[k,nps];
      t2=Table[0,{lg}];
      For[j=1,j<=lp,j=j+1,
        If[t1[[j]] > 0,
          t2 = Mod[t2+pptosy[pprs[ps[[j,t1[[j]]]]],g],2];
          ,
          Null;
        ];
      ];
      t3=bintonum[t2]+1;
      t4=minwl[t1,ps,wps,lmg];
      If[t4<ws[[t3]],
        etos[t3]={};ws[[t3]]=t4;
        ,
        Null;  
      ];
      (**)
      If[t4==ws[[t3]],
        etos[t3] = Append[etos[t3],t1];
        ,
        Null;
      ];
    ];

    (* Distribute error probability to minimal error sets. 
     * iee is to contain the distributed error probabilities. *)
    iee = Table[0,{np}];
    For[k=0,k<2^lg,k=k+1,
      If[Length[etos[k+1]] <= 0,
        Print["nialgl: Syndrome without associated errors found."];
        Return[Null]
        ,
        Null
      ];
      t1 = Length[etos[k+1]];
      t2 = x[[k+1]][[1]];
      For[j=1,j<=t1,j=j+1,
        iee[[strtonum[etos[k+1][[j]],nps]+1]] = t2/t1;
      ];
    ];

    (* Determine minimum independent error parameters for which
     * likelihood ratios are bounded. *)
    oerp = Table[Table[0,{Length[ps[[j]]]}],{j,1,lp}];
    For[k=0,k<np,k=k+1,
      If[progressdisplay && (lp>7) && (Mod[k,2^14] == 0),
        Print["nialgl: Compiling oerp at ", k, "..."]
        ,
        Null
      ];
      t1=numtostr[k,nps];
      For[j=1,j<=lp,j=j+1,
        t2 = t1[[j]];
        If[!(t2 == 0),
          t4 = strtonum[ReplacePart[t1,0,j],nps];
          t5 = If[iee[[t4+1]] > 0, iee[[k+1]]/iee[[t4+1]],
                  If[iee[[k+1]] > 0, Infinity,0]];
          If[oerp[[j,t2]]< t5, oerp[[j,t2]]=t5, Null];
          ,
          Null;
        ];
      ];
    ];

    ie[oerp,ps,wps]
  ];


numtostr[k_,nps_] := 
  Block[{j,ck,sk,m1},
    ck=k;sk={};
    For[j=1,j<=Length[nps],j=j+1,
      m1=Mod[ck,nps[[j]]];
      sk=Append[sk,m1];
      ck=(ck-m1)/nps[[j]];
    ];
    sk
  ];

strtonum[str_,nps_] :=
  Block[{j,k,n},
    k=0;n=1;
    For[j=1,j<=Length[str],j=j+1,
      k=k+str[[j]]*n;
      n=n*nps[[j]];
    ];
    k
  ];


minwl[x_] :=
  Block[{w,j},
    w=0;
    For[j=1,j<=Length[x],j=j+1,
      If[x[[j]]>0,w=w+1,Null];
    ];
    w
  ];

minwl[x_,wx_] :=
  Block[{w,j},
    w=0;
    For[j=1,j<=Length[x],j=j+1,
      If[x[[j]]>0,w=w+wx[[j]],Null];
    ];
    w
  ];

minwl[x_,ps_,wx_] :=
  Block[{w,j},
    w=0;
    For[j=1,j<=Length[x],j=j+1,
      If[x[[j]]>0,
        w=w+(wx[[j]]*minw[pprs[ps[[j,x[[j]]]]]]);
        ,
        Null];
    ];
    w
  ];

minwl[x_,ps_,wx_, lmg_] :=
  Block[{w,j},
    w=0;
    For[j=1,j<=Length[x],j=j+1,
      If[x[[j]]>0,
        w=w+(wx[[j]]*minwm[pprs[ps[[j,x[[j]]]]],lmg]);
        ,
        Null];
    ];
    w
  ];

minwm[pp_,lmg_] :=
  Block[{j,t1,t4},
      t4 = 0;
      For[j=1,j<=Length[pp]/2,j=j+1,
        t1 = bintonum[pp[[{2j-1,2j}]]];
        t4 = t4+lmg[[t1+1]];
      ];
      t4
  ];



(* 
 * Algorithm for inferring next level error model.
 *)
nextlevel[eep_, eem_, ec_, spect_, level_] :=
   nextlevel[eep, eem, ec, spect, level, "purify", "margec"]; 

nextlevel[eep_, eem_, ec_, spect_, level_, method_, marginal_] :=
  (*
   * eep, eem: prep and measure error model. Format as for errone[].
   * ec: cnot error mode. Format as for errtwo[].
   * spect = "Z" or "X", the state to be used for the spectator qubit.
   * level: Current level. 0=physical. This affects the cnot induced error rate
   *        computation.
   * method: "purify" or "catspurr".
   * Note: For simplicity, this is going to be uniform in each level,
   *       Rather than optimized using odd/even techniques.
   * 
   * Output: {newperr, newmerr, newcnerr, localerr, intrnl}.
   *   localerr is the lower level
   *   model use to bound the one applicable to 
   *   purified encoded Bell states. It is a list of
   *   a first half and second half model.
   *   intrnl: A list of values of internal variables for analysis.
   *
  *)
  Block[{fep,tep,npl,npr,npt,npte,nptre,ps,
          trnsfl,trnsfr,np,
          nppaa,inde,indeaa,maxindeaa,minsf,
          marg,
          berr,
          mx, mz, px, pz,
          t1,t2,t3,t4,t5,
          k,l,n,m,
          newperr, newmerr, newcnerr, intrnl},
    (* Initial states for the two halves of the encoded Bell state. *)
    If[spect == "X",
      fep = eprepn[eprepn[eprepn[eprepn["Z",eep],"Z",eep],"X",eep],"X",eep];
      fep = ecnot[ecnot[fep,3,1,ec],4,2,ec];
      tep = eprepn[eprepn[eprepn[eprepn["Z",eep],"X",eep],"X",eep],"X",eep];
      tep = ecnot[ecnot[ecnot[tep,2,1,ec],4,1,ec],3,2,ec];
      trnsfl = btr[
       {{0,0,1,1,0,0,0,0},
        {1,1,0,0,1,0,0,0},
        {0,0,0,1,0,0,0,0},
        {0,0,1,1,0,0,1,1},
        {0,0,0,0,1,0,0,0},
        {0,0,0,1,0,1,0,1},
        {0,0,0,0,0,1,0,0},
        {1,0,0,0,0,0,0,0}}
      ];
      trnsfr = btr[
       {{0,0,1,1,0,0,1,1},
        {1,0,0,0,0,0,0,0},
        {0,1,0,1,0,0,0,1},
        {0,0,0,0,0,0,1,1},
        {1,0,0,0,1,1,0,0},
        {0,0,0,0,0,0,0,1},
        {0,1,0,0,0,0,0,0},
        {0,0,0,0,1,0,0,0}}
      ];
      ,
      fep = eprepn[eprepn[eprepn[eprepn["X",eep],"Z",eep],"Z",eep],"Z",eep];
      fep = ecnot[ecnot[ecnot[fep,1,2,ec],1,4,ec],2,3,ec];
      tep = eprepn[eprepn[eprepn[eprepn["Z",eep],"X",eep],"Z",eep],"X",eep];
      tep = ecnot[ecnot[tep,2,1,ec],4,3,ec];
      trnsfl = btr[
       {{1,0,0,0,0,0,0,0},
        {0,0,1,1,1,0,1,0},
        {0,1,1,1,0,0,1,0},
        {1,0,0,0,0,1,0,1},
        {0,0,0,0,1,0,1,0},
        {0,0,0,0,0,0,1,0},
        {0,0,0,0,0,1,0,0},
        {0,1,1,0,0,0,0,0}}
      ];
      trnsfr = btr[
       {{0,1,0,1,1,0,0,0},
        {1,0,1,0,0,0,0,0},
        {0,0,1,0,0,0,0,0},
        {0,0,0,0,1,0,0,0},
        {1,0,1,0,0,0,1,1},
        {0,0,1,0,0,1,1,1},
        {0,1,0,0,0,0,0,0},
        {0,0,0,0,0,1,1,0}}
      ];
    ];
      
    (* Combine the two halves. *)
    npl = ecnot[ecnot[ecnot[ecnot[epjoin[fep,tep],5,1,ec],
                                 6,2,ec],7,3,ec],8,4,ec];
    npr = ecnot[ecnot[ecnot[ecnot[epjoin[tep,fep],1,5,ec],
                                 2,6,ec],3,7,ec],4,8,ec];

    (* Change bases to match and be in standard form. *)
    npl=eptimes[npl,trnsfl];
    npr=eptimes[npr,trnsfr];

    (* Average the two. *)
    np=eptimes[epplusep[npl,npr],1/2];

    If[method == "purify", 
      (* Purify. *)
      If[spect == "X",
        (* Purify X errors first. *)
        nppaa = purify[purify[
                  purify[purify[np,"Z",eem,ec],"X",eem,ec],
                  "Z",eem,ec],"X", eem,ec];
        ,
        (* Purify Z errors first. *)
        nppaa = purify[purify[
                  purify[purify[np,"X",eem,ec],"Z",eem,ec],
                  "X",eem,ec],"Z", eem,ec];
      ];
      nppaa[[1]] = nppaa[[1]]/nppaa[[1,1,1]];

      (* Determine bounding parameters for
       * independent qubit error mode. *)

      If[marginal == "marged", 
        marg = marged[eep,eem,ec];
        ,
        marg = margec[ec];
      ];
      ps =  {
             "XIIII","ZIIII",
             "IXIII","IZIII",
             "IIXII","IIZII",
             "IIIXI","IIIZI"
            };
      If[spect == "X",
        npt = tmeas0[tmeas0[tmeas0[nppaa,6,"X"],6,"Z"],6,"X"];
        ,
        npt = tmeas0[tmeas0[tmeas0[nppaa,6,"X"],6,"Z"],6,"Z"];
      ];
      minsf = minws[ps,marg,npt[[2]]];
      minsf = Delete[Map[(StringDrop[#1,-1])&,minsf],1];
  
      inde = ie[{},
             {
              Map[(StringJoin[#1,"IIII"])&, minsf],
              Map[(StringJoin["IIII",#1])&, minsf]
             },
             {1,1}
           ];
      indeaa = nialgl[inde,nppaa,marg];
  
      berr = {Table[{minsf[[k]],{indeaa[[1,1,k]]}},{k,1,Length[minsf]}],
              Table[{minsf[[k]],{indeaa[[1,2,k]]}},{k,1,Length[minsf]}]};
      ,
      (* Cat's purr. *)
      If[progressdisplay, 
         Print["nextlevel: purr..."], Null];
      nppaa = catspurr[catspurr[np,eep,eem,ec,spect],
                                eep,eem,ec,spect];
      If[progressdisplay, 
         Print["nextlevel:   purr..."], Null];
      nppaa = catspurr[catspurr[nppaa,eep,eem,ec,spect],
                                eep,eem,ec,spect];
      nppaa[[1]] = nppaa[[1]]/nppaa[[1,1,1]];

      (* Estimate the independent model. *)

      (* Construct the likelihood matrices. *)
      (* Full matrix: 32x32, index scheme from 8 bit syndromes
       * associated with nppaa:
       * aaa000aa, 000bbbbb, then reduce on last two bits *)
      Block[ {fmx,tmx,
              q,r,s,t,br,bs,bt,
              s0,u0,v0,l0,r0,
              lh,rh,lhrh,berrs,berrd,greds,gredd,fill,
              tl,tl0,tl1,tl2,tl3,tl4,tl5},

        For[r=0,r<=3,r=r+1, fmx[r]=Table[0,{8},{8}]];
        For[r=0,r<=3,r=r+1, tmx[r]=fmx[r]];

        For[r=1,r<=8,r=r+1,
          For[s=1,s<=8,s=s+1,
            br=numtobin[r-1,3]; bs=numtobin[s-1,3];
            For[t=0,t<=3,t=t+1,
              tl = numtobin[t,2];
              bt[t] = bintonum[Join[br,bs,tl]]+1;
              tmx[t] = ReplacePart[tmx[t], nppaa[[1,bt[t],1]],{r,s}];
            ];
            (**)
            For[t=0,t<=3,t=t+1,
              tl = numtobin[t,2];
              fmx[t] = ReplacePart[fmx[t], 
                (nppaa[[1,bt[0],1]]
                 +((-1)^tl[[1]])*nppaa[[1,bt[1],1]]
                 +((-1)^tl[[2]])*nppaa[[1,bt[2],1]]
                 +((-1)^tl[[1]])*((-1)^tl[[2]])*nppaa[[1,bt[3],1]]), {r,s}];
            ];
          ];
        ];
        (* Extract maximum singular values. *)
        For[t=0,t<=3,t=t+1,
          {u0[t],s0[t],v0[t]} = SingularValues[fmx[t]];
          l0[t]=Sqrt[s0[t][[1]]]*u0[t][[1]];r0[t]=Sqrt[s0[t][[1]]]*v0[t][[1]];
        ];

        (* Make first block exact. *)

        For[t=0,t<=3,t=t+1,
          If[(l0[t][[1]]*r0[t][[1]]) == 0, 
            Print["nextlevel: Attempt to divide by 0."];
            tl1 = fmx[t][[1,1]];
            l0[t]=ReplacePart[l0[t],
              Sqrt[tl1],1]; 
            r0[t]=ReplacePart[r0[t],
              Sqrt[tl1],1];
            ,
            tl1 = fmx[t][[1,1]]/(l0[t][[1]]*r0[t][[1]]);
            l0[t]=ReplacePart[l0[t],
              Sqrt[tl1]*l0[t][[1]],1]; 
            r0[t]=ReplacePart[r0[t],
              Sqrt[tl1]*r0[t][[1]],1];
          ];
        ];
        
        (* Invert to 32 element likelihood distributions. *)
        lh = Table[tl=numtobin[t,2];
                   (l0[0]+((-1)^tl[[1]])*l0[1]
                         +((-1)^tl[[2]])*l0[2]
                         +((-1)^tl[[1]])*((-1)^tl[[2]])*l0[3])/4, {t,0,3}];
        lh = Flatten[Transpose[lh]];

        rh = Table[tl=numtobin[t,2];
                   (r0[0]+((-1)^tl[[1]])*r0[1]
                         +((-1)^tl[[2]])*r0[2]
                         +((-1)^tl[[1]])*((-1)^tl[[2]])*r0[3])/4, {t,0,3}];
        rh = Flatten[Transpose[rh]];
        (* Adjust to ensure bounding and positivity property. *)

        (* Set all negative values to zero. *)
        If[(Min[lh[[{1,2,3,4}]]] < 0) || (Min[rh[[{1,2,3,4}]]]),
          Print["nextlevel: Negative value in first blocks of lh or rh."];
          ,
          Null;
        ];
        For[r=1,r<=32,r=r+1,
          If[lh[[r]]<0, lh[[r]] = 0, Null];
          If[rh[[r]]<0, rh[[r]] = 0, Null];
        ];

        (* Adjust rest of first row, column. *)
        For[r=2,r<=8,r=r+1,
          For[t=0,t<=3,t=t+1,
            tl = Transpose[{lh}].{rh};
            tl0 = fblock[tl,1,r];
            tl2 = ftr[tl0,t];
            If[tl2 < tmx[t][[1,r]],
              tl4 = fxvec[lh,1,0];
              tl3 = (tmx[t][[1,r]]-tl2)/tl4;
              rh = favec[rh,r,4*tl3,0];
              ,
              Null;
            ];
            tl0 = fblock[tl,r,1];
            tl2 = ftr[tl0,t];
            If[tl2 < tmx[t][[r,1]],
              tl4 = fxvec[rh,1,0];
              tl3 = (tmx[t][[r,1]]-tl2)/tl4;
              lh = favec[lh,r,4*tl3,0];
              ,
              Null;
            ];
          ];
        ];

        (* Adjust the rest. *)
        For[r=2,r<=8,r=r+1,
          For[s=2,s<=8,s=s+1,
            t = Mod[s+r-4,7]+2;
            For[q=0,q<=3,q=q+1,
              tl = Transpose[{lh}].{rh};
              tl0 = fblock[tl,s,t];
              tl4 = fxvec[lh,s,0];
              tl5 = fxvec[rh,t,0];
              tl2 = ftr[tl0,q];
              If[tl2 < tmx[q][[s,t]],
                tl3 = tmx[q][[s,t]]-tl2;
                If[tl4>tl5,
                  rh = favec[rh,t,4*tl3/tl4,0];
                  ,
                  lh = favec[lh,s,4*tl3/tl5,0];
                ];
                ,
                Null;
              ];
            ];
          ];
        ];

        minwsx = 
          {"IIII", "IIZI", "IXII", "IXZI", 
           "ZZII", "IIIZ", "ZYII", "IIIY", 
           "ZIZI", "ZIII", "IYIZ", "ZXII", 
           "IZZI", "IZII", "IYZI", "IYII", 
           "XXII", "IXYI", "XIII", "IIYI", 
           "YYII", "XIIY", "YZII", "XIIZ", 
           "YXZI", "YXII", "YIZI", "YIII",
           "IYYI", "XYII", "IZYI", "XZII"};
        minwsz = 
          {"IIII", "IIZI", "IXII", "IXZI", 
           "XIXI", "XIYI", "IIIX", "IIIY", 
           "ZIZI", "ZIII", "IYZI", "IYII", 
           "YIYI", "YIXI", "ZIIY", "ZIIX", 
           "XXII", "IIYX", "XIII", "XIZI", 
           "IXXI", "IXYI", "IIXI", "IIYI", 
           "YXZI", "YXII", "YIZI", "YIII",
           "IYYI", "IYXI", "ZIYI", "ZIXI"};
        greds = {nppaa[[2,1]],nppaa[[2,2]],nppaa[[2,3]],
                nppaa[[2,7]],nppaa[[2,8]]};
        gredd = {nppaa[[2,4]],nppaa[[2,5]],nppaa[[2,6]],
                nppaa[[2,7]],nppaa[[2,8]]};
        fill = Table[0,{8}];

        (* Construct error model suitable for localref input. *)
        If[spect == "X",
          berrs = Table[{minwsx[[r]],
            {lh[[1+bintonum[
              tl1 = pptosy[Join[pprs[minwsx[[r]]],fill],greds];
              {tl1[[4]],tl1[[5]],tl1[[1]],tl1[[2]],tl1[[3]]}
                    ]]]/lh[[1]]}},
                    {r,2,32}];
          berrd = Table[{minwsx[[r]],
            {rh[[1+bintonum[
              tl1 = pptosy[Join[fill,pprs[minwsx[[r]]]],gredd];
              {tl1[[4]],tl1[[5]],tl1[[1]],tl1[[2]],tl1[[3]]}
                    ]]]/rh[[1]]}},
                    {r,2,32}];
          ,
          berrs = Table[{minwsz[[r]],
            {lh[[1+bintonum[
              tl1 = pptosy[Join[pprs[minwsz[[r]]],fill],greds];
              {tl1[[4]],tl1[[5]],tl1[[1]],tl1[[2]],tl1[[3]]}
                    ]]]/lh[[1]]}},
                    {r,2,32}];
          berrd = Table[{minwsz[[r]],
            {rh[[1+bintonum[
              tl1 = pptosy[Join[fill,pprs[minwsz[[r]]]],gredd];
              {tl1[[4]],tl1[[5]],tl1[[1]],tl1[[2]],tl1[[3]]}
                    ]]]/rh[[1]]}},
                    {r,2,32}];
        ];
        berr = {berrs,berrd};

      ];


    ];
    intrnl = {nppaa};

    (* Apply error model to error free states, npt and nptr: *)
    {npte, nptref,nptrel} = localref[spect,berr];
    intrnl = Append[intrnl,npte]; 
      (* To allow comparison between inferred error model
       * and the independent, bounding one. *)

    (* State preparation and measurement error, 
     * possibly different rates. *)
    (* Prepare X: *)
    px = emeas0[emeas0[emeas0[emeas0[nptref,
         1,"X",eem],1,"X",eem],1,"X",eem],1,"X",eem];

    (* Prepare Z: *)
    pz = emeas0[emeas0[emeas0[emeas0[nptref,
         1,"Z",eem],1,"Z",eem],1,"Z",eem],1,"Z",eem];

    (* Measure X: *)
    mx = emeas0[emeas0[emeas0[emeas0[nptrel,
         1,"X",eem],1,"X",eem],1,"X",eem],1,"X",eem];

    (* Measure Z: *)
    mz = emeas0[emeas0[emeas0[emeas0[nptrel,
         1,"Z",eem],1,"Z",eem],1,"Z",eem],1,"Z",eem];

    (* Figure out combined error model. *)
    t1 = pz[[1,2,1]]/pz[[1,1,1]];
    t2 = px[[1,2,1]]/px[[1,1,1]];
    newperr =  {{"X",{t1}},
                 {"Z",{t2}},
                 {"Y",{0}}};   
    t1 = mz[[1,2,1]]/mz[[1,1,1]];
    t2 = mx[[1,2,1]]/mx[[1,1,1]];
    newmerr =  {{"X",{t1}},
                 {"Z",{t2}},
                 {"Y",{0}}};   
    If[progressdisplay, 
       Print["nextlevel: newperr = ", newperr//InputForm];
       Print["           newmerr = ", newmerr//InputForm],
       Null];

    (* cnot error.*)
    If[level == 0,
     (* Efficient ordering at physical level. *)
      t1 = emeas0[ecnot[epjoin[nptref,nptrel],1,6,ec],6,"Z",eem];
       t1 = emeas0[ecnot[t1,2,6,ec],6,"Z",eem];
        t1 = emeas0[ecnot[t1,3,6,ec],6,"Z",eem];
         t1 = emeas0[ecnot[t1,4,6,ec],6,"Z",eem];
      t2 = emeas0[ecnot[epjoin[nptref,nptrel],6,1,ec],6,"X",eem];
       t2 = emeas0[ecnot[t2,6,2,ec],6,"X",eem];
        t2 = emeas0[ecnot[t2,6,3,ec],6,"X",eem];
         t2 = emeas0[ecnot[t2,6,4,ec],6,"X",eem];
      t3 = epjoin[t1,t2];
       t3 = emeas0[emeas0[ecnot[t3,1,7,ec],7,"Z",eem],1,"X",eem];
        t3 = emeas0[emeas0[ecnot[t3,1,6,ec],6,"Z",eem],1,"X",eem];
         t3 = emeas0[emeas0[ecnot[t3,1,5,ec],5,"Z",eem],1,"X",eem];
          t3 = emeas0[emeas0[ecnot[t3,1,4,ec],4,"Z",eem],1,"X",eem];
      (* Undo cnot on source to get error model mapped onto syndromes.*)
      t4 = tcnot[t3,1,3];
      (* Standardize the syndrome, the map goes from qubits 1,3 to qubits
       * 2,4. The cnot was from 2 to 4. *)
      t5 = eptimes[t4, btr[{{0,1,0,0},{1,0,0,0},{0,1,1,0},{0,0,0,1}}]];
      ,
     (* Because the implementation is full transversal through the
      * levels rather than strictly concatenated, much of the error
      * is already accounted for. Do a virtual less efficient
      * implementation with only actual cnot subject to error. *)
      t1 = ecnot[ecnot[ecnot[ecnot[epjoin[nptref,nptref],1,6,ec],2,7,ec],
                   3,8,ec],4,9,ec];
      t2 = tmeas0[tmeas0[tcnot[epjoin[t1,nptrel],1,11],11,"Z"],1,"X"];
       t2 = tmeas0[tmeas0[tcnot[t2,1,10],10,"Z"],1,"X"];
        t2 = tmeas0[tmeas0[tcnot[t2,1,9],9,"Z"],1,"X"];
         t2 = tmeas0[tmeas0[tcnot[t2,1,8],8,"Z"],1,"X"];
      (**)
      t3 = tmeas0[tmeas0[tcnot[epjoin[t2,nptrel],2,8],8,"Z"],2,"X"];
       t3 = tmeas0[tmeas0[tcnot[t3,2,7],7,"Z"],2,"X"];
        t3 = tmeas0[tmeas0[tcnot[t3,2,6],6,"Z"],2,"X"];
         t3 = tmeas0[tmeas0[tcnot[t3,2,5],5,"Z"],2,"X"];
      (**)
      t5 = tcnot[t3,1,2]; (* Note, this is from 1,2 to 3,4 *)
     ];

    (* Compute new error model from syndrome. *)
    newcnerr = {};l=Length[t5[[2]]];
    For[k=1,k<2^l,k=k+1,
      t1=numtobin[k,l];
      t1=xyzs[{t1[[2]],t1[[1]],t1[[4]],t1[[3]]}];
      newcnerr = Append[newcnerr,{t1,{t5[[1,k+1,1]]/t5[[1,1,1]]}}];
    ];

    If[progressdisplay, Print["nextlevel: newcnerr = ", newcnerr//InputForm],
       Null];
    {newperr, newmerr, newcnerr, berr, intrnl}
  ];

decerr[ep_, em_, ec_, spect_, codeerr_, decerr_] :=
  (* Compute the local error by decoding at this point.
   * em, ep and ec are the *physical* error rates for measurement and
   *   cnots.
   * spect is the current state of the spectator qubit.
   * codeerr is the currently active (bounding) error model for
   *   the bell pair. 
   * decerr is the induced decoding error for each qubit
   *   from below.
   * Output:
   *  newlocalerr
   *)
  Block[{npte,nptref,nptrel,t1,t2},
    {npte, nptref, nptrel} = localref[spect,codeerr];

    (* Apply qubit by qubit, prior decoding error to nptrel. *)
    eptimes[eptimes[eptimes[eptimes[nptrel,
      errone[1,decerr]],errone[2,decerr]],errone[3,decerr]],
      errone[4,decerr]];    

    If[spect == "X",
      t1=emeas0[ecnot[emeas0[ecnot[emeas0[ecnot[nptrel,
           4,2,ec],4,"X",eem],
           3,1,ec],3,"X",eem],
           1,2,ec],2,"Z",eem];
      ,
      t1=emeas0[ecnot[emeas0[ecnot[emeas0[ecnot[nptrel,
           3,4,ec],4,"Z",eem],
           1,2,ec],2,"Z",eem],
           2,1,ec],2,"X",eem];
    ];

    t2 = t1[[1,1,1]];
    newlocalerr = {{"X",{t1[[1,3,1]]/t2}},
           {"Z",{t1[[1,2,1]]/t2}},
           {"Y",{t1[[1,4,1]]/t2}}};

    newlocalerr
  ];

haderr[eep_, eem_, eec_, eeh_, spect_, localerr_, level_] :=
  (* Compute the error in encoded Hadamard gates. *)
  Block[{npte,nptref, nptrel,t1,t2},
    {npte, nptref, nptrel} = localref[spect,localerr];
    If[level == 0,
      (* Apply Hadamard and Bell measure. *)
      t1 = emeas0[emeas0[ecnot[ehad[epjoin[nptref,nptrel],1,eeh],1,6,eec],
             6,"Z",eem],1,"X",eem];
      t1 = emeas0[emeas0[ecnot[ehad[t1,1,eeh],1,6,eec],
             6,"Z",eem],1,"X",eem];
      t1 = emeas0[emeas0[ecnot[ehad[t1,1,eeh],1,4,eec],
             4,"Z",eem],1,"X",eem];
      t1 = emeas0[emeas0[ecnot[ehad[t1,1,eeh],1,3,eec],
             3,"Z",eem],1,"X",eem];
      (* Undo at source: *)
      t2 = thad[t1,1];      
       ,
      (* Apply Hadamard and assume ideal Bell measurement. *)
      t1 = tmeas0[tmeas0[tcnot[ehad[epjoin[nptref,nptrel],1,eeh],1,6],
             6,"Z"],1,"X"];
      t1 = tmeas0[tmeas0[tcnot[ehad[t1,1,eeh],1,6],
             6,"Z"],1,"X"];
      t1 = tmeas0[tmeas0[tcnot[ehad[t1,1,eeh],1,4],
             4,"Z"],1,"X"];
      t1 = tmeas0[tmeas0[tcnot[ehad[t1,1,eeh],1,3],
             3,"Z"],1,"X"];
      t2 = thad[t1,1];      
     ];
    (**)
    (* New error model computation. *)
    {{"X", {t2[[1,3,1]]/t2[[1,1,1]]}}, {"Z", {t2[[1,2,1]]/t2[[1,1,1]]}}, 
     {"Y", {t2[[1,4,1]]/t2[[1,1,1]]}}}
  ];

(* Auxiliary, for preparing states with reference entanglement
 * and the local error model applied. *)
localref[spect_,localerr_] := 
  Block[{fept,tept,trnsfl,npt,nptr,npte,nptref,nptrel},
    (* The following could be made global (no need to recompute),
     * but doesn't take much time. *)
    If[spect == "X",
      fept = tprepn[tprepn[tprepn[tprepn["Z"],"Z"],"X"],"X"];
      fept = tcnot[tcnot[fept,3,1],4,2];
      tept = tprepn[tprepn[tprepn[tprepn["Z"],"X"],"X"],"X"];
      tept = tcnot[tcnot[tcnot[tept,2,1],4,1],3,2];
      trnsfl = btr[
       {{0,0,1,1,0,0,0,0},
        {1,1,0,0,1,0,0,0},
        {0,0,0,1,0,0,0,0},
        {0,0,1,1,0,0,1,1},
        {0,0,0,0,1,0,0,0},
        {0,0,0,1,0,1,0,1},
        {0,0,0,0,0,1,0,0},
        {1,0,0,0,0,0,0,0}}
      ];
      ,
      fept = tprepn[tprepn[tprepn[tprepn["X"],"Z"],"Z"],"Z"];
      fept = tcnot[tcnot[tcnot[fept,1,2],1,4],2,3];
      tept = tprepn[tprepn[tprepn[tprepn["Z"],"X"],"Z"],"X"];
      tept = tcnot[tcnot[tept,2,1],4,3];
      trnsfl = btr[
       {{1,0,0,0,0,0,0,0},
        {0,0,1,1,1,0,1,0},
        {0,1,1,1,0,0,1,0},
        {1,0,0,0,0,1,0,1},
        {0,0,0,0,1,0,1,0},
        {0,0,0,0,0,0,1,0},
        {0,0,0,0,0,1,0,0},
        {0,1,1,0,0,0,0,0}}
      ];
    ];
    (* Combine the two halves. *)
    npt = tcnot[tcnot[tcnot[tcnot[epjoin[fept,tept],5,1],
                                 6,2],7,3],8,4];
    (**)
    (* Change bases to match and be in standard form. *)
    npt=eptimes[npt,trnsfl];
    (**)
    (* Reduced version of error-free version, npt, with reference
     * entanglement, reference qubit is qubit 5.*)
    If[spect == "X",
       nptr = eptimes[tmeas0[tmeas0[tmeas0[npt,8,"X"],7,"Z"],6,"X"],
         btr[{{1,0,0,0,0},{0,1,0,0,0},{0,0,1,0,0},{0,0,0,1,0},{0,0,0,0,1}}]];
       ,
       nptr = eptimes[tmeas0[tmeas0[tmeas0[npt,8,"Z"],7,"Z"],6,"X"],
         btr[{{1,0,0,0,0},{0,1,0,0,0},{0,0,1,0,0},{0,0,0,1,0},{0,0,0,0,1}}]];
    ];       
    (**)
    (* Apply error model to error free states, npt and nptr: *)
    npte = eptimes[eptimes[npt,
      errgen[Map[({StringJoin[#1[[1]],"IIII"],#1[[2]]})&,localerr[[1]]]]],
      errgen[Map[({StringJoin["IIII",#1[[1]]],#1[[2]]})&,localerr[[2]]]]];
    nptref = eptimes[nptr,
      errgen[Map[({StringJoin[#1[[1]],"I"],#1[[2]]})&,localerr[[1]]]]];
    nptrel = eptimes[nptr,
      errgen[Map[({StringJoin[#1[[1]],"I"],#1[[2]]})&,localerr[[2]]]]];
    (**)
    If[Length[localerr] == 3,
      (* An independent "encoded" error has been provided. This gets
       * applied to the full state and the first half. *)
      npte = eptimes[npte,
         errgen[Map[({StringJoin[#1[[1]],"IIII"],#1[[2]]})&,localerr[[3]]]]];
      nptref = eptimes[nptr,
         errgen[Map[({StringJoin[#1[[1]],"I"],#1[[2]]})&,localerr[[3]]]]];
      ,
      Null;
    ];
    {npte,nptref,nptrel}
  ];

(* "Cat state" subroutines. *)
catx2[eep_, eem_, eec_] := 
    ecnot[eprepn[eprepn["Z",eep],"X",eep],2,1,eec];
catz2[eep_, eem_, eec_] := 
    ecnot[eprepn[eprepn["X",eep],"Z",eep],1,2,eec];

catx4[eep_, eem_, eec_] :=
  Block[{c1,c2,c3,c4},
    c1=eprepn[eprepn[eprepn[eprepn["Z",eep],"X",eep],"Z",eep],"Z",eep];
    c1=ecnot[ecnot[ecnot[c1,2,3,eec],2,1,eec],3,4,eec];
    (**)
    c2=catx2[eep,eem,eec];
    c3=ecnot[ecnot[epjoin[c1,c2],2,5,eec],3,6,eec];
    c3=emeas0[emeas0[c3,5,"Z",eem],5,"Z",eem];
    (**)
    c4=ecnot[ecnot[epjoin[c3,c2],1,5,eec],4,6,eec];
    c4=emeas0[emeas0[c4,5,"Z",eem],5,"Z",eem];
    (**)
    c4
  ];

catz4[eep_, eem_, eec_] :=
  Block[{c1,c2,c3,c4},
    c1=eprepn[eprepn[eprepn[eprepn["X",eep],"Z",eep],"X",eep],"X",eep];
    c1=ecnot[ecnot[ecnot[c1,3,2,eec],1,2,eec],4,3,eec];
    (**)
    c2=catz2[eep,eem,eec];
    c3=ecnot[ecnot[epjoin[c1,c2],5,2,eec],6,3,eec];
    c3=emeas0[emeas0[c3,5,"X",eem],5,"X",eem];
    (**)
    c4=ecnot[ecnot[epjoin[c3,c2],5,1,eec],6,4,eec];
    c4=emeas0[emeas0[c4,5,"X",eem],5,"X",eem];
    (**)
    c4
  ];

catspurr[epstr_, eep_, eem_, eec_, spect_] :=
  (* Purifies an encoded Bell state using the cat's scheme, one cycle *)
  Block[{t1,t2,t3,t4,trnsflx},
    (**)
    If[spect == "X",
      If[!(epstr[[2]] === 
         {{1,0,1,0,1,0,1,0,0,0,0,0,0,0,0,0},{0,1,0,1,0,1,0,1,0,0,0,0,0,0,0,0},
          {0,0,1,0,0,0,1,0,0,0,0,0,0,0,0,0},{0,0,0,0,0,0,0,0,1,0,1,0,1,0,1,0},
          {0,0,0,0,0,0,0,0,0,1,0,1,0,1,0,1},{0,0,0,0,0,0,0,0,0,0,1,0,0,0,1,0},
          {1,0,1,0,0,0,0,0,1,0,1,0,0,0,0,0},{0,1,0,0,0,1,0,0,0,1,0,0,0,1,0,0}}
        ),
        Print["catspurr: Not the expected stabilizer."];
        Return[Null];
        ,
        Null;
      ];
      (* Entangled "X": *)
      t3=emeas0[ecnot[epjoin[epstr,catx4[eep,eem,eec]],9,1,eec],9,"X",eem];
       t3 = emeas0[ecnot[t3,9,2,eec],9,"X",eem];
       t3 = emeas0[ecnot[t3,9,5,eec],9,"X",eem];
       t3 = emeas0[ecnot[t3,9,6,eec],9,"X",eem];
      t3=emeas0[ecnot[epjoin[t3,catx4[eep,eem,eec]],9,3,eec],9,"X",eem];
       t3 = emeas0[ecnot[t3,9,4,eec],9,"X",eem];
       t3 = emeas0[ecnot[t3,9,7,eec],9,"X",eem];
       t3 = emeas0[ecnot[t3,9,8,eec],9,"X",eem];
      (* Entangled "Z": *)
      t4=emeas0[ecnot[epjoin[t3,catz4[eep,eem,eec]],1,9,eec],9,"Z",eem];
       t4 = emeas0[ecnot[t4,3,9,eec],9,"Z",eem];
       t4 = emeas0[ecnot[t4,5,9,eec],9,"Z",eem];
       t4 = emeas0[ecnot[t4,7,9,eec],9,"Z",eem];
      t4=emeas0[ecnot[epjoin[t4,catz4[eep,eem,eec]],2,9,eec],9,"Z",eem];
       t4 = emeas0[ecnot[t4,4,9,eec],9,"Z",eem];
       t4 = emeas0[ecnot[t4,6,9,eec],9,"Z",eem];
       t4 = emeas0[ecnot[t4,8,9,eec],9,"Z",eem];
      (* Spectator "X" and company: *)
      t1=emeas0[ecnot[epjoin[t4,catx2[eep,eem,eec]],9,1,eec],9,"X",eem];
       t1=emeas0[ecnot[t1,9,3,eec],9,"X",eem];
      t1=emeas0[ecnot[epjoin[t1,catx2[eep,eem,eec]],9,2,eec],9,"X",eem];
       t1=emeas0[ecnot[t1,9,4,eec],9,"X",eem];
      t1=emeas0[ecnot[epjoin[t1,catx2[eep,eem,eec]],9,5,eec],9,"X",eem];
       t1=emeas0[ecnot[t1,9,7,eec],9,"X",eem];
      t1=emeas0[ecnot[epjoin[t1,catx2[eep,eem,eec]],9,6,eec],9,"X",eem];
       t1=emeas0[ecnot[t1,9,8,eec],9,"X",eem];
      (* Error detecting "Z": *)
      t2=emeas0[ecnot[epjoin[t1,catz4[eep,eem,eec]],1,9,eec],9,"Z",eem];
       t2 = emeas0[ecnot[t2,2,9,eec],9,"Z",eem];
       t2 = emeas0[ecnot[t2,3,9,eec],9,"Z",eem];
       t2 = emeas0[ecnot[t2,4,9,eec],9,"Z",eem];
      t2=emeas0[ecnot[epjoin[t2,catz4[eep,eem,eec]],5,9,eec],9,"Z",eem];
       t2 = emeas0[ecnot[t2,6,9,eec],9,"Z",eem];
       t2 = emeas0[ecnot[t2,7,9,eec],9,"Z",eem];
       t2 = emeas0[ecnot[t2,8,9,eec],9,"Z",eem];
      trnsflx = btr[
       {{0,1,1,0,0,0,0,0},
        {0,0,0,1,0,0,0,0},
        {0,1,0,0,0,0,0,0},
        {0,0,0,0,0,1,1,0},
        {0,0,0,0,0,0,0,1},
        {0,0,0,0,1,1,0,0},
        {0,1,1,0,1,1,1,0},
        {1,0,0,0,0,0,0,1}}
      ]; 
      t5 = eptimes[t2,trnsflx];
      ,
      If[!(epstr[[2]] === 
         {{1,0,1,0,1,0,1,0,0,0,0,0,0,0,0,0},{0,1,0,1,0,1,0,1,0,0,0,0,0,0,0,0},
          {0,0,0,0,0,1,0,1,0,0,0,0,0,0,0,0},{0,0,0,0,0,0,0,0,1,0,1,0,1,0,1,0},
          {0,0,0,0,0,0,0,0,0,1,0,1,0,1,0,1},{0,0,0,0,0,0,0,0,0,0,0,0,0,1,0,1},
          {1,0,1,0,0,0,0,0,1,0,1,0,0,0,0,0},{0,1,0,0,0,1,0,0,0,1,0,0,0,1,0,0}}
        ),
        Print["catspurr: Not the expected stabilizer."];
        Return[Null];
        ,
        Null;
      ];
      (* Entangled "Z": *)
      t3=emeas0[ecnot[epjoin[epstr,catz4[eep,eem,eec]],1,9,eec],9,"Z",eem];
       t3 = emeas0[ecnot[t3,3,9,eec],9,"Z",eem];
       t3 = emeas0[ecnot[t3,5,9,eec],9,"Z",eem];
       t3 = emeas0[ecnot[t3,7,9,eec],9,"Z",eem];
      t3=emeas0[ecnot[epjoin[t3,catz4[eep,eem,eec]],2,9,eec],9,"Z",eem];
       t3 = emeas0[ecnot[t3,4,9,eec],9,"Z",eem];
       t3 = emeas0[ecnot[t3,6,9,eec],9,"Z",eem];
       t3 = emeas0[ecnot[t3,8,9,eec],9,"Z",eem];
      (* Entangled "X": *)
      t4=emeas0[ecnot[epjoin[t3,catx4[eep,eem,eec]],9,1,eec],9,"X",eem];
       t4 = emeas0[ecnot[t4,9,2,eec],9,"X",eem];
       t4 = emeas0[ecnot[t4,9,5,eec],9,"X",eem];
       t4 = emeas0[ecnot[t4,9,6,eec],9,"X",eem];
      t4=emeas0[ecnot[epjoin[t4,catx4[eep,eem,eec]],9,3,eec],9,"X",eem];
       t4 = emeas0[ecnot[t4,9,4,eec],9,"X",eem];
       t4 = emeas0[ecnot[t4,9,7,eec],9,"X",eem];
       t4 = emeas0[ecnot[t4,9,8,eec],9,"X",eem];
      (* Spectator "Z" and company: *)
      t1=emeas0[ecnot[epjoin[t4,catz2[eep,eem,eec]],1,9,eec],9,"Z",eem];
       t1=emeas0[ecnot[t1,2,9,eec],9,"Z",eem];
      t1=emeas0[ecnot[epjoin[t1,catz2[eep,eem,eec]],3,9,eec],9,"Z",eem];
       t1=emeas0[ecnot[t1,4,9,eec],9,"Z",eem];
      t1=emeas0[ecnot[epjoin[t1,catz2[eep,eem,eec]],5,9,eec],9,"Z",eem];
       t1=emeas0[ecnot[t1,6,9,eec],9,"Z",eem];
      t1=emeas0[ecnot[epjoin[t1,catz2[eep,eem,eec]],7,9,eec],9,"Z",eem];
       t1=emeas0[ecnot[t1,8,9,eec],9,"Z",eem];
      (* Error detecting "X": *)
      t2=emeas0[ecnot[epjoin[t1,catx4[eep,eem,eec]],9,1,eec],9,"X",eem];
       t2 = emeas0[ecnot[t2,9,2,eec],9,"X",eem];
       t2 = emeas0[ecnot[t2,9,3,eec],9,"X",eem];
       t2 = emeas0[ecnot[t2,9,4,eec],9,"X",eem];
      t2=emeas0[ecnot[epjoin[t2,catx4[eep,eem,eec]],9,5,eec],9,"X",eem];
       t2 = emeas0[ecnot[t2,9,6,eec],9,"X",eem];
       t2 = emeas0[ecnot[t2,9,7,eec],9,"X",eem];
       t2 = emeas0[ecnot[t2,9,8,eec],9,"X",eem];
      trnsflx = btr[
       {{0,0,0,1,0,0,0,0},
        {0,0,1,0,0,0,0,0},
        {0,1,0,0,0,0,0,0},
        {0,0,0,0,0,0,0,1},
        {0,0,0,0,0,1,1,0},
        {0,0,0,0,1,0,0,0},
        {1,0,0,0,0,0,0,1},
        {0,1,1,0,1,0,1,0}}
      ]; 
      t5 = eptimes[t2,trnsflx];
    ];
    t5
  ];
  

(* Marginal cnot error probability computations. *)
margec[ec_] :=
  Block[{k,j,t1,t2,margl,margr,marg},
    margl = {{"X",{0}},{"Z",{0}},{"Y",{0}}};
    margr = margl;
    For[k=1,k<=Length[ec],k=k+1,
       t1=pprs[ec[[k,1]]];
       t2 = t1[[{1,2}]];
       If[! (t2 === {0,0}),
         margl[[bintonum[t2],2,1]] += ec[[k,2,1]];
         ,
         Null;
       ];
       t2 = t1[[{3,4}]];
       If[! (t2 === {0,0}),
         margr[[bintonum[t2],2,1]] += ec[[k,2,1]];
         ,
         Null;
       ];
    ];
    marg = margl;
    For[k=1,k<=3,k=k+1,
      If[margr[[k,2,1]] > margl[[k,2,1]], marg[[k,2,1]] = margr[[k,2,1]],Null];
    ];
  
    marg
  ];

marged[eep_, eem_, eec_] :=
  (* Compute the error distribution after purifying
   * a perfectly encoded Bell state twice. *)
  Block[{fep,fepa,fepb,fepp},
    fep = tcnot[tprepn[tprepn["X"], "Z"],1,2];
    fepa =  purify[purify[
                  purify[purify[fep,"Z",eem,eec],"X",eem,eec],
                  "Z",eem,eec],"X", eem,eec];
    fepb =  purify[purify[
                  purify[purify[fep,"X",eem,eec],"Z",eem,eec],
                  "X",eem,eec],"Z", eem,eec];
    (* Average: *)
    fepp = eptimes[epplusep[fepa,fepb],1/2];
    fepp[[1]] = fepp[[1]]/fepp[[1,1,1]];
    (* Error distribution: *)
    {{"X",{fepp[[1,3,1]]}},{"Z",{fepp[[1,2,1]]}},{"Y",{fepp[[1,4,1]]}}}
  ];

minws[p_,marg_,g_] :=
  (* Compute minimum weight errors. *)
  Block[{lg,lp,fp,ws,lmg,k,j,t1,t2,t3,t4,etos},
    lg = Length[g]; lp=Length[p];
    If[!(lg == StringLength[p[[1]]]),
      Print["minws: Length mismatch between g and p[[1]]."]; Return[Null];
      ,
      Null;
    ];
    fp = Map[pprs,p];
    ws = Table[Infinity,{2^lg}];
    (**)
    (* Construct etos and ws:
     * etos[i] is the list of minimum weight error combinations
     * for the i'th syndrome, ws[[i]] is the minimum weight. *)    
    lmg = Prepend[Map[(-Log[2,#1[[2,1]]])&, marg],0];
    For[k=0,k<2^(2*lp),k=k+1,
      If[progressdisplay && (lp>7) && (Mod[k,2^14] == 0),
        Print["minws: Compiling etos at ", k, "..."]
        ,
        Null
      ];
      (**)
      t1=numtobin[k,lp];
      t2=Mod[t1.fp,2];
      t3=bintonum[pptosy[t2,g]]+1;
      t4 = minwm[t2,lmg];
      (**)
      If[t4<ws[[t3]],
        etos[t3]={};ws[[t3]]=t4;
        ,
        Null;  
      ];
      (**)
      If[t4==ws[[t3]],
        etos[t3] = Append[etos[t3],t1];
        ,
        Null;
      ];
    ];
    (**)
    mins = Table["",{2^lg}];
    For[k=0,k<2^lg,k=k+1,
      t1 = etos[k+1][[1]];
      t2 = Table[0,{Length[fp[[1]]]}];
      For[j=1,j<=lp,j=j+1,
        If[t1[[j]] == 1,
          t2 = t2+fp[[j]];
          ,
          Null;
        ];
      ];
      mins[[k+1]] = xyzs[Mod[t2,2]];
    ];
    mins
  ];


(* Auxiliary to check the ratio range between two
 * syndrome likelihood distributions. *)
ratiorange[np1_,np2_] :=
  {1/(Max[(np1[[1]]/np1[[1,1,1]])/(np2[[1]]/np2[[1,1,1]])]),
   Max[(np2[[1]]/np2[[1,1,1]])/(np1[[1]]/np1[[1,1,1]])]};
diffrange[np1_,np2_] :=
  {Min[(np2[[1]]/np2[[1,1,1]])-(np1[[1]]/np1[[1,1,1]])],
   Max[(np2[[1]]/np2[[1,1,1]])-(np1[[1]]/np1[[1,1,1]])]};

fblock[mx_,r_,s_] :=
  Table[mx[[4*(r-1)+i+1,4*(s-1)+j+1]],{i,0,3},{j,0,3}];
ftr[mx_,t_] :=
  Block[{tl},
    tl=numtobin[t,2];
    Apply[Plus,Flatten[Table[mx[[1+(i)+(j)*2,
      1+(Mod[i+tl[[1]],2])+(Mod[j+tl[[2]],2])*2]],{i,0,1},{j,0,1}]]]
  ];

fxr[mx_,t_] :=
  Block[{xv,tl},
    tl=numtobin[t,2];
    xv = {1,(-1)^(tl[[1]]),(-1)^(tl[[2]]),(-1)^(tl[[1]])*(-1)^(tl[[2]])};
    xv.(mx.xv)
  ];

fxvec[lh_, r_, t_] :=
  Block[{tl},
    tl=numtobin[t,2];
    lh[[(r-1)*4+1]]
     + ((-1)^(tl[[1]]))*lh[[(r-1)*4+2]]
     + ((-1)^(tl[[2]]))*lh[[(r-1)*4+3]]
     + ((-1)^(tl[[1]]))*((-1)^(tl[[2]]))*lh[[(r-1)*4+4]]
  ];
  
favec[lh_, r_, s_, t_] :=
  Block[{th,m,tl}, th=lh;
          tl=numtobin[t,2];
          th[[(r-1)*4+1]] = th[[(r-1)*4+1]]+(s/4);
          th[[(r-1)*4+2]] = th[[(r-1)*4+2]]+((-1)^(tl[[1]]))*(s/4);
          th[[(r-1)*4+3]] = th[[(r-1)*4+3]]+((-1)^(tl[[2]]))*(s/4);
          th[[(r-1)*4+4]] = th[[(r-1)*4+4]]
             +((-1)^(tl[[1]]))*((-1)^(tl[[2]]))*(s/4);
          m = Min[th[[{(r-1)*4+1,(r-1)*4+2,(r-1)*4+3,(r-1)*4+4}]]];
          If[m<0,
            th[[(r-1)*4+1]] = th[[(r-1)*4+1]]-m;
            th[[(r-1)*4+2]] = th[[(r-1)*4+2]]-m;
            th[[(r-1)*4+3]] = th[[(r-1)*4+3]]-m;
            th[[(r-1)*4+4]] = th[[(r-1)*4+4]]-m;
            ,
            Null;
          ];
          th
  ];


(*
 * Auxiliary encapsulated blocks with references to
 * global variables used for compiling 
 * data for graphs etc.
 *)

runlevel[lregion_, sched_ ,probuBnd_, dname_, lvl_, jlow_, jhigh_] :=
  (* See analysis.tex for how this is used for
   * compiling data on a two-dimensional grid of
   * error probabilities for memory/preparation and cnot. *)
  (* Requires global definition of:
   * noerr
   *)
Block[{j,k,elp,elh,elc,eepls,eemls,eehls,eecls,llsize},
 llsize = Length[lregion];
 For[k=jlow,k<=Min[jhigh,Length[lregion[[1]]]],k=k+1,
  For[j=1,j<=llsize,j=j+1,
    If[ (lvl == 1) || (eeMaxProb[j,k,lvl-1] < probuBnd),
     (* Total error likelihoods: *)
     elp = lregion[[j,k,2]]/(1-2*lregion[[j,k,2]]);
     elh = lregion[[j,k,2]]/(2-3*lregion[[j,k,2]]);
     elc = lregion[[j,k,1]]/(1-lregion[[j,k,1]]);
     (**)
     (* Compute error model specifications: *)
     If[lvl == 1,
       eep[j,k,0] = {{"X", {elp}}, {"Z", {elp}}, {"Y", {0}}};
       eem[j,k,0] = eep[j,k,0]; (* Measurement error is the same. *)
       eeh[j,k,0] = {{"X", {elh}}, {"Z", {elh}}, {"Y", {elh}}};
       eec[j,k,0] = symerrstwo[{elc/15},{elc/15},{elc/15}]; 
         (* cnot error. *)
       dloc[j,k,0] = noerr;
       ,
       Null;
     ];
     (**)
     (* Compute level 1 logical and terminal error error parameters.
      * Use spectators in state |+>. *)
     {eep[j,k,lvl],eem[j,k,lvl],eec[j,k,lvl],loc[j,k,lvl-1],intrnl[j,k,lvl-1]} = 
         nextlevel[eep[j,k,lvl-1], eem[j,k,lvl-1], eec[j,k,lvl-1], 
                   sched[[lvl]], lvl-1];
     dloc[j,k,lvl] = decerr[eep[j,k,0], eem[j,k,0], eec[j,k,0], sched[[lvl]], 
                          loc[j,k,lvl-1], dloc[j,k,lvl-1]];
     eeh[j,k,lvl] = haderr[eep[j,k,0],eem[j,k,0],eec[j,k,0],eeh[j,k,lvl-1],
                       sched[[lvl]], loc[j,k,lvl-1], lvl-1];
     (* Probabilities of error: *)
     eepls = Table[eep[j,k,lvl][[x,2,1]], {x,1,3}];
     eepProb[j,k,lvl] = Max[eepls]/(1+Apply[Plus, eepls]);
     eemls = Table[eem[j,k,lvl][[x,2,1]], {x,1,3}];
     eemProb[j,k,lvl] = Max[eemls]/(1+Apply[Plus, eemls]);
     eehls = Table[eeh[j,k,lvl][[x,2,1]], {x,1,3}];
     eehProb[j,k,lvl] = Apply[Plus,eehls]/(1+Apply[Plus, eehls]);
     eecl = Apply[Plus,Table[eec[j,k,lvl][[x,2,1]], {x,1,15}]];
     eecProb[j,k,lvl] = eecl/(1+eecl);
     (* Comparison between independent and computed error model for
      * the encoded Bell state. Gives the range of the ratios between
      * the likelihoods. *)
     locRange[j,k,lvl-1] = ratiorange[intrnl[j,k,lvl-1][[1]], 
                                      intrnl[j,k,lvl-1][[2]]];
     (**)
     eeMaxProb[j,k,lvl] = Max[eepProb[j,k,lvl],eemProb[j,k,lvl],
                              eehProb[j,k,lvl],eecProb[j,k,lvl]];
     If[eeMaxProb[j,k,lvl] > probuBnd,
       For[l=j+1,l<=llsize,l=l+1,
         eeMaxProb[l,k,lvl] = 1;
       ];
       Break[];
       ,
       Null;
     ];
     ,
     eeMaxProb[j,k,lvl] = 1;
    ];
  ];
  DumpSave[StringJoin["analysis", ToString[lvl], 
                     "_", ToString[k-1], ":", dname, ".mx"]];
 ];
 DumpSave[StringJoin["analysis", ToString[lvl], "_f", ":", dname, ".mx"]];
];

eemProbs[eem_] :=
  Block[{eml},
     eeml = 1+Apply[Plus,Table[eem[[x,2,1]], {x,1,3}]];
     {eem[[1,2,1]],eem[[2,2,1]],eem[[3,2,1]]}/eeml
  ];
eecProbs[eec_] :=
  Block[{eml},
     eeml = 1+Apply[Plus,Table[eec[[x,2,1]], {x,1,15}]];
     Table[eec[[x,2,1]],{x,1,15}]/eeml
  ];

(*\end{verbatim}*)

}

}

\ignore{
\pagebreak
\section{Instructions for generating the data needed for analysis}

{\footnotesize
\begin{verbatim}
(* Mathematica code. *)
SetOptions["stdout", PageWidth->124];
<<retestate_fns.math (* Functions needed. *)

eptruncate = -1;
epmaxe = 1/400; (* Maximum error likelihood, used for truncating
 * likelihood polynomials in one variable. *)
eprec = 48; (* Number of digits of precision for real numbers. *)

noerr = {{"X", {0}},{"Z", {0}}, {"Y", {0}}}; 
  (* One qubit error model with zero error. *)

(*
 * Task:
 * Determine error behavior after one level for a large number
 * of points parametrized by error probability for
 * preparation/measurement and error probability for cnot.
 * Unform distributions over Pauli products are assumed.
*)

llsize = 10;
maxp = SetPrecision[1/10,eprec]; pratio = 2; 
minp = SetPrecision[1/10,eprec]*pratio^(-llsize);
lregion = Table[Table[{minp*pratio^(j-1), minp*pratio^(k-1)},{j,1,llsize}],{k,1,llsize}];
probuBnd = .25;
sched = {"X","Z","Z","Z","X"};

(* Run the following sets independently. *)

(* Set: ...an3_b1.math...*)
runlevel[lregion, sched, probuBnd, "b1", 1, 1, 2];
(* <<analysis1_f:b1.mx *)
runlevel[lregion, sched, probuBnd, "b1", 2, 1, 2];
(* <<analysis2_f:b1.mx *)
runlevel[lregion, sched, probuBnd, "b1", 3, 1, 2];
(* <<analysis3_f:b1.mx *)
runlevel[lregion, sched, probuBnd, "b1", 4, 1, 2];
(* <<analysis4_f:b1.mx *)
runlevel[lregion, sched, probuBnd, "b1", 5, 1, 2];
(* Full data set in analysis5_f:b1.mx *)

(* Set: ...an3_b2.math...*)
runlevel[lregion, sched, probuBnd, "b2", 1, 3, 4];
(* <<analysis1_f:b2.mx *)
runlevel[lregion, sched, probuBnd, "b2", 2, 3, 4];
(* <<analysis2_f:b2.mx *)
runlevel[lregion, sched, probuBnd, "b2", 3, 3, 4];
(* <<analysis3_f:b2.mx *)
runlevel[lregion, sched, probuBnd, "b2", 4, 3, 4];
(* <<analysis4_f:b2.mx *)
runlevel[lregion, sched, probuBnd, "b2", 5, 3, 4];
(* Full data set in analysis5_f:b2.mx *)

(* Set: ...an3_b3.math...*)
runlevel[lregion, sched, probuBnd, "b3", 1, 5, 7];
(* <<analysis1_f:b3.mx *)
runlevel[lregion, sched, probuBnd, "b3", 2, 5, 7];
(* <<analysis2_f:b3.mx *)
runlevel[lregion, sched, probuBnd, "b3", 3, 5, 7];
(* <<analysis3_f:b3.mx *)
runlevel[lregion, sched, probuBnd, "b3", 4, 5, 7];
(* <<analysis4_f:b3.mx *)
runlevel[lregion, sched, probuBnd, "b3", 5, 5, 7];(* Full data set in analysis5_f:b3.mx *)

(* Set: ...an3_b4.math...*)
runlevel[lregion, sched, probuBnd, "b4", 1, 8, 10];
(* <<analysis1_f:b4.mx *)
runlevel[lregion, sched, probuBnd, "b4", 2, 8, 10];
(* <<analysis2_f:b4.mx *)
runlevel[lregion, sched, probuBnd, "b4", 3, 8, 10];
(* <<analysis3_f:b4.mx *)
runlevel[lregion, sched, probuBnd, "b4", 4, 8, 10];
(* <<analysis4_f:b4.mx *)
runlevel[lregion, sched, probuBnd, "b4", 5, 8, 10];
(* Full data set in analysis5_f:b4.mx *)

(* Fill in near the threshold for cnot error. *)
lregion = Table[Join[lregion[[k]],Table[{maxp*pratio^(-3+1/2+j), 
                     minp*pratio^(k-1)},{j,1,2}]],{k,1,llsize}];
(* Set: ...an3_b5.math...*)
runlevel[lregion, sched, probuBnd, "b5", 1, 11, 12];
(* <<analysis1_f:b5.mx *)
runlevel[lregion, sched, probuBnd, "b5", 2, 11, 12];
(* <<analysis2_f:b5.mx *)
runlevel[lregion, sched, probuBnd, "b5", 3, 11, 12];
(* <<analysis3_f:b5.mx *)
runlevel[lregion, sched, probuBnd, "b5", 4, 11, 12];
(* <<analysis4_f:b5.mx *)
runlevel[lregion, sched, probuBnd, "b5", 5, 11, 12];
(* Full data set in analysis5_f:b5.mx *)


(*
 * Task:
 * Special cases. 
 * elp = .01, elc = .03
 * elp = .002, elc = .008
 *)

lregion = {{{SetPrecision[3/100,eprec],SetPrecision[1/100,eprec]}}};
sched = {"X","Z","Z","Z","X"};
For[lvl=1,lvl<=5,lvl=lvl+1,
  runlevel[lregion, sched, 1, "sp1", lvl, 1, 1];
];
(* Full data set in analysis5_f:sp1.mx *)

lregion = {{{SetPrecision[8/1000,eprec],SetPrecision[2/1000,eprec]}}};
sched = {"X","Z","Z","X","Z"};
For[lvl=1,lvl<=5,lvl=lvl+1,
  runlevel[lregion, sched, 1, "sp2", lvl, 1, 1];
];
(* Full data set in analysis5_f:sp2.mx *)


\end{verbatim}
}}

\ignore{
Some experiences in comparing different approaches, for the record:
\begin{itemize}
  \item I tried several variations of the method for approximating
with an upper-bounding independent error model. The first
is the one described in the text. Modifying it by computing
the marginals based on decoding a perfectly encoded state
improved the approxmation in some ranges, but not others.
It did a little, but not much, better in the range where
preparation/measurement errors dominate. It did a little
worse in the range where cnot errors dominate.

A more significantly different method was to use the singular value
decomposition to obtain the error model. This is a little more
sophisticated. It is based on analyzing the Hadamard transform of the
sets of four cosets that give rise to the same syndrome.  If the error
model is exactly independent, it obtains a solution. Some work (using
heurstics) had to be done to ensure it is bounding if not. My
implementation approximates fairly well, though not quite as
well as the original method in the range where cnot errors dominate.
If preparation/measurement errors dominate, it (sometimes?) does better
than the original method and worse than the original method
with the second marginal distribution computation. I only
spot checked the behaviors, so I can't be more specific on how
these methods compare. One note: The singular-value-based method
was designed to be strictly bounding, unlike the weight based (original)
method.
\end{itemize}
}

\ignore{
Batch execution instructions:
1. Create file with mathematica instructions (say an1.math).
2. # nice +20 at now
3. Enter on terminal, followed by ^d:
math5 < an3_b1.math > an3_b1.out
4. You can watch the output with # tail -f an1.out.
}

\ignore{

(* Contour plot instructions. *)
(* 
 * 1. Make an "interpolation" object from each data set. 
 *)

maxlvl = 5;
Get[StringJoin["analysis", ToString[maxlvl], "_f:b1.mx"]];
Get[StringJoin["analysis", ToString[maxlvl], "_f:b2.mx"]];
Get[StringJoin["analysis", ToString[maxlvl], "_f:b3.mx"]];
Get[StringJoin["analysis", ToString[maxlvl], "_f:b4.mx"]];
Get[StringJoin["analysis", ToString[maxlvl], "_f:b5.mx"]]; 

For[lvl=1,lvl<=maxlvl,lvl=lvl+1,
  (* cnot error varies with second index in eeMaxProb[j,k,lvl]. *)
  tbl[lvl] = Flatten[Table[
      {Log[10,lregion[[j,k,1]]], 
       Log[10,lregion[[j,k,2]]], Log[10,eeMaxProb[j,k,lvl]]},
      {k,1,Length[lregion[[1]]]}, {j,1,Length[lregion]}],1];
];

For[lvl=1,lvl<=maxlvl,lvl=lvl+1,
  fndt[lvl] = Interpolation[
    tbl[lvl],
    InterpolationOrder->1];
];

xvls = Table[lregion[[1,k,1]], {k,1,Length[lregion[[1]]]}];
oxvls = Ordering[xvls];
yvls = Table[lregion[[j,1,2]], {j,1,Length[lregion]}];
oyvls = Ordering[yvls];
fx = Log[10,Min[xvls]]; lx = Log[10,Max[xvls]];
fy = Log[10,Min[yvls]]; ly = Log[10,Max[yvls]];

locRangeCut = 100;
llocRangeMin = Log[10,1-Min[Table[If[eeMaxProb[i,j,maxlvl] < 1, 
                                    locRange[i,j,l-1][[1]],
                                    1],
                   {i,1, Length[lregion]},
                  {j,1,Length[lregion[[1]]]}, {l,1,maxlvl}]]];
(* 04/01/04: llocRangeMin = -6.0378 *)
locRangeMax = Table[If[eeMaxProb[i,j,maxlvl] < 1, 
                                    locRange[i,j,l-1][[2]],
                                    1],
                   {i,1, Length[lregion]},
                  {j,1,Length[lregion[[1]]]}, {l,1,maxlvl}];
locRangeMax = SetPrecision[locRangeMax,5];

whitepnts = {};
For[k=1,k<=Length[xvls],k=k+1,
 For[j=1,j<=Length[yvls],j=j+1,
   dowhite = 0;
   For[lvl=1,lvl<=maxlvl,lvl=lvl+1,
     If[(eeMaxProb[oyvls[[j]],oxvls[[k]],lvl] < 1) && 
        (locRange[oyvls[[j]],oxvls[[k]],lvl-1][[2]] > locRangeCut),
       dowhite = 1; Break[];
       ,
       Null;
     ];
   ];
   If[dowhite > 0,
     xpos = Log[10,xvls[[oxvls[[k]]]]];
     If[k>1, xlpos = (xpos+Log[10,xvls[[oxvls[[k-1]]]]])/2, xlpos = xpos];
     If[k<Length[xvls], xrpos = (xpos+Log[10,xvls[[oxvls[[k+1]]]]])/2, xrpos=xpos];
     ypos = Log[10,yvls[[oyvls[[j]]]]];
     If[j>1, ypos = (ypos+Log[10,yvls[[oyvls[[j-1]]]]])/2, ylpos = ypos];
     whitepnts = Append[Append[whitepnts, 
        {xlpos, ypos}], 
        {xrpos, ypos}];
     Break[];
     ,
     Null;
   ];
 ];
];
grdpts = Flatten[Table[
      Point[{Log[10,lregion[[j,k,1]]], 
       Log[10,lregion[[j,k,2]]]}],
      {k,1,Length[lregion[[1]]]}, {j,1,Length[lregion]}],1];

(*
 * 2. Contour plot.
 *)
cntrlist = {-8,-7,-6,-5,-4,-3,-2,-1, -.6021,0};
cntrlvls = {1,.975,.95,.925,.9,.875,.85,.825,.8,0,0};
grlv[x_] := Block[{j,l},
    l = 11;
    For[j=1,j<=Length[cntrlist],j=j+1,
      If[x<cntrlist[[j]],
        l=j;
        Break[];
        ,
        Null;
      ];
    ];
    GrayLevel[cntrlvls[[l]]]
  ];
cntrs[1] = {-8,-7,-6,-5,-4,-3,-2,-1, -.6021};
cntrs[2] = {-16,-14,-12,-10,-8,-6,-4,-2, -.6021};
cntrs[3] = {-24,-21,-18,-15,-12,-9,-6,-3, -.6021};
cntrs[4] = {-40,-35,-30,-25,-20,-15,-10,-5, -.6021};
cntrs[5] = {-64,-56,-48,-40,-32,-24,-16,-8, -.6021};
ticks = {{-4,"-4",{.02,.0}},{-3,"-3",{.02,.0}},{-2,"-2",{.02,.0}},
                    {-3.5,"",{.015,.0}},{-2.5,"",{.015,.0}},
                    {-1.5,"",{.015,.0}}};
ticks = Join[ticks, Table[{j,"",{.005,.0}}, {j,-4.1,0,.1}]];
For[lvl=1,lvl<=maxlvl,lvl=lvl+1,
  cplot[lvl]=ContourPlot[Evaluate[fndt[lvl]][x,y],{x,fx,lx},{y,fy,ly},
    ContourShading -> True,
    ColorFunction -> Evaluate[(grlv[#1])&],
    ColorFunctionScaling -> False,
    Contours -> cntrs[lvl],
    ContourStyle ->{{AbsoluteThickness[4]}},
    ImageSize -> {76*3,76*3},
    Frame -> True,
    FrameTicks -> {ticks,ticks, Automatic, Automatic}
    ];
  Export[StringJoin["cplotb", ToString[lvl], ".eps"],cplot[lvl]];
  cplot[lvl] = Show[cplot[lvl], 
    Graphics[{GrayLevel[.5], AbsoluteThickness[2], Line[whitepnts]}]];
  cplot[lvl] = Show[cplot[lvl], 
    Graphics[{GrayLevel[.5], AbsolutePointSize[10], 
                Point[{Log[10,.03], Log[10,.01]}]}]];
  cplot[lvl] = Show[cplot[lvl], 
    Graphics[{GrayLevel[.5], AbsolutePointSize[10], 
                Point[{Log[10,.008], Log[10,.002]}]}]];
  cplot[lvl] = Show[cplot[lvl],
    Graphics[{GrayLevel[1], AbsolutePointSize[2], grdpts}]];
  Export[StringJoin["cplotw", ToString[lvl], ".eps"],cplot[lvl]];
];

(* Puzzle: Why does the error improve at high preparation/measurement
 * error as cnot error is small but increasing?
 * Conjecture: Better independence approximation.
 * Indeed, that is what happens. The approximation is in fact
 * terrible when preparation/measurement error dominates.
 * this shouldn't be surprising, since the heuristic
 * used to get the approximation is cued to cnot error.
 * For visualization: There is no reason to cover the
 * part where prep/measurement >> cnot errors.
 *)

}

\ignore{
(* analysis5_f:sp1.mx data.
 ***
 * Level 0:
 ***
eep[1,1,0]//InputForm
  {{"X", {0.0102040816326530612244897959183673469387755102040816326530612244898`47.991226075692495}},
   {"Z", {0.0102040816326530612244897959183673469387755102040816326530612244898`47.991226075692495}}, {"Y", {0}}}
eem[1,1,0]//InputForm
  {{"X", {0.0102040816326530612244897959183673469387755102040816326530612244898`47.991226075692495}},
   {"Z", {0.0102040816326530612244897959183673469387755102040816326530612244898`47.991226075692495}}, {"Y", {0}}}
eeh[1,1,0]//InputForm
  {{"X", {0.0050761421319796954314720812182741116751269035532994923857868020304`47.993436230497615}},
   {"Z", {0.0050761421319796954314720812182741116751269035532994923857868020304`47.993436230497615}},
   {"Y", {0.0050761421319796954314720812182741116751269035532994923857868020304`47.993436230497615}}}
margec[eec[1,1,0]]//InputForm
  {{"X", {0.008247422680412371134020618556701030927835051546391752577319587629`47.98677173426625}},
   {"Z", {0.008247422680412371134020618556701030927835051546391752577319587629`47.98677173426625}},
   {"Y", {0.008247422680412371134020618556701030927835051546391752577319587629`47.98677173426625}}}
 ***
 * Level 1:
 ***
locRange[1,1,0]//InputForm
  {0.99999956954544654414807073183148382540755609207659793103`44.96964290801767,
   3.199702954906327516046572388650711775998042181062925216976`44.793128014244}
eepProb[1,1,1]//InputForm
  0.004143760382413792300227014579552590892257271663500369704`45.28522188568683
eemProb[1,1,1]//InputForm
  0.004143760382413792300227014579552590892257271663500369704`45.28522188568683
eehProb[1,1,1]//InputForm
  0.024362656418140047065091541213176368533021780988011435062`45.26224246225716
eecProb[1,1,1]//InputForm
  0.029815038089576523433537591475995351939585243309505025797`45.20679351191002
dloc[1,1,1]//InputForm
  {{"X", {0.005741066482047630609488150406562441826911376579355834084`45.808976429708856}},
   {"Z", {0.01333962649571216283693044642797952317450076178822979471`46.26076355102456}},
   {"Y", {0.0048661923280948285600927744241886925484151582829033410133473148449`46.06936317184592}}}
eep[1,1,1]//InputForm
  {{"X", {0.004165108739926061880835223564534422697836417752820528041`45.287413917092664}},
   {"Z", {0.0009868198326990509371716260439596140468522790243579148380558230456`45.31736450782705}}, {"Y", {0}}}
eem[1,1,1]//InputForm
  {{"X", {0.004165108739926061880835223564534422697836417752820528041`45.287413917092664}},
   {"Z", {0.0009868198326990509371716260439596140468522790243579148380603352601`45.31736450782705}}, {"Y", {0}}}
eeh[1,1,1]//InputForm
  {{"X", {0.011299313190647923564056476919890519601381065990715696215`45.27704427568225}},
   {"Z", {0.011299313190647923564056476919890519601381065990715696215`45.27704427568223}},
   {"Y", {0.002372390337577748152373290129483570591538212296564636818`45.23332432287745}}}
margec[eec[1,1,1]]//InputForm
  {{"X", {0.018735887731946405727005153241966279149678574280820412763`45.18737745049856}},
   {"Z", {0.006266654690966731489367514115021890630409092809227336863`45.30787557310199}},
   {"Y", {0.0003797315479978968836204832806293894736257837680653182571731939565`45.096052416093706}}}
eec[1,1,1]//InputForm
  {{"ZI", {0.005120052681986838322289712635897987513775279267470900087`45.344730009543696}},
   {"XI", {0.004579425318455346336826885358085002301052665842419519717`45.217964881635226}},
   {"YI", {0.000212358832014465311935234027583191651660060247952895999483683801`45.12268850613883}},
   {"IZ", {0.0012956790392236884928330906186513556402839917524892569951436684764`45.33640688313759}},
   {"ZZ", {0.0005522481101212086830846822484758280510616401376633166222543838864`45.21839378814136}},
   {"XZ", {0.0000112470662175449716257852497355234125467565788093888560105547892`44.967864138622545}},
   {"YZ", {6.7835070069405615234168455217121102926391759856062102003780487`44.93001329386263*^-6}},
   {"IX", {0.015650211631923683050637599202148469463196500795270136951`45.2023143418987}},
   {"ZX", {0.00052306174559936059164087687291053642816258806016321110314761307`45.13930792204877}},
   {"XX", {0.002407088850395084281422622662568260992000353898053013685`45.11701626228654}},
   {"YX", {0.0001555255040282778033040545043390122663191315273340510244302306133`45.07799249631432}},
   {"IY", {0.0001331989075366446642672000368063833804261726473039689643955744732`45.14219875715595}},
   {"ZY", {0.00007129215325932389235224235773753863740958534392990905067852508`45.10612725779307}},
   {"XY", {8.0557008494011908186661012489374336512379696219849919274510029`44.92824890930977*^-6}},
   {"YY", {5.0637049482132068577779031854734453539528167927650230589014935`44.90103103479929*^-6}}}
 ***
 * Level 2:
 ***
locRange[1,1,1]//InputForm
  {0.999999972103567512287604097113351901886071135554717132141`43.98183456148802,
   8.59079172535126095276898303601219723493086813972546293668`43.59976061144932}
eepProb[1,1,2]//InputForm
  0.000486760361820963149680136869388465333884943544435188328`44.08358626681283
eemProb[1,1,2]//InputForm
  0.000486760361820963149680136869388465333884943544435188328`44.08358626681283
eehProb[1,1,2]//InputForm
  0.012158368413350817092256516062620660090962881614752916691`43.98240184877827
eecProb[1,1,2]//InputForm
  0.009528698787593667604729351993152218212836162871805756547`43.976723237807285
dloc[1,1,2]//InputForm
  {{"X", {0.01325867652296955079984180374576767210358333243680695375`45.29181888583016}},
   {"Z", {0.004975051604920249340674906060737170666928731780186089379`45.028873121225864}},
   {"Y", {0.004666088605670124125151986034685389963195562880661096215`45.18602166638236}}}
eep[1,1,2]//InputForm
  {{"X", {0.000487209071991267799543768070851964896736040897577404102`44.08395049039164}},
   {"Z", {0.000434620652651946583764109578760033223295328962214842077`44.17481314989139}}, {"Y", {0}}}
eem[1,1,2]//InputForm
  {{"X", {0.000487209071991267799543768070851964896736040897577404102`44.08395049039164}},
   {"Z", {0.000434620652651946583764109578760033223295328962214842077`44.17481314989139}}, {"Y", {0}}}
eeh[1,1,2]//InputForm
  {{"X", {0.006091471258833871296748986288542526533580197578054847444`43.98841017576801}},
   {"Z", {0.006091471258833871296748986288542526533580197578054847444`43.98841017576801}},
   {"Y", {0.000125071261648600944369521183677313484575129994580289256`43.91936073808654}}}
margec[eec[1,1,2]]//InputForm
  {{"X", {0.005070960460442039810619683708133765086684963233143550475`43.947311379239295}},
   {"Z", {0.003416577572944896876385817882746294122751900973862965329`44.01670056840416}},
   {"Y", {8.4040391693402258730348128358097274453343234865200625222563619`43.83796238906846*^-6}}}
eec[1,1,2]//InputForm
  {{"ZI", {0.002946340653066843828547642008167062822403010521316258902`44.009245705460614}},
   {"XI", {0.000618807632018730346410296590058305926876470890716481606`43.980705664663}},
   {"YI", {3.7731700548188479396500126158752139606503815033098014322881586`43.85296057977735*^-6}},
   {"IZ", {0.000524315680785012028913489157337005500880062395711031208`44.09952725430032}},
   {"ZZ", {0.000447051127174710385689424940419109309950680896131328142`44.09129752888513}},
   {"XZ", {3.851618692526812910191919904956616742363246225603694985734585`43.74657582440318*^-7}},
   {"YZ", {3.304555327110239010241713817556741138445524273690132815158235`43.743314413736286*^-7}},
   {"IX", {0.004625941456104323969064997175500199868142152747301093089`43.949549979431126}},
   {"ZX", {0.0000195904365235491226898296289179129075589448014101409064563621837`43.73625298683079}},
   {"XX", {0.000422655959736237711778634739898356728056149067513485516`43.936408764095496}},
   {"YX", {2.7726080779290070862221638172955829277166169188309636457415296`43.83970407721837*^-6}},
   {"IY", {4.2986112099218218791707347107153097083447617721541189240462257`43.84944969504721*^-6}},
   {"ZY", {3.5953561797935394589213052422090828392647550052373785061370618`43.843124827726356*^-6}},
   {"XY", {2.746746244199061694284586569745879259997238502801331227332483`43.72434609458103*^-7}},
   {"YY", {2.35397155204958365514314225910746971725082858848431969339826`43.721271729962204*^-7}}}
 ***
 * Level 3:
 ***
locRange[1,1,2]//InputForm
  {1.`44.000193396945036, 2.635012757903696076097798571127943420494550207005904915484`42.97875347724326}
eepProb[1,1,3]//InputForm
  0.000241512889842071165786332457892029122656165384479197748`43.12154788996729
eemProb[1,1,3]//InputForm
  0.000241512889842071165786332457892029122656165384479197748`43.12154788996729
eehProb[1,1,3]//InputForm
  0.001453897601884872972126551877048548051453638408166810886`43.16432599446505
eecProb[1,1,3]//InputForm
  0.001571543672364227261917850439655729872441129535112742439`43.09115827170774
dloc[1,1,3]//InputForm
  {{"X", {0.012609265423285113669715840835448049730403712785683980902`44.98955978213129}},
   {"Z", {0.004651752245667568342209580983447750404322070643932347581`44.220742339626945}},
   {"Y", {0.004431920881977543956683797200871612690677363001605963271`44.59480352797065}}}
eep[1,1,3]//InputForm
  {{"X", {5.098518930790848225420740910291757796373149412593950529946255`43.12648555398236*^-6}},
   {"Z", {0.000241572464064014425602223104511429128102540858004646865`43.121654953710895}}, {"Y", {0}}}
eem[1,1,3]//InputForm
  {{"X", {5.0985189307908482254207409102917577963731494125939505298174704`43.12648555398236*^-6}},
   {"Z", {0.000241572464064014425602223104511429128102540858004646865`43.121654953710895}}, {"Y", {0}}}
eeh[1,1,3]//InputForm
  {{"X", {0.000727637193257167894653028546502548973810055628602356475`43.16508610069041}},
   {"Z", {0.000727637193257167894653028546502548973810055628602356475`43.16508610069041}},
   {"Y", {7.401113573023650151935436232021066850338327674306994757979417`42.965030660998096*^-7}}}
margec[eec[1,1,3]]//InputForm
  {{"X", {0.000093822673015261787457312601862286968079804910178914905`43.20150846930617}},
   {"Z", {0.00122038825610895652275724893567661049362260922542960135`43.08096472344877}},
   {"Y", {4.19771761234672255764501314353060415408467033359678832385252`42.910693578914014*^-8}}}
eec[1,1,3]//InputForm
  {{"ZI", {0.000970025186632786177224916758564111148498528488129225803`43.07446994703092}},
   {"XI", {5.9932590351910279266308468252542515081124986549564272021510858`43.12705364841102*^-6}},
   {"YI", {4.7227568894547968338170609880008474647802437738348474611964`42.888040015389684*^-9}},
   {"IZ", {0.000253873551570063638218462728492541903940774807834922862`43.1083856775013}},
   {"ZZ", {0.000250252916553387752456146085842724163873410590352741981`43.10723779347772}},
   {"XZ", {1.5224449457983702774699447011891408213824475232202076663204`42.81990796410679*^-9}},
   {"YZ", {1.5001222111211489862717553025957638647899320453040894422017`42.81932516963561*^-9}},
   {"IX", {0.000088925278055914608811927765586228498565180499365239564`43.207216094786375}},
   {"ZX", {9.05902405688181384742376486197574081553901777352116254645307`42.8434596237671*^-8}},
   {"XX", {4.8032591388326422461563463835538342502507067646843399544509578`43.11630450315519*^-6}},
   {"YX", {3.545579945718260754252243884877856218313871255789728728604`42.880187582236736*^-9}},
   {"IY", {1.99951386248214267840422991222442266219747781672104267057154`42.91805639481377*^-8}},
   {"ZY", {1.95626822137749377118536211554238425147567698983537590765328`42.91658873946219*^-8}},
   {"XY", {1.2186340525255640165373799340441719919970353907217514948689`42.81477810496857*^-9}},
   {"YY", {1.2007212323452970640168312235938004121181198796819459614081`42.81420976100313*^-9}}}
 ***
 * Level 4:
 ***
locRange[1,1,3]//InputForm
  {1.`43.70793981389, 1.896403618176583622206917867356977505576186811716144674978`42.179395280629365}
eepProb[1,1,4]//InputForm
  0.000026961657139311234969472562352980013942135285211786865`42.32352203433315
eemProb[1,1,4]//InputForm
  0.000026961657139311234969472562352980013942135285211786865`42.32352203433315
eehProb[1,1,4]//InputForm
  0.000071930167349915200956028661103764320752330478813776804`42.33998463679237
eecProb[1,1,4]//InputForm
  0.000159290710590523520264192285957980166974121907207002071`42.27605579092628
dloc[1,1,4]//InputForm
  {{"X", {0.01247199702610099760056677664264321964879939371393937883`44.69835445358396}},
   {"Z", {0.004333661183451201176071162516457094984238601674081441343`44.084878425143344}},
   {"Y", {0.004310699986861043416216758103830404872760634185777978074`44.24198514004032}}}
eep[1,1,4]//InputForm
  {{"X", {3.113507075878409811114544849900896866510385112021180682111`42.36695358338233*^-10}},
   {"Z", {0.000026962384098261481740931340668474325257567675684392417`42.32353374359656}}, {"Y", {0}}}
eem[1,1,4]//InputForm
  {{"X", {3.113507075878409811114544849900896866510385112021180682021`42.36695358338233*^-10}},
   {"Z", {0.000026962384098261481740931340668474325257567675684392417`42.32353374359656}}, {"Y", {0}}}
eeh[1,1,4]//InputForm
  {{"X", {0.000035967014577997313819321899772731876513079579353052009`42.340023678170205}},
   {"Z", {0.000035967014577997313819321899772731876513079579353052009`42.340023678170205}},
   {"Y", {1.3125150853474285984806638209758804540458800879534149729506`42.04230393303154*^-9}}}
margec[eec[1,1,4]]//InputForm
  {{"X", {2.08162075365488552845310821522609370081999124216367986189714`42.68471368051316*^-8}},
   {"Z", {0.000131824212465229029466555751560674146517480639268542746`42.26798753901471}},
   {"Y", {1.0743278688086800347434880486983145483917770730428250746`42.19675369496549*^-12}}}
eec[1,1,4]//InputForm
  {{"ZI", {0.000104412975729751741835990982615442270647319249727134534`42.25595899303527}},
   {"XI", {3.53334346811126189373928428567349747287379570380967462429`42.38358186518365*^-10}},
   {"YI", {2.86967065250008542279034561575179695790908432122497594`42.10417133619436*^-14}},
   {"IZ", {0.000027470707720079891003513250763335218279580578437550079`42.31722531744653}},
   {"ZZ", {0.000027411234049303817310730438147686050992132194935263613`42.31712400671571}},
   {"XZ", {9.6966624868679323874472271907239739331520908731455759`42.04909630729029*^-15}},
   {"YZ", {9.6755252718357914363768519608781385081315426190714184`42.04904251969748*^-15}},
   {"IX", {2.05150296885453732044168852387251734451708945331649021403514`42.69179627071091*^-8}},
   {"ZX", {2.1578987581990174263509402765573014390517921704758821045`42.121526145017384*^-12}},
   {"XX", {2.989965338005168428697260401482755085656644334078422289909`42.36446753517042*^-10}},
   {"YX", {2.34154447662198181199331109307530243016628935783675245`42.09960843934826*^-14}},
   {"IY", {5.296634186381509571228178436326035566060271235884599725`42.19972631414277*^-13}},
   {"ZY", {5.282747121208169044466055483207277555543524289173338008`42.19964126329561*^-13}},
   {"XY", {8.2038064374046912410510506300898620608237527511394493`42.040218762106726*^-15}},
   {"YY", {8.185931612307481933013606114893374170573767785891852`42.04016639047615*^-15}}}
 ***
 * Level 5:
 ***
locRange[1,1,4]//InputForm
  {1.`43.85757665419749, 1.942763116087148315138026684406998124175295508351444902985`41.55451855871383}
eepProb[1,1,5]//InputForm
  7.3704269298149762105797175529986849286768011719176806707539`41.73427054657455*^-9
eemProb[1,1,5]//InputForm
  7.3704269298149762105797175529986849286768011719176806621632`41.73427054657455*^-9
eehProb[1,1,5]//InputForm
  3.51462606258179672832149013081242583560947379971051129960024`41.72248099012232*^-8
eecProb[1,1,5]//InputForm
  6.86900710400226838910818672173196732340109579861032441887203`41.727088734118645*^-8
dloc[1,1,5]//InputForm
  {{"X", {0.0042506986590404329315325004782369002603193851399186495215255604905`46.636717574831785}},
   {"Z", {0.01241302699683422048571297707767051878686422607758443228`46.06160018567665}},
   {"Y", {0.0042506986590368785182436207326304948074799611395884726906595985624`46.636717611780675}}}
eep[1,1,5]//InputForm
  {{"X", {3.6536732162757472248168583912244660877546226889377763`41.63011483384364*^-15}},
   {"Z", {7.3704269841381966678383267674849881706467250346450808971268`41.73427054977549*^-9}}, {"Y", {0}}}
eem[1,1,5]//InputForm
  {{"X", {3.6536732162757472248168583912244660877546226889377763`41.63011483384364*^-15}},
   {"Z", {7.3704269841381966678383267674849881706467250346450808885361`41.73427054977549*^-9}}, {"Y", {0}}}
eeh[1,1,5]//InputForm
  {{"X", {1.75731307755392302030572591367478525349412006883988880987542`41.72248100920408*^-8}},
   {"Z", {1.75731307755392302030572591367478525349412006883988881572313`41.72248100920408*^-8}},
   {"Y", {3.099991862697813887969961923096406828509955177069513`41.42216935160926*^-16}}}
margec[eec[1,1,5]]//InputForm
  {{"X", {2.35164398418560148383348602419943519347132317276909946`41.66342978732542*^-14}},
   {"Z", {6.12069844404563663161767669480986257846061450749211138847443`41.72630169451627*^-8}},
   {"Y", {3.7419751712724315701007339371157593258058049673487243652`41.43842183401628*^-22}}}
eec[1,1,5]//InputForm
  {{"ZI", {5.37370008683664462079903601986306341802991907949321975733804`41.72531366818599*^-8}},
   {"XI", {3.6613168592547997753346363892310306595998919075938393`41.630330040959194*^-15}},
   {"YI", {1.60030536403201080925807610409945561732456283023743608`41.44974057450576*^-22}},
   {"IZ", {7.4830641401364789377916650376772947462739759251344056518265`41.733580910003155*^-9}},
   {"ZZ", {7.4699835720887310910984980906073043016613047044194316171398`41.73347631235671*^-9}},
   {"XZ", {2.739718077062459335027769678005384955185065391776298096`41.37788432416996*^-23}},
   {"YZ", {2.734928756918166131532963888357603132111689089507020515`41.37783822637195*^-23}},
   {"IX", {1.9864756451464307888477933839251868513077178549689152`41.66984454931077*^-14}},
   {"ZX", {1.06876795400549743974953630737889244397760217935894556994`41.39585970077305*^-21}},
   {"XX", {3.6516821620833754817249348970495189060620168205330409`41.630102816650464*^-15}},
   {"YX", {1.5954037746263455175615665713668159237986662235672043748`41.44964411404685*^-22}},
   {"IY", {1.205003735669521860860196090296888357396878751448351195`41.48019143538578*^-22}},
   {"ZY", {1.2024913390316142093776633827068312550709204481718556405`41.48017404296856*^-22}},
   {"XY", {2.73250825631425943530922689497634761407523678717149211`41.37775720310112*^-23}},
   {"YY", {2.727731569222586301277948728137274714714070045933818588`41.37771111974207*^-23}}}
*)

(* analysis5_f:sp2.mx data.
 ***
 * Level 0:
 ***
eep[1,1,0]//InputForm
  {{"X", {0.002008032128514056224899598393574297188755020080321285140562248996`47.9982593384237}},
   {"Z", {0.002008032128514056224899598393574297188755020080321285140562248996`47.9982593384237}}, {"Y", {0}}}
eem[1,1,0]//InputForm
  {{"X", {0.002008032128514056224899598393574297188755020080321285140562248996`47.9982593384237}},
   {"Z", {0.002008032128514056224899598393574297188755020080321285140562248996`47.9982593384237}}, {"Y", {0}}}
eeh[1,1,0]//InputForm
  {{"X", {0.001003009027081243731193580742226680040120361083249749247743229689`47.99869515831166}},
   {"Z", {0.001003009027081243731193580742226680040120361083249749247743229689`47.99869515831166}},
   {"Y", {0.001003009027081243731193580742226680040120361083249749247743229689`47.99869515831166}}}
margec[eec[1,1,0]]//InputForm
  {{"X", {0.0021505376344086021505376344086021505376344086021505376344086021505`47.99651167215418}},
   {"Z", {0.0021505376344086021505376344086021505376344086021505376344086021505`47.99651167215418}},
   {"Y", {0.0021505376344086021505376344086021505376344086021505376344086021505`47.99651167215418}}}
 ***
 * Level 1:
 ***
locRange[1,1,0]//InputForm
  {1.`45.275162520728024, 1.107842530092498734539217097178882541583986142007270738434`44.80333937734433}
eepProb[1,1,1]//InputForm
  0.0001709610375146190653102477333614649168877199068841857089905399054`45.310246669269084
eemProb[1,1,1]//InputForm
  0.0001709610375146190653102477333614649168877199068841857089905909959`45.310246669269084
eehProb[1,1,1]//InputForm
  0.0009591926252090960195053901354149910709213557835818243810505132591`45.34603693690272
eecProb[1,1,1]//InputForm
  0.0012670397853021707434671541390798939479444812882927754740438772126`45.306946188804055
dloc[1,1,1]//InputForm
  {{"X", {0.0011597206585051100379689307249132725488359118741854843911544092657`46.29347304326884}},
   {"Z", {0.0032738130891607365567582620684012732695044971915021665501224059478`46.61594655314276}},
   {"Y", {0.0011147224220950402675010488531577023675239939571124106609209464267`46.50274521432766}}}
eep[1,1,1]//InputForm
  {{"X", {0.0001709964505834145852748048274933022901351734208471788611903500926`45.31033467070901}},
   {"Z", {0.0000361447162598888446694066229157533630155264047768792027336426983`45.367363227052735}}, {"Y", {0}}}
eem[1,1,1]//InputForm
  {{"X", {0.0001709964505834145852748048274933022901351734208471788611904012797`45.31033467070901}},
   {"Z", {0.0000361447162598888446694066229157533630155264047768792027340946839`45.367363227052735}}, {"Y", {0}}}
eeh[1,1,1]//InputForm
  {{"X", {0.0004469254875620506554884758060161459385195880429714202447787459373`45.34604818832187}},
   {"Z", {0.0004469254875620506554884758060161459385195880429714202447799844017`45.34604818832187}},
   {"Y", {0.000066262583930202854982404922547955565275548781895332895210454486`45.351955456924486}}}
margec[eec[1,1,1]]//InputForm
  {{"X", {0.0008102353821787173279497437451759032290127753916149797595374648641`45.25846008044019}},
   {"Z", {0.000235708669099418290757448797977989432254694427760545973940255058`45.495045042030526}},
   {"Y", {2.7657841877894668585130361721491794777732439757128980989614341`45.20393929838675*^-6}}}
eec[1,1,1]//InputForm
  {{"ZI", {0.0002100501423811829362997602364794089876723900260309245949381438533`45.52361180822297}},
   {"XI", {0.0001780966879119469763995969665137449658119817012237472831002644114`45.30930908628308}},
   {"YI", {1.6096286026205696748075630670175526361231236702422825003965314`45.22815211934781*^-6}},
   {"IZ", {0.0000466189540737213083281572726857580247771361779953994948383286262`45.45693294349431}},
   {"ZZ", {0.0000203941667761623709714644727491072241139952044557608689683305624`45.328086485181885}},
   {"XZ", {1.64928028208994886918208332754836723989173283673724378357439`45.04238021810717*^-8}},
   {"YZ", {9.6928368854206902410036963142982645066549685433850803095483`45.00289369439031*^-9}},
   {"IX", {0.0007165921189346423539829695939295782040940196301481300015459446552`45.262299693004216}},
   {"ZX", {4.7087214050635156258177456638942469464696879032445143875414646`45.267136205722856*^-6}},
   {"XX", {0.0000877949721033461640011965095576137149259207735408929733636095293`45.229003703658876}},
   {"YX", {1.1395697356652943397598960248170630463653000227122702403692149`45.175685794997555*^-6}},
   {"IY", {1.0421539459800180631117072468361977449115209855585670078380904`45.224462052823846*^-6}},
   {"ZY", {5.556385370094678604063430855789735218395093706159956462391777`45.19152752204449*^-7}},
   {"XY", {1.13787333596977416723374739442083938705008167144508346636384`44.99869451866078*^-8}},
   {"YY", {6.8930126181821537045733840002655307781653142149602778861395`44.96787975995433*^-9}}}
 ***
 * Level 2:
 ***
locRange[1,1,1]//InputForm
  {0.999999999999942554162293511232626210547862108103047567014`44.59658045350525,
   1.14318143791905166504208816902845883545931549420424220521`44.059504326634894}
eepProb[1,1,2]//InputForm
  5.315251499480068709553677396659504187743576252069557012863066`44.815297410056274*^-7
eemProb[1,1,2]//InputForm
  5.31525149948006870955367739665950418774357625206955701286312`44.815297410056274*^-7
eehProb[1,1,2]//InputForm
  7.6555944907584717819987334794366414156409236971458354461698917`44.550362584829855*^-6
eecProb[1,1,2]//InputForm
  7.0142044852353781793556505904571221805137391610946551416817703`44.58454450794309*^-6
dloc[1,1,2]//InputForm
  {{"X", {0.0032325604376852864174272832450494651087763191560725484179356158642`46.69949079072594}},
   {"Z", {0.0010873445703331823887219147899428996832189210724745384039966082147`46.629268396653686}},
   {"Y", {0.001086924789532411307137406779083498378068969672217641654579770242`46.638293004890976}}}
eep[1,1,2]//InputForm
  {{"X", {2.576884234206528775437651838373150707039000409957433143829877`44.5625456087433*^-7}},
   {"Z", {5.315255694350927646707936052273770049413862303886550064752916`44.81529784117152*^-7}}, {"Y", {0}}}
eem[1,1,2]//InputForm
  {{"X", {2.576884234206528775437651838373150707039000409957433143829926`44.5625456087433*^-7}},
   {"Z", {5.315255694350927646707936052273770049413862303886550064752969`44.81529784117152*^-7}}, {"Y", {0}}}
eeh[1,1,2]//InputForm
  {{"X", {3.8224619365509015435249769259960688932081383658374220108258137`44.550067204075496*^-6}},
   {"Z", {3.8224619365509015435249769259960688932081383658374220108959818`44.550067204075496*^-6}},
   {"Y", {1.07292262323591155804554671785711470403755328650504902316743`44.843004313018184*^-8}}}
margec[eec[1,1,2]]//InputForm
  {{"X", {2.5570922736965131816543342817892131804136699585817224781028313`44.534011363599326*^-6}},
   {"Z", {3.6298078080049472054491205102240816931223852570468302178681921`44.59937766794925*^-6}},
   {"Y", {1.85228377862911200775741007065087428175894416704978758182`44.49027173683732*^-11}}}
eec[1,1,2]//InputForm
  {{"ZI", {3.1483061095543864281306572988282172735513529817597724677147219`44.573407354799635*^-6}},
   {"XI", {2.661301096313404881750906611076867136391552363850323538089923`44.55532553645983*^-7}},
   {"YI", {9.2213436152871827436744686071232357285599139324190410359`44.428968897610694*^-12}},
   {"IZ", {5.612474455153884323739358843295735013481110601422858641381114`44.790821294923845*^-7}},
   {"ZZ", {4.814497014526702521070417911134384775160750256286341000961906`44.823426507788454*^-7}},
   {"XZ", {1.624901627518890496851002075217849290483663203717811576`44.3531870029713*^-13}},
   {"YZ", {1.381231704826952588966947795754925048449758356328923667`44.363488784816674*^-13}},
   {"IX", {2.4084514311287493882666225638427834438694481431032551271398302`44.532807089841505*^-6}},
   {"ZX", {4.39657229456887079649361091786912171389698026405024502021`44.37615828913918*^-11}},
   {"XX", {1.485910122114034838719588485149897102306174440614913712197435`44.55407556842515*^-7}},
   {"YX", {5.8646334146208077879333222613350964654016143354772930555`44.42707624448988*^-12}},
   {"IY", {1.03210543942918271135468861681959186297494774500214622034`44.48339661649441*^-11}},
   {"ZY", {8.0312749448365034564841732472508378182798557831476070775`44.502838124174936*^-12}},
   {"XY", {9.23611279463378995997989859586194961759764711915992641`44.346129529627284*^-14}},
   {"YY", {7.81473192164516079432423051033668733841319661372072733`44.3572200973633*^-14}}}
 ***
 * Level 3:
 ***
locRange[1,1,2]//InputForm
  {0.999999999999999999991838464246464075052552426273200585171`43.98894315277017,
   1.516802122551022127195933308548600279146236351378630250435`43.44822701208532}
eepProb[1,1,3]//InputForm
  1.163607811384738770356740657275279038280256450242651924746`44.175818926039476*^-10
eemProb[1,1,3]//InputForm
  1.163607811384738770356740657275279038280256450242651924746`44.17581892603946*^-10
eehProb[1,1,3]//InputForm
  7.042685555017418958553957772267703879938646306402475527591`43.9553065960545*^-10
eecProb[1,1,3]//InputForm
  1.0402013054197336886682909015442310871849861285349076408807`43.90608046074788*^-9
dloc[1,1,3]//InputForm
  {{"X", {0.0032287296177652700196408487401926597118042884824815407672854960126`46.76470974917555}},
   {"Z", {0.0010839811536032107839636970066696826107532945270367249428941180993`46.75631397610238}},
   {"Y", {0.0010839810635528342021046710199720671119002488161430316290249393168`46.75632451479937}}}
eep[1,1,3]//InputForm
  {{"X", {5.951633516336626990191044751852731141532296129410459491`43.83725866857438*^-13}},
   {"Z", {1.163607811520829620969145326692654834666225022057150529525`44.17581892609057*^-10}}, {"Y", {0}}}
eem[1,1,3]//InputForm
  {{"X", {5.951633516336626990191044751852731141532296129410459491`43.83725866857438*^-13}},
   {"Z", {1.163607811520829620969145326692654834666225022057150529525`44.17581892609056*^-10}}, {"Y", {0}}}
eeh[1,1,3]//InputForm
  {{"X", {3.521341541983618184662763125830657646065111323280564301905`43.95530649777881*^-10}},
   {"Z", {3.521341541983618184662763125830657646065111323280564301972`43.9553064977788*^-10}},
   {"Y", {2.4760101245754066637506428236285661128319790473927856916911526`44.40587410206958*^-16}}}
margec[eec[1,1,3]]//InputForm
  {{"X", {1.86366351048471363379324116948052348593770259223905244116`43.91980088156824*^-11}},
   {"Z", {9.03058628341891976210680024645864605755661255584199999501`43.88098143707589*^-10}},
   {"Y", {9.28526541232751206777596972984435619540609981489868176278`43.8144911075394*^-21}}}
eec[1,1,3]//InputForm
  {{"ZI", {7.892363689817959655619212438319194027008664347708239635277`43.85086056106122*^-10}},
   {"XI", {6.144947285042168773561080042579229421402405190039789899`43.83453978926002*^-13}},
   {"YI", {1.59153955516832352049884210023155465694627547209326782694`43.69941636393814*^-21}},
   {"IZ", {1.178915483473976838311408959460798818780546560002822883293`44.1671063909827*^-10}},
   {"ZZ", {1.138222593281875569048968459111379005952567656737668493721`44.180259180780375*^-10}},
   {"XZ", {7.2839054836228185038462124580517098456317971373952888`43.669425768971415*^-23}},
   {"YZ", {7.024205514674240868847680815347594382627376758458933858`43.67365349939354*^-23}},
   {"IX", {1.82658980748511581652351148238940256950757116910255820068`43.92129143218286*^-11}},
   {"ZX", {2.751963947133288902879135227316190908330238693525209741793`43.64088970142694*^-20}},
   {"XX", {3.707370014646867166703733339416293771547092624791156383`43.85205781429115*^-13}},
   {"YX", {1.01165198469403450817822751398469588558343983116637498732`43.71466393532193*^-21}},
   {"IY", {4.81017211961477350395169884685377244008598226752131791888`43.811333062425156*^-21}},
   {"ZY", {4.38881427252904587401595018637614605630679966599260611114`43.82102738452907*^-21}},
   {"XY", {4.392260455078755484252530064392485621741090271275772794`43.68103346065706*^-23}},
   {"YY", {4.235641563290513496579539597051284279590697867200000482`43.68547037218291*^-23}}}
*)

}

\end{document}